\documentclass[tradiabstract]{aa} 

\usepackage{graphicx,txfonts}
\usepackage{amsmath}
\usepackage{mathtools}
\usepackage{relsize}

\newcommand{\simgreat} {\mathbin{\lower 3pt\hbox{$\rlap{\raise
        5pt\hbox{$\char'076$}}\mathchar"7218$}}}

\newcommand{\simless}{\mathbin{\lower 3pt\hbox {$\rlap{\raise
        5pt\hbox{$\char'074$}}\mathchar"7218$}}}

\begin{document}

\title{Large Interstellar Polarisation Survey}
\subtitle{II. UV/optical study of cloud-to-cloud variations of dust in the diffuse ISM}

\authorrunning{R.~Siebenmorgen et al.}
\titlerunning{Cloud--to--cloud variations of dust in the diffuse ISM}

\author {R.~Siebenmorgen\inst{1}, N.V. Voshchinnikov\inst{2}, S. Bagnulo\inst{3}, N.L.J.~Cox\inst{4,7}, J. Cami\inst{5,6}, C. Peest\inst{1}}

\institute{European Southern Observatory, Karl-Schwarzschild-Str. 2, D-85748
  Garching b. M\"unchen, Germany, \newline email: Ralf.Siebenmorgen@eso.org
\and
Sobolev Astronomical Institute,
St.~Petersburg University, Universitetskii prosp. 28,
           St.~Petersburg, 198504 Russia
\and
Armagh Observatory and Planetarium, College Hill, Armagh BT61 9DG, UK
\and
Anton Pannekoek Institute for Astronomy, University of Amsterdam,
NL-1090 GE Amsterdam, The Netherlands
\and
Department of Physics and Astronomy and Centre for Planetary Science
and Exploration (CPSX), The University of Western Ontario, London, ON
N6A 3K7, Canada
\and 
SETI Institute, 189 Bernardo Ave, Suite 100, Mountain View, CA 94043,
USA
\and
ACRI-ST, 31000, Toulouse, France
}

\date{Received xxx, 2017 / Accepted xxx, xxx}

\abstract{It is well known that the dust properties of the diffuse
  interstellar medium exhibit variations towards different sight-lines
  on a large scale. We have investigated the variability of the dust
  characteristics on a small scale, and from cloud-to-cloud. We use
  low-resolution spectro-polarimetric data obtained in the context of
  the Large Interstellar Polarisation Survey (LIPS) towards 59
  sight-lines in the Southern Hemisphere, and we fit these data using
  a dust model composed of silicate and carbon particles with sizes
  from the molecular to the sub-micrometre domain. Large ($\geq
  6$\,nm) silicates of prolate shape account for the observed
  polarisation. For 32 sight-lines we complement our data set with
  UVES archive high-resolution spectra, which enable us to establish
  the presence of single-cloud or multiple-clouds towards individual
  sight-lines. We find that the majority of these 35 sight-lines
  intersect two or more clouds, while eight of them are dominated by a
  single absorbing cloud. We confirm several correlations between
  extinction and parameters of the Serkowski law with dust parameters,
  but we also find previously undetected correlations between these
  parameters that are valid only in single-cloud sight-lines.  We find
  that interstellar polarisation from multiple-clouds is smaller than
  from single-cloud sight-lines, showing that the presence of a second
  or more clouds depolarises the incoming radiation. We find large
  variations of the dust characteristics from cloud-to-cloud. However,
  when we average a sufficiently large number of clouds in
  single-cloud or multiple-cloud sight-lines, we always retrieve
  similar mean dust parameters. The typical dust abundances of the
  single-cloud cases are [C]/[H] = 92\,ppm and [Si]/[H] = 20\,ppm.}

\maketitle

\section{Introduction}

In the transition regions between the diffuse and dense interstellar
medium (ISM) of the Milky Way, large variations of the dust properties
are observed \citep{CK}. These variations are theoretically explained
by the fact that dust coagulation and accretion depend on the ambient
density \citep{Koehler12}. However, one does not expect variations of
dust properties within the diffuse ISM, and the Milky Way is often
assumed to be characterised by a ``standard'' extinction curve, which
is due to a ``typical dust mixture'' in the diffuse ISM.  This
standard extinction curve is represented by a constant
total-to-selective extinction ratio of $R_{\rm V} \sim 3.1$
\citep{Morgan53, Cardelli}, and is widely applied for purposes of
de-reddening and foreground removal.

There is observational evidence that the extinction curves of
the diffuse ISM change from sight-line to sight-line
\citep{FM90,FM07}. Such variations can be interpreted in terms of
changes of the chemical composition and sizes of the grains, and of
the various chemical and physical processes that are altering the
dust.  The distribution of $R_{\rm V}$ in a local volume of the
diffuse ISM reveals variations of dust properties on scales larger
than individual clouds \citep{Schlaffy17}. By observing the spectral
variation and spatial morphology of dust extinction curves in the
Milky Way one may reveal the nature of the processes that are
responsible for the variation of the dust properties in such local
environments, and ultimately may help to understand the evolution of
the dust in the Milky Way.

Variations of dust properties in the diffuse ISM have also been
detected at large scales in the far infrared (IR) by the Planck
Collaboration \citep{PlanckXVII,PlanckXXIV,PlanckXXIX}. \citet{Bot09}
found that { in the regions of cirrus clouds where the
  60\,$\mu$m/100\,$\mu$m flux ratio decreases, the
  160\,$\mu$m/100\,$\mu$m flux ratio increases.} These
colour variations cannot be explained by changing the interstellar
radiation field \citep[ISRF,][]{Mathis83}, but are due to significant
changes of the dust properties such as their size distribution, grain
emissivity or mixing of clouds in different physical
conditions. Understanding such variations is important
{ also for an accurate removal of foreground contamination in extra-galactic studies.}

\citet{PlanckXXIX} found that { in certain regions of the diffuse
  ISM, dust extinction decreases, temperature increases, but the
  luminosity per H atom is constant. A decrease of the dust extinction
  and increase of dust temperature could be explained by a change of
  the strength of the ISRF, but this would produce also a change in
  the luminosity, which is not observed.  Therefore,
  a change of the dust properties has been suggested as an explanation.}
Recently, \citet{Ysard15}, who applied the \citet{Jones13}
dust model, could demonstrate that the Planck observation of the
diffuse ISM dust emission $I_{\nu}$ from $100 - 850 \ \mu$m may be fit
by $I_{\nu} = \tau_{\nu_0} \ B_{\nu}(T) \ ({\nu}/{\nu_{0}})^{\beta}$,
where $\tau_{\nu_0}$ is the optical depth at $\nu_0 = 353$\,GHz
(850\,$\mu$m), $T$ is the dust colour temperature, and $\beta$ is the
submillimeter slope.

In this paper we present new information that support the hypothesis
that dust properties are varying within the diffuse ISM and at small
scales from cloud-to-cloud. We use the data of our recent
spectro-polarimetric survey of the interstellar medium
\citep{Bagnulo17}, combined with data from the literature which
provide extinction measurements (Sect.~\ref{obs.sec}). For those stars
for which high-resolution spectroscopic data are available in
astronomical archives, we can disentangle sight-lines with a
single-cloud from sight-lines with multiple component dust clouds. We
then consider the former data set, and we apply the \citet{S14} dust
model to fit simultaneously the extinction and polarisation curve
(Sect.~\ref{dust.sec}). By using the derived dust parameters towards
individual sources of the Large Interstellar Polarisation Survey
(LIPS, Sect.~\ref{res.sec}) we successfully search for correlations
between dust and extinction or polarisation parameters when { either} one
single- or multiple-cloud sight-lines are considered
(Sect.~\ref{res.sec}). Finally, in Sect.~\ref{concl.sec} we summarise
our main findings.

\section{Observational data}\label{obs.sec}

{ We consider the targets observed with FORS2 in
  spectropolarimetric mode (see Sect.~\ref{Sect_Specpol} below)
  for which there exist also measurements of the
  extinction curve (Sect.~\ref{Sect_Ext}).
  This subsample includes 59 sight-lines. For this
subsample we also searched in the ESO archive high-resolution UVES
spectra (Sect.~\ref{Sect_UVES}), which may be used to distinguish single-cloud from
multiple-cloud sight-lines. }

\subsection{Spectro-polarimetry}\label{Sect_Specpol}
To study the properties of interstellar dust in the diffuse ISM, we
have recently obtained spectro-polarimetry data of a large sample of
more than one hundred early-type OB stars in both hemispheres, using the
FORS2 instrument \citep{Appenzeller98} of the ESO VLT for the Southern
Hemisphere, and the ISIS instrument of the William Herschel Telescope
for the survey in the Northern Hemisphere. { The targets of this
Large Intestellar Polarisation Survey (LIPS) are not associated to clouds and are widely
distributed in the galactic disk, except { for} the two high galactic
lattitude stars HD~203532 and HD~210121.} In this paper we use the
polarisation-spectra published by \citet{Bagnulo17} in
the context of the Southern part of LIPS, which { includes} 101 targets
observed in the { wavelength} range 380 --
950\,nm at a resolving power of $\lambda/\Delta\lambda \sim 880$.
{ For the targets with maximum polarisation higher than 0.7\,\%, (76 out of 101),
\citet{Bagnulo17} report the best-fit parameters obtained} using the empirical formula given by
\citet{Serkowski}
\begin{equation}
{p(\lambda)} = {p_{\max}} \ \exp \left[ -k_{\rm p} \ \ln^2
    \left( \frac{\lambda_{\max}}{\lambda} \right) \right]\,,
\label{serk.eq}
\end{equation}
which includes three free parameters: the maximum
polarisation $p_{\max}$, the wavelength $\lambda_{\max}$ at
$p_{\max}$, and the width of the spectrum $k_{\rm p}$.

\subsection{Extinction}\label{Sect_Ext}
For 59 sight-lines of the LIPS { targets observed by \citet{Bagnulo17}}
we have retrieved the extinction
curves in the range 2\,$\mu$m -- 90\,nm { using} the sample assembled by
\citet{Valencic} and \cite{Gordon}. These { common} sight-lines will be
hereafter referred to as the LIPS sample. The dust extinction is
derived using the so-called standard pair method \citep{Stecher},
i.e., by measuring the ratio of the fluxes of pairs of reddened and
unreddened stars with the same spectral type. With this method, the
accuracy of the dust extinction estimate depends critically on how
well the distance to the star is known. Unfortunately, distances to
hot, early-type stars are subject to large errors; therefore one often
prefers to rely on relative measurements by considering the extinction
curve normalised to the value of the extinction in the V band. The
extinction curve $\tau/\tau_{\rm V}$ is then usually reproduced by a
third-order polynomial and a Drude profile, which accounts for the
217\,nm extinction bump. \citet{FM90, FM07, Valencic, Gordon} express
the extinction curve in the range $x = 1/\lambda \geq
3.3\mu\rm{m}^{-1}$ by:
\begin{equation}
\frac{\tau (x)}{\tau_{\rm V}} = c_1 + c_2 \ x + c_3 \ D(x,\gamma, x_0)
+ c_4 \ F(x) \/ ,
\label{ext.eq}
\end{equation}
where $F(x)$, which describes the non-linear UV part of the curve, { is defined as}
\begin{align*}
F(x) & = 0.5392 (x - 5.9)^2+ 0.05644 (x - 5.9)^3&; x \ge
5.9 \mu\rm{m}^{-1}\\
F(x) & = 0 &; x < 5.9 \mu\rm{m}^{-1}
\end{align*}
{ {and $ D(x,\gamma, x_0)$ is the Drude profile
\begin{equation}
  D(x,\gamma, x_0) = \frac{x^2}{(x^2-x_0^2)^2 + (x \ \gamma)^2} \ ,
\end{equation}
with damping constant $\gamma$ and central wavelength $1/x_0$}}. We note
that for 75 sight-lines, \citet{Gordon} have refined the extinction
parametrisation by supplementing data from the International
Ultraviolet Explorer (IUE) at $3.3\,\mu\rm{m}^{-1} < \lambda^{-1} <
8.6\,\mu\rm{m}^{-1}$ with Far Ultraviolet Spectroscopic Explorer
(FUSE) spectra at $3.3\,\mu\rm{m}^{-1} < \lambda^{-1} <
11\,\mu\rm{m}^{-1}$. Following their work, we adopt for the $c_4$
coefficient a value which is $\sim 8$\,\% smaller than that estimated
from the IUE data alone by \citet{Valencic} for stars with no
available FUSE data. At longer wavelengths, specifically for the
$UBVJHK$ bands, we apply the $R_{\rm{V}}$ parametrisation presented by
\citet{Fitzpatrick04}, where $R_{\rm{V}} = A_{\rm{V}} /$E(B-V) is the
ratio of total-to-selective extinction.

\begin{figure*} [h!tb]
\includegraphics[width=18cm,clip=true,trim=2.7cm 3.cm 1cm 3.5cm]{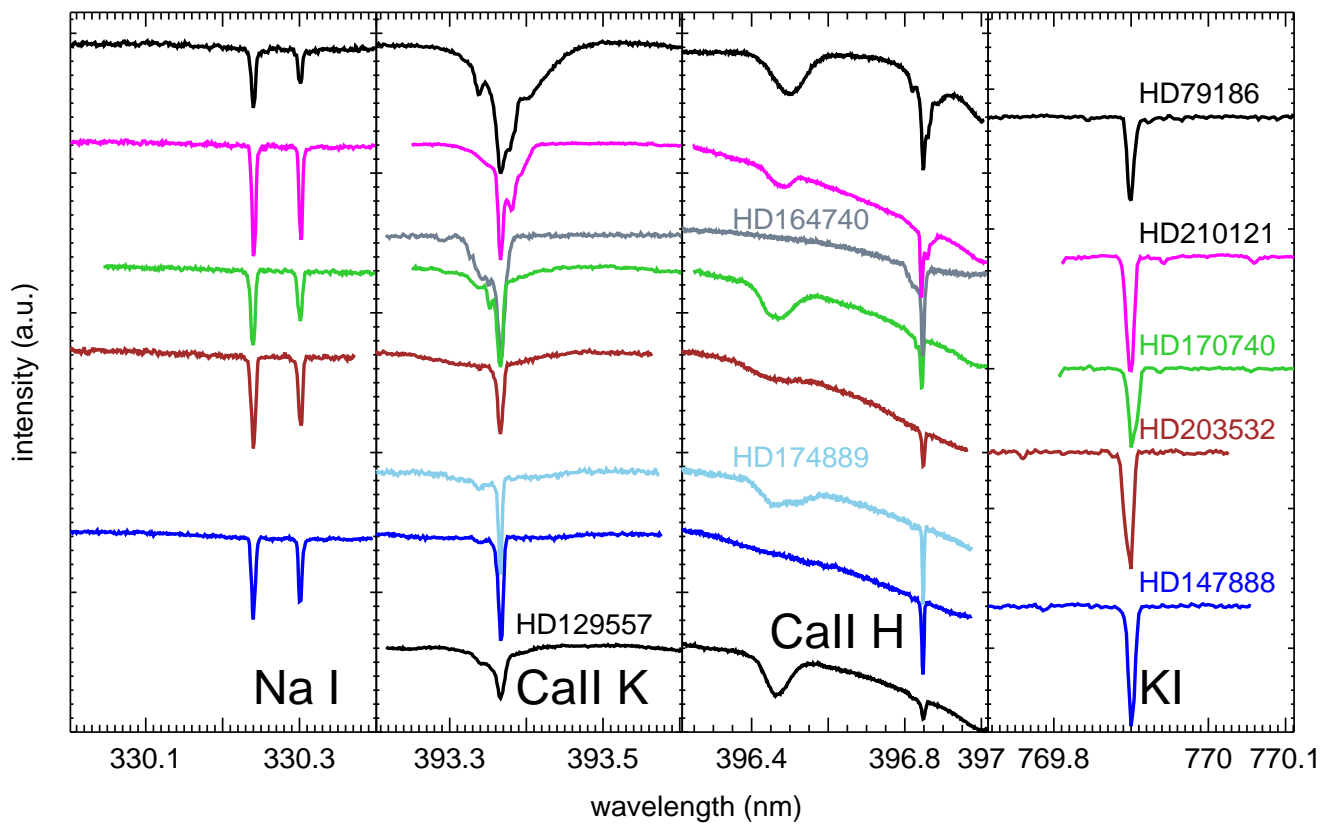}
\caption{Relative intensities of UVES spectra in the Na\,I, Ca\,II~H, Ca\,II~K and K\,I absorption lines for sight-lines with a single-cloud in the
  observing beam \label{1cloud.fig}.}
\end{figure*}

\subsection{Optical high-resolution spectra}\label{Sect_UVES}

We retrieved UVES high-resolution spectra ($\lambda/\Delta \lambda
\sim 10^5$) from the ESO archive for { {32}} of the LIPS sample
stars. UVES (Dekker et al 2000) is an instrument that provides spectra
in the range 300 -- 1100\,nm with a spectral resolution up to
110\,000. For this project, we were primarily interested in the
profiles and shapes of interstellar absorption lines arising from the
diffuse ISM, such as the K and H component of Ca\,II at 393.366 and
396.847\,nm\footnote{all wavelengths in air}; of the Na\,I doublet at
330.237 and 330.298\,nm; and the K\,I line at 769.9\,nm. Velocity
profiles of Ca\,II are usually more complex than those of neutral
species (Na\,I, K\,I). The former traces warmer, more widespread ISM,
while the latter lines probe colder regions \citep{Pan05}.

We followed the method proposed by \citet{Krelowski} and
\citet{Sonnentrucker}, which is also applied to studies of diffuse
interstellar bands \citep{Cami97,Ensor}: we measured the radial
velocities from various spectral lines in the line of sight toward a
given target. If all radial velocities are identical, it means that
the star is observed through just a single-cloud; if instead the
interstellar lines are broadened or split in multiple red or blue
Doppler shifted velocity components, then between us and the target
star there must be two or more interstellar clouds.  Finally we assume
that the gas and dust in the diffuse ISM is well-mixed, so that the
gas distribution traced by the atomic lines is a good proxy of the
dust distribution. In total, we found suitable UVES spectra for { {32}}
stars in our LIPS sample that display one or more interstellar
absorption lines of Ca\,II, Na\,I, and K\,I; { the currently ongoing
  EDIBLES program \citet{Cox} will result in more good candidates in the
  near future.} We report our results in
Table~\ref{Cloud.tab}, where we distinguish the lines into single
(''S''), dominated by a single (''dS''), and multiple component
(''M'') profiles.  Our analysis of UVES spectra confirm the result
previously found by \citet{Krelowski} that the vast majority of
reddened OB stars are observed through two or more interstellar
clouds. However, we found that eight out of the { {59}} LIPS
sight-lines are crossing just a single absorbing dust cloud.
Figure~\ref{1cloud.fig} shows the UVES spectra of these eight stars
observed through a single-cloud { {sight-line}}. These stars will
be subject to our special analysis in Sect.~\ref{res.sec}.

\begin{table} [h!tb]
\begin{center}
  \caption{\label{cloud.tab} Intersections of clouds towards targets.
    We distinguish Na\,I, Ca\,II and K\,I lines into single (''S''),
    dominated by a single (''dS''), and multiple component ( ''M'')
    profiles. { Stars observed through} sight-lines with a single
    dust cloud are { highlighted} with boldface fonts. \label{Cloud.tab}}
\begin{tabular}{|l|ccc|}
\hline
\hline
Target       & Na\,I  & Ca\,II  & K\,I  \\ 
\hline
HD~36982      &S   &M    &S  \\ 
HD~37367      &   -  &dS    & - \\ 
HD~37903      &$M^a$& -   &M  \\ 
HD~38087      &   -  &M    &-  \\ 
HD~75309      &dS    &dS    &dS  \\ 
{\bf HD~79186}      &S   &dS    &S  \\ 
HD~91824      &dS    &M    &dS  \\
HD~91983      &    - &M    & - \\
HD~93205      &M    &M    &M  \\
HD~94493      &dS    &M    &dS  \\
HD~103779     &$M^a$&M    &M  \\
HD~104705     &M   &M   &M  \\
HD~112272     & M   &M    &M \\
HD~122879     &    - &M    &  -\\
{\bf HD~129557}     &   -  &S    &S \\
{\bf HD~147888}    &    - &S    &S  \\
{\bf HD~147889}     &    - &S    & - \\
HD~148379     & M   &     - &M  \\
HD~149404     & M   &    -  &M \\
HD~151804     & M   &M    &M  \\
HD~152235     &  M  &dS    &M  \\
HD~152249     &dS    &M    &dS  \\
HD~152408     & M   &   - &M  \\
HD~152424     &-     &dS    & - \\
HD~154368     &dS    &dS    &M \\
{\bf HD~164740}     &  -   &S    & - \\
HD~169454     & -    &M    &dS \\
{\bf HD~170740}     &S   &dS    &S  \\
{\bf HD~203532}     &S   &S    &S  \\
{\bf HD~210121}     &S   &dS    &S  \\
HD~303308     &  -   &M    & - \\
CPD~63 2495&  -   &M    &M  \\
\hline
\end{tabular}
\end{center}
{\bf{Notes:}} $^a$ Low signal-noise spectra.
\end{table}

\section{Dust model}\label{dust.sec}

To fit the wavelength-dependence of extinction and linear polarisation
of the LIPS sample, we used the dust model from \citet{S14} in which
silicate and carbonaceous dust particles are considered. Extinction
data cannot be reproduced by grains with a single size, therefore we
adopted the well-known power-law size distribution $n(r) \propto
r^{-q}$ \citep[the so-called MRN distribution; see][]{Mathis77} in
addition to polycyclic aromatic hydrocarbon molecules (PAHs) with 150
C atoms and 60 H atoms. Significant contributors to the extinction are
small silicates (sSi) in the FUV, small graphite (gr) and PAH in the
217.5\,nm extinction bump region, large amorphous carbon (aC) with a
nearly constant extinction at $2 \leq x = 1/\lambda \leq 7 \, \mu
\rm{m}^{-1}$, and large silicates (Si) that show a linear increase of
the extinction with $x$ in that range.

\citet{Draine16} pointed out that the rapid fall-off of the
polarisation of starlight in the far-UV suggests that small ($r <
6$\,nm) grains are nearly spherical and do not contribute to the
far-UV polarisation.  On the other hand, interstellar polarisation
cannot be explained by spherical particles made of optically isotropic
material. There must be large, partly aligned, and non-spherical
grains. We consider spheroids as simple examples. They come in two
flavours: prolates, which are obtained by a rotation of an ellipse
around the major axis $a$ (e.g. like a rugbyball); or oblates, which
are obtained by a rotation of an ellipse around the minor axis $b$
(i.e. more disk-like). Unless the degree of the polarisation is very
high, prolates provide better fits to polarisation spectra than
oblates \citep{V14, S14, V16}. We use prolates with $a/b =2$; their
volume is the same of a sphere with radius $r=(ab^2)^{1/3}$. Following
\citet{Hirashita} we employ different upper size limits $r_{+}^{\rm
  {aC}}$ and $r_{+}^{\rm {Si}}$ for large amorphous carbon and
silicate grains.

To compute the scattering and absorption efficiencies of the
spheroidal dust grains we have used a solution to the light scattering
problem given by \citet{vf93}. We have computed cross-sections of
prolate aC and Si particles for 100 values of their radius, in the
interval from 6 to 800\,nm. For aC grains, we have adopted optical
constants of the ACH2 hydrogenated amorphous carbon particle mixture
by \citet{Zubko} with bulk density of 1.6\,g\,cm$^{-3}$
\citep{Furton99, Robertson, Casiraghi}. For silicates, we have
considered optical constants by \citet{Draine03} and a density of
3.5\,g\,cm$^{-3}$.  Formulas for computing the various extinction and
polarisation cross sections $K_{\rm ext}$ and $K_{\rm p}$ are given by
\citet{S14}.  The extinction curve for each star is fit by

\begin{equation}
  {\tau(x) \over \tau_{\rm V}}  = {K_{\rm ext}(x) \over K_{\rm ext,\,V}} \ ,
\end{equation}

\noindent
where $K_{\rm ext}$ is the extinction cross section of the dust
averaged over sizes and rotations in units cm$^2/$g-ISM dust.
Distances of our sample stars are small enough (generally $\simless
1$\,kpc) so that systematic uncertainties due to incoherent scattering
can be ignored \citep{Scicluna}. The observed linear polarisation is
fit by

\begin{equation}
  {p(\lambda) \over p_{\rm {max}}} = {K_{\rm {p}}(\lambda) \over K_{\rm p}(\lambda_{\rm {max}})} \ ,
\label{ptau.eq}
\end{equation}

\noindent
where $K_{\rm p}$ is the linear polarisation cross section of aligned
silicates averaged over sizes and rotations in units cm$^2/$g-ISM
dust. { {We do not need to consider alignment of carbon particles
    when fitting the data.  The polarisation strength depends
    critically on the axial ratio of the spheroids $a/b$, the
    alignment efficiency $\delta_0$ of the assumed imperfect
    Davis--Greenstein alignment, and the magnetic field orientation
    $\Omega$. However, none of these parameters have a significant
    impact on the spectral shape of the polarization curve \citep{V12,
      S14, V16}. Therefore we can simplify our modeling efforts and do
    not fit the absolute value of $p_{\max}$. We scale the
    polarisation spectrum of the dust model to the data using $p_{\rm
      {max}}$/$ K_{\rm p}(\lambda_{\rm {max}})$ as scaling parameter
    (Eq.~\ref{ptau.eq}).  We apply the same choice of
    parameters as in \citet{S14}, namely prolate spheroids with axial
    ratio $a/b =2$, $\delta_0=10\mu$m, and $\Omega=90^{\rm {o}}$. We
    need to introduce a minimum radius of aligned silicate $r_{-}^{\rm
      {pol}}$ when fitting a polarisation curve \citep{Draine09,
      Das}.}}

For each dust component, its specific mass (or its relative abundance)
needs to be specified.  Scaling these abundances up or down by a
constant factor does not change the predicted extinction or
polarisation curve.  For direct comparison with cosmic abundance
constraints \citep{Draine11} we set as normalisation [Si]/[H] =
15\,ppm in large silicate grains, which is at the low end of $15 \leq
\rm{[Si]/[H]} \leq 31.4$\,(ppm) reported by \citet{VH10}. The rest of
Si is locked up in small silicates.

The amount of [C]/[H] depleted in dust is difficult to estimate and
there are various numbers in the literature.  The interstellar
absorption feature at 3.4\,$\mu$m can be explained by hydrogenated
amorphous carbon with $72 \leq \rm{[C]/[H]} \leq 97$\,(ppm)
\citep{Duley, Furton99}, when using a gas phase abundance by
\citet{Sofia04} $ \rm{[C]/[H]} = 84 \pm 23$\,ppm is estimated by
\citet{Nieva}, for the same reference abundance a median $\rm{[C]/[H]}
= 102 \pm 47$\,ppm is derived by \citet{Parvathi}, and assuming that
half of C is depleted into grains [C]/[H] as low as $25 \leq
\rm{[C]/[H]} \leq 120$\,(ppm) is found by \citet{Gerin}. We will see
below that the carbon abundance used in the dust models is in
good agreement with such estimates.

In addition depletion of Fe and O on dust is openly discussed
\citep{Dwek16, Jenkins, Koehler14} as well as the details of the
silicate mineralogy \citep{H10}.


\begin{figure*} [h!tb]
\includegraphics[width=17cm,clip=true,trim=1.8cm 2.5cm 1.2cm 1.5cm]{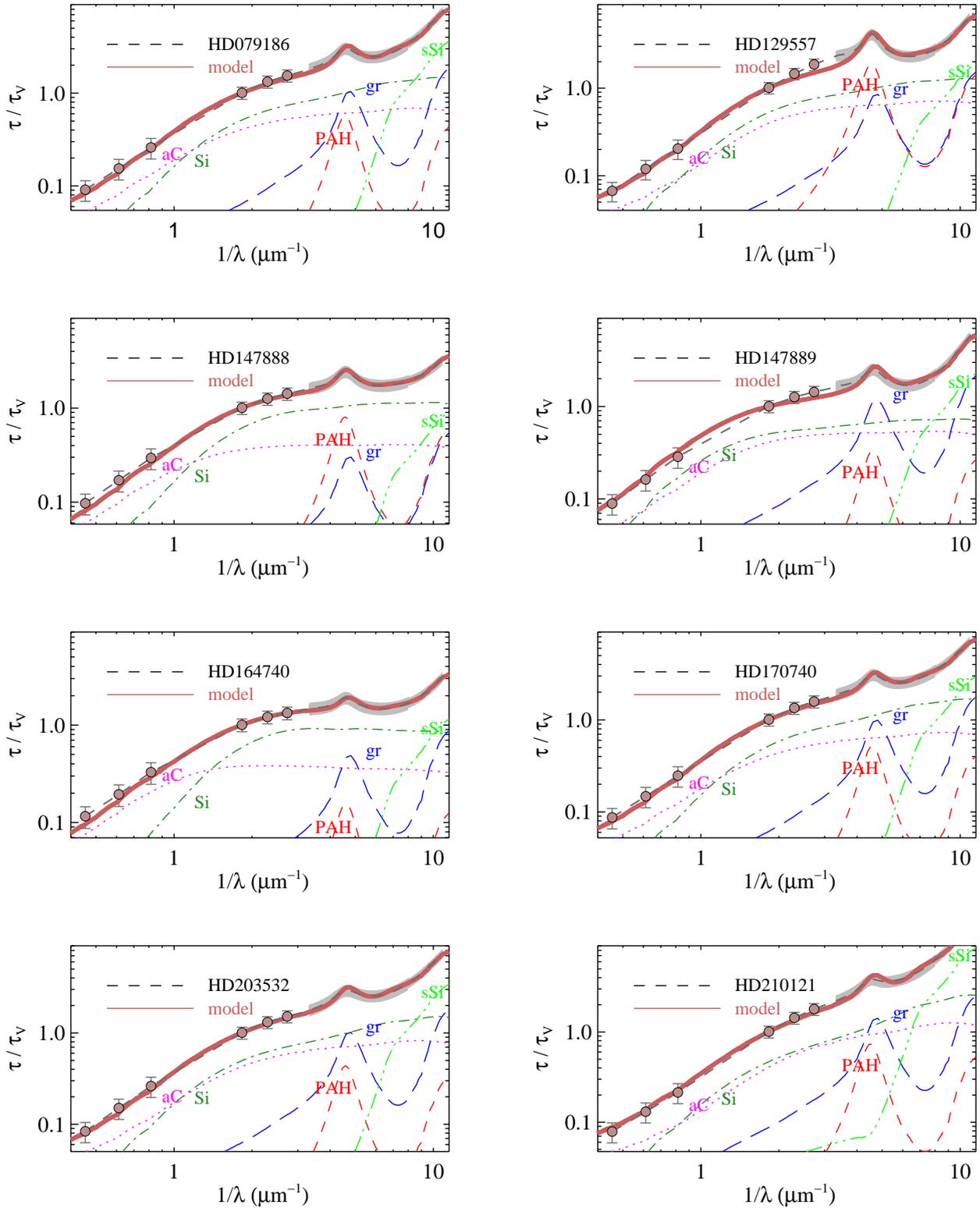}
\caption{Observed and modeled extinction curve of single-cloud
  sight-lines. { {The extinction curve by
  \citet{Fitzpatrick04}, \citet{Valencic} and \citet{Gordon} is shown as dashed line,
  $UBVJHK$ photometry as filled circles, IUE/FUSE spectra are
  represented by the grey shaded area, and uncertainties are
  1\,$\sigma$.}}  The model is shown (brown solid lines), as well as the contribution
  of the different dust populations to the extinction (solid
  lines { of various colours as labelled in the panels}.) \label{1cl_ext.fig}}
\end{figure*}

\begin{figure*} [h!tb]
\includegraphics[width=17cm,clip=true,trim=1.8cm 2.5cm 1.2cm 1.5cm]{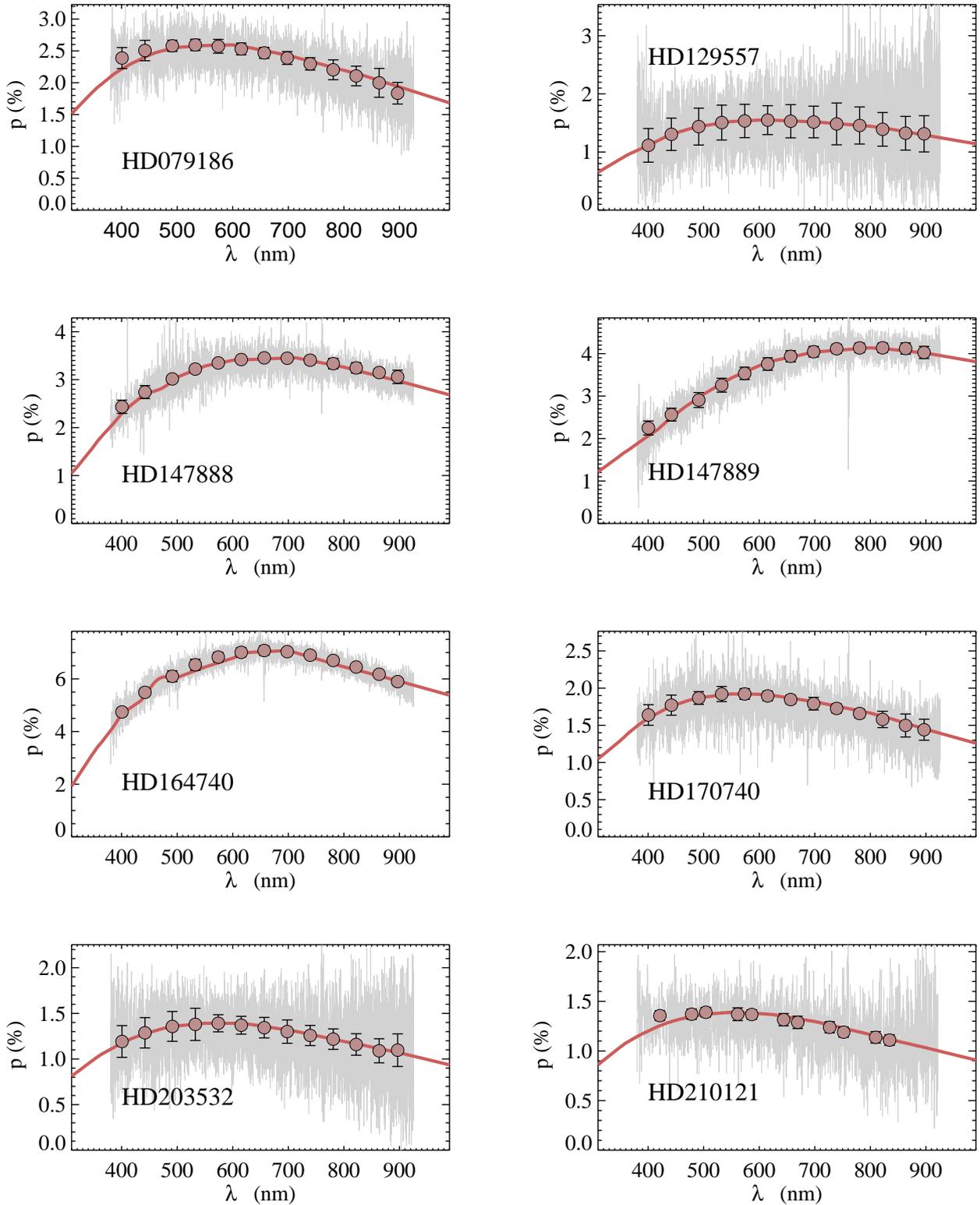}

\caption{Polarisation curves of single-cloud sight-lines. { The polarisation
  spectra obtained with FORS2 are shown with grey lines; the filled circles
  (with 1\,$\sigma$ error bars) show the same data rebinned to a spectral
  resolution of $\lambda/\Delta \lambda \sim 50$. The best-fit obtained with
  our dust model is shown with a brown solid line.\label{1cl_pol.fig}}}
\end{figure*}

\begin{table*}[!htb]
\scriptsize
\begin{center}
  \caption {Best-fit dust parameters with 1\,$\sigma$ uncertainty {
      {and polarisation scaling parameter (Eq.~\ref{ptau.eq})}.
The stars corresponding to single-cloud sight-lines are highlighted with bold face fonts.}
    For each parameter the median with $1\,\sigma$ error of the
    single-cloud and LIPS sample is given at the
    bottom. \label{para.tab}}
 \begin{tabular}{l c c c c c c c c c }
\hline\hline
 & & & & & & & & & \\
Target & [Si]/[H]$_{\rm {tot}}$& [C]/[H]$_{\rm {tot}}$ & [C]/[H]$_{\rm {gr}}$& [C]/[H]$_{\rm {PAH}}$ & $q$ & $r_{\rm{pol}}^{-}$ & $r^{+}_{\rm {Si}}$ & $r^{+}_{\rm {aC}}$  & $\frac{p_{\rm {max}}}{K_{\rm p}(\lambda_{\rm {max}})}$ \\
       & (ppm)& (ppm)& (ppm)& (ppm)& & (nm)& (nm)& (nm) & \\
\hline 
BD~134920& $   20 \pm   0.2$&$   98 \pm    13$&$   16 \pm   1.5$&$     13 \pm   1.3$&$ 3.02 \pm  0.02$&$  118 \pm     6$&$  309 \pm    29$&$   339 \pm    32$&$  0.85$  \\
CD~285205& $   22 \pm   0.3$&$   80 \pm    10$&$   11 \pm   0.9$&$     12 \pm   1.0$&$ 2.79 \pm  0.02$&$  174 \pm     9$&$  231 \pm    13$&$   328 \pm    19$&$  0.57$  \\
CP~632495& $   21 \pm   0.2$&$   81 \pm     8$&$    9 \pm   0.6$&$      8 \pm   0.6$&$ 2.40 \pm  0.02$&$  166 \pm     8$&$  220 \pm    11$&$   314 \pm    16$&$  1.01$  \\
HD~36982& $   18 \pm   0.1$&$   84 \pm     7$&$   10 \pm   0.4$&$      4 \pm   0.3$&$ 1.82 \pm  0.03$&$  158 \pm    45$&$  237 \pm    12$&$   403 \pm    20$&$  0.14$  \\
HD~37021& $   15 \pm   0.1$&$  211 \pm     9$&$    1 \pm   0.1$&$     10 \pm   0.4$&$ 2.80 \pm  0.01$&$  166 \pm    18$&$  438 \pm    22$&$   308 \pm    15$&$  0.15$  \\
HD~37367& $   18 \pm   0.2$&$  106 \pm    10$&$    8 \pm   0.6$&$     18 \pm   0.9$&$ 3.11 \pm  0.01$&$  107 \pm     5$&$  422 \pm    21$&$   294 \pm    15$&$  0.32$  \\
HD~37903& $   20 \pm   0.2$&$   80 \pm     7$&$   14 \pm   0.6$&$      6 \pm   0.4$&$ 2.56 \pm  0.02$&$  166 \pm     8$&$  268 \pm    19$&$   364 \pm    25$&$  0.56$  \\
HD~38087& $   17 \pm   0.1$&$   75 \pm     6$&$    7 \pm   0.4$&$     11 \pm   0.6$&$ 2.31 \pm  0.02$&$  107 \pm     5$&$  296 \pm    15$&$   421 \pm    21$&$  0.44$  \\
HD~45314& $   18 \pm   0.1$&$   70 \pm     9$&$    3 \pm   0.3$&$      6 \pm   0.5$&$ 2.75 \pm  0.01$&$  183 \pm     9$&$  258 \pm    13$&$   462 \pm    23$&$  0.52$  \\
HD~73882& $   20 \pm   0.2$&$   86 \pm     8$&$   13 \pm   0.6$&$      4 \pm   0.3$&$ 3.13 \pm  0.01$&$  130 \pm     6$&$  415 \pm    21$&$   288 \pm    14$&$  0.72$  \\
HD~75309& $   19 \pm   0.2$&$  139 \pm    13$&$   10 \pm   0.6$&$     18 \pm   0.9$&$ 3.27 \pm  0.01$&$   84 \pm    20$&$  424 \pm    21$&$   312 \pm    16$&$  0.18$  \\
{\bf {HD~79186}}& $   23 \pm   0.4$&$   94 \pm     9$&$   18 \pm   1.0$&$     10 \pm   0.6$&$ 3.04 \pm  0.02$&$  107 \pm    10$&$  311 \pm    26$&$   317 \pm    26$&$  0.73$  \\
HD~89137& $   18 \pm   0.2$&$   95 \pm    12$&$   15 \pm   1.1$&$      8 \pm   0.7$&$ 3.37 \pm  0.01$&$  158 \pm    16$&$  397 \pm    20$&$   266 \pm    13$&$  0.22$  \\
HD~91824& $   23 \pm   0.2$&$   78 \pm     7$&$   15 \pm   0.6$&$      5 \pm   0.3$&$ 2.27 \pm  0.02$&$  130 \pm     6$&$  202 \pm    10$&$   307 \pm    15$&$  0.32$  \\
HD~91983& $   22 \pm   0.3$&$   89 \pm     9$&$   19 \pm   1.2$&$     10 \pm   0.7$&$ 2.91 \pm  0.02$&$  130 \pm     6$&$  247 \pm    23$&$   266 \pm    25$&$  0.34$  \\
HD~93160& $   20 \pm   0.2$&$   74 \pm    14$&$    7 \pm   0.9$&$      6 \pm   0.9$&$ 2.82 \pm  0.02$&$   84 \pm     5$&$  328 \pm    17$&$   424 \pm    21$&$  0.44$  \\
HD~93205& $   21 \pm   0.3$&$   78 \pm     8$&$   13 \pm   0.7$&$      7 \pm   0.5$&$ 2.99 \pm  0.02$&$  124 \pm     6$&$  263 \pm    16$&$   313 \pm    19$&$  0.54$  \\
HD~93222& $   19 \pm   0.1$&$   75 \pm     9$&$    5 \pm   0.4$&$      7 \pm   0.6$&$ 2.53 \pm  0.02$&$   88 \pm    11$&$  265 \pm    13$&$   420 \pm    21$&$  0.13$  \\
HD~93632& $   18 \pm   0.1$&$   76 \pm     7$&$    6 \pm   0.4$&$      9 \pm   0.5$&$ 2.90 \pm  0.01$&$  233 \pm    12$&$  305 \pm    16$&$   412 \pm    21$&$  0.95$  \\
HD~93843& $   25 \pm   0.3$&$  141 \pm    19$&$   20 \pm   0.8$&$      0 \pm   0.1$&$ 1.99 \pm  0.04$&$  118 \pm    18$&$  179 \pm     9$&$   266 \pm    13$&$  0.14$  \\
HD~94493& $   20 \pm   0.2$&$   77 \pm     8$&$    8 \pm   0.6$&$      9 \pm   0.6$&$ 2.54 \pm  0.02$&$   84 \pm    57$&$  233 \pm    13$&$   330 \pm    18$&$  0.09$  \\
HD~96715& $   27 \pm   0.5$&$  125 \pm    12$&$   33 \pm   1.4$&$      8 \pm   0.6$&$ 3.37 \pm  0.01$&$  124 \pm     6$&$  353 \pm    18$&$   243 \pm    12$&$  0.97$  \\
HD~97484& $   26 \pm   0.4$&$   88 \pm     9$&$   21 \pm   1.2$&$      9 \pm   0.7$&$ 2.79 \pm  0.02$&$  143 \pm     8$&$  212 \pm    35$&$   204 \pm    34$&$  0.43$  \\
HD~99953& $   20 \pm   0.2$&$   84 \pm     9$&$   12 \pm   0.9$&$      9 \pm   0.7$&$ 2.79 \pm  0.02$&$  107 \pm     6$&$  263 \pm    22$&$   371 \pm    31$&$  0.41$  \\
HD~103779& $   23 \pm   0.3$&$  108 \pm    10$&$    9 \pm   0.6$&$     17 \pm   0.9$&$ 2.51 \pm  0.03$&$  130 \pm    10$&$  199 \pm    10$&$   317 \pm    16$&$  0.18$  \\
HD~104705& $   22 \pm   0.3$&$   87 \pm     8$&$   11 \pm   0.7$&$     17 \pm   0.9$&$ 2.67 \pm  0.02$&$  192 \pm    10$&$  202 \pm    10$&$   324 \pm    16$&$  1.09$  \\
HD~111934& $   19 \pm   0.2$&$   87 \pm     9$&$   10 \pm   0.7$&$     20 \pm   1.0$&$ 3.12 \pm  0.01$&$  124 \pm     6$&$  280 \pm    14$&$   210 \pm    10$&$  0.73$  \\
HD~112272& $   20 \pm   0.2$&$   86 \pm     8$&$   11 \pm   0.7$&$     17 \pm   1.0$&$ 2.73 \pm  0.02$&$  143 \pm    14$&$  238 \pm    15$&$   328 \pm    20$&$  0.25$  \\
HD~116852& $   16 \pm   0.1$&$  146 \pm    30$&$    3 \pm   0.5$&$     23 \pm   1.2$&$ 3.64 \pm  0.01$&$  107 \pm    17$&$  319 \pm    16$&$   225 \pm    11$&$  0.43$  \\
HD~122879& $   20 \pm   0.2$&$   77 \pm     8$&$    5 \pm   0.4$&$     14 \pm   0.8$&$ 2.49 \pm  0.02$&$  166 \pm     8$&$  202 \pm    10$&$   331 \pm    17$&$  0.73$  \\
{\bf {HD~129557}}& $   19 \pm   0.2$&$  111 \pm    10$&$   15 \pm   0.9$&$     35 \pm   1.5$&$ 2.87 \pm  0.01$&$  136 \pm     7$&$  291 \pm   113$&$   228 \pm    89$&$  0.46$  \\
HD~134591& $   22 \pm   0.2$&$   87 \pm     8$&$    7 \pm   0.5$&$     11 \pm   0.6$&$ 2.34 \pm  0.03$&$  150 \pm    68$&$  164 \pm     8$&$   245 \pm    12$&$  0.16$  \\
{\bf {HD~147888}}& $   17 \pm   0.1$&$   77 \pm     7$&$    5 \pm   0.3$&$     14 \pm   0.8$&$ 2.49 \pm  0.02$&$  174 \pm     9$&$  263 \pm    18$&$   368 \pm    25$&$  0.99$  \\
{\bf {HD~147889}}& $   20 \pm   0.2$&$   96 \pm     7$&$   23 \pm   0.8$&$      6 \pm   0.3$&$ 2.61 \pm  0.02$&$  143 \pm     7$&$  409 \pm    20$&$   293 \pm    15$&$  1.31$  \\
HD~148379& $   19 \pm   0.2$&$  107 \pm    16$&$   12 \pm   1.1$&$     13 \pm   1.2$&$ 3.30 \pm  0.01$&$  107 \pm    19$&$  441 \pm    22$&$   294 \pm    15$&$  0.63$  \\
HD~149404& $   20 \pm   0.2$&$   97 \pm    10$&$   15 \pm   0.9$&$     12 \pm   0.7$&$ 3.03 \pm  0.01$&$  102 \pm    20$&$  365 \pm    18$&$   273 \pm    14$&$  0.80$  \\
HD~151804& $   18 \pm   0.1$&$   79 \pm     9$&$    8 \pm   0.6$&$     10 \pm   0.7$&$ 2.81 \pm  0.01$&$  107 \pm     6$&$  325 \pm    16$&$   417 \pm    21$&$  0.24$  \\
HD~151805& $   20 \pm   0.2$&$   87 \pm     8$&$    9 \pm   0.5$&$     18 \pm   0.9$&$ 2.89 \pm  0.02$&$  130 \pm     6$&$  269 \pm    18$&$   351 \pm    24$&$  0.17$  \\
HD~152235& $   25 \pm   0.4$&$  152 \pm    13$&$   23 \pm   1.1$&$      8 \pm   0.5$&$ 2.55 \pm  0.03$&$  136 \pm    13$&$  167 \pm     8$&$   246 \pm    12$&$  0.45$  \\
HD~152248& $   21 \pm   0.2$&$  115 \pm    11$&$    5 \pm   0.4$&$     16 \pm   0.8$&$ 2.69 \pm  0.02$&$  102 \pm    10$&$  210 \pm    10$&$   304 \pm    15$&$  0.10$  \\
HD~152249& $   19 \pm   0.2$&$   79 \pm     9$&$    9 \pm   0.7$&$     11 \pm   0.8$&$ 2.96 \pm  0.01$&$  192 \pm    10$&$  263 \pm    18$&$   359 \pm    24$&$  0.14$  \\
HD~152408& $   21 \pm   0.2$&$   78 \pm     8$&$    9 \pm   0.6$&$     11 \pm   0.7$&$ 2.25 \pm  0.02$&$  183 \pm    12$&$  236 \pm    12$&$   358 \pm    18$&$  0.34$  \\
HD~152424& $   22 \pm   0.2$&$  108 \pm    11$&$   11 \pm   0.8$&$     10 \pm   0.7$&$ 2.04 \pm  0.03$&$   62 \pm    10$&$  181 \pm     9$&$   260 \pm    13$&$  0.04$  \\
HD~153919& $   21 \pm   0.2$&$   86 \pm     9$&$   15 \pm   0.9$&$      9 \pm   0.6$&$ 2.80 \pm  0.02$&$  102 \pm    25$&$  342 \pm    35$&$   311 \pm    32$&$  0.62$  \\
HD~154368& $   23 \pm   0.3$&$   82 \pm     8$&$   15 \pm   0.9$&$      9 \pm   0.6$&$ 2.69 \pm  0.02$&$   69 \pm    21$&$  239 \pm    12$&$   353 \pm    18$&$  0.07$  \\
HD~163181& $   20 \pm   0.3$&$  129 \pm    14$&$   11 \pm   0.6$&$     18 \pm   1.2$&$ 3.28 \pm  0.01$&$   92 \pm     5$&$  434 \pm    22$&$   297 \pm    15$&$  0.49$  \\
HD~164073& $   16 \pm   0.1$&$  104 \pm    11$&$    5 \pm   0.4$&$     13 \pm   0.7$&$ 3.11 \pm  0.01$&$  124 \pm     9$&$  362 \pm    18$&$   257 \pm    13$&$  0.31$  \\
{\bf {HD~164740}}& $   18 \pm   0.1$&$   85 \pm     8$&$    9 \pm   0.4$&$      3 \pm   0.2$&$ 1.95 \pm  0.02$&$  183 \pm     9$&$  235 \pm    12$&$   386 \pm    19$&$  2.55$  \\
HD~167838& $   21 \pm   0.2$&$   82 \pm     8$&$   13 \pm   0.7$&$      9 \pm   0.6$&$ 2.39 \pm  0.02$&$  136 \pm    16$&$  202 \pm    10$&$   316 \pm    16$&$  0.08$  \\
HD~168076& $   17 \pm   0.1$&$   90 \pm     8$&$    6 \pm   0.4$&$     13 \pm   0.7$&$ 3.19 \pm  0.01$&$  124 \pm     6$&$  312 \pm    29$&$   389 \pm    36$&$  0.87$  \\
HD~169454& $   22 \pm   0.3$&$   86 \pm     9$&$   17 \pm   0.8$&$      6 \pm   0.4$&$ 3.16 \pm  0.01$&$  102 \pm    25$&$  413 \pm    21$&$   277 \pm    14$&$  0.65$  \\
{\bf {HD~170740}}& $   21 \pm   0.3$&$   87 \pm     8$&$   17 \pm   1.0$&$      9 \pm   0.6$&$ 3.10 \pm  0.01$&$  124 \pm     6$&$  291 \pm    27$&$   303 \pm    28$&$  0.55$  \\
{\bf {HD~203532}}& $   23 \pm   0.3$&$   94 \pm     9$&$   18 \pm   0.8$&$      8 \pm   0.6$&$ 3.12 \pm  0.01$&$  107 \pm     5$&$  328 \pm    90$&$   278 \pm    77$&$  0.40$  \\
{\bf {HD~210121}}& $   31 \pm   0.6$&$   95 \pm    10$&$   21 \pm   1.2$&$     11 \pm   0.8$&$ 3.51 \pm  0.01$&$  107 \pm     5$&$  335 \pm    17$&$   224 \pm    11$&$  0.68$  \\
HD~251204& $   22 \pm   0.3$&$  106 \pm    11$&$   20 \pm   1.1$&$     10 \pm   0.8$&$ 3.18 \pm  0.01$&$  102 \pm     5$&$  387 \pm    19$&$   269 \pm    13$&$  1.54$  \\
HD~252325& $   19 \pm   0.2$&$  113 \pm    14$&$   12 \pm   1.0$&$     13 \pm   1.0$&$ 3.01 \pm  0.01$&$  102 \pm     6$&$  381 \pm    19$&$   268 \pm    13$&$  1.41$  \\
HD~303308& $   21 \pm   0.2$&$   84 \pm    10$&$   13 \pm   0.9$&$     11 \pm   0.9$&$ 3.09 \pm  0.02$&$  124 \pm     6$&$  269 \pm    22$&$   339 \pm    28$&$  0.78$  \\
HD~315023& $   19 \pm   0.2$&$  112 \pm    11$&$    8 \pm   0.5$&$     15 \pm   0.9$&$ 2.95 \pm  0.01$&$  130 \pm     6$&$  414 \pm    21$&$   311 \pm    16$&$  0.44$  \\
LS~-~908& $   22 \pm   0.3$&$   84 \pm     7$&$   11 \pm   0.6$&$     15 \pm   0.8$&$ 2.75 \pm  0.02$&$  150 \pm    10$&$  218 \pm    16$&$   257 \pm    19$&$  0.69$  \\
\hline
single-cloud &$20 \pm 2$   &$94 \pm 10$&$17\pm 6$ &$ 10 \pm 9$&$3.0 \pm 0.4 $&$136 \pm 28 $&$303 \pm 50 $&$316 \pm 157 $&$ 0.96 \pm 0.66$ \\
all &$20 \pm 3$ &$87 \pm 24$&$11\pm 6$ &$10 \pm 6$  &$2.8 \pm 0.4 $&$124 \pm 36 $&$312 \pm 58 $&$374 \pm 162 $ & $0.55 \pm 0.43$\\
\hline
\end{tabular}
\end{center} 
\end{table*}

\begin{figure*} [!htb]
\centering
\includegraphics[width=17cm,clip=true,trim=2.2cm 3.8cm 1.95cm 4cm]{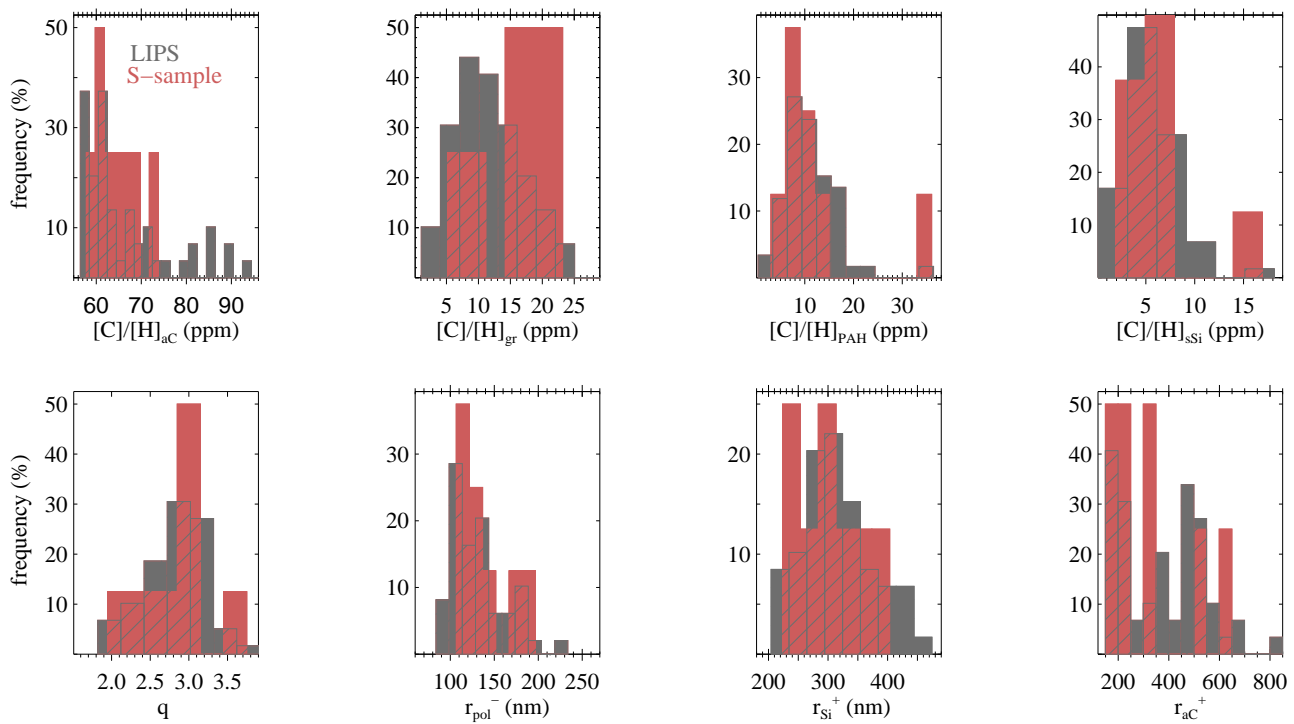}
\caption{Histograms of best-fit parameters for all stars of the LIPS
  sample { {(shaded area in grey) and the single-cloud sight-lines
      (shaded area in red). The intersection of both samples is shown
      as red-grey hatched area. }} In the top panels we show (from
  left to right) the dust abundances [C]/[H]$_{\rm {aC}}$,
  [C]/[H]$_{\rm {gr}}$, [C]/[H]$_{\rm {PAH}}$, and [Si]/[H]$_{\rm
    {sSi}}$, and in the bottom panels $q$, $r_{-}^{\rm {pol}}$,
  $r_{+}^{\rm {Si}}$, and $r_{+}^{\rm {aC}}$,
  respectively. \label{param.fig}}
\end{figure*}

\begin{figure} [!htbp]
\centering
\includegraphics[width=8.8cm,clip=true,trim=4.cm 1.5cm 3cm 2cm]{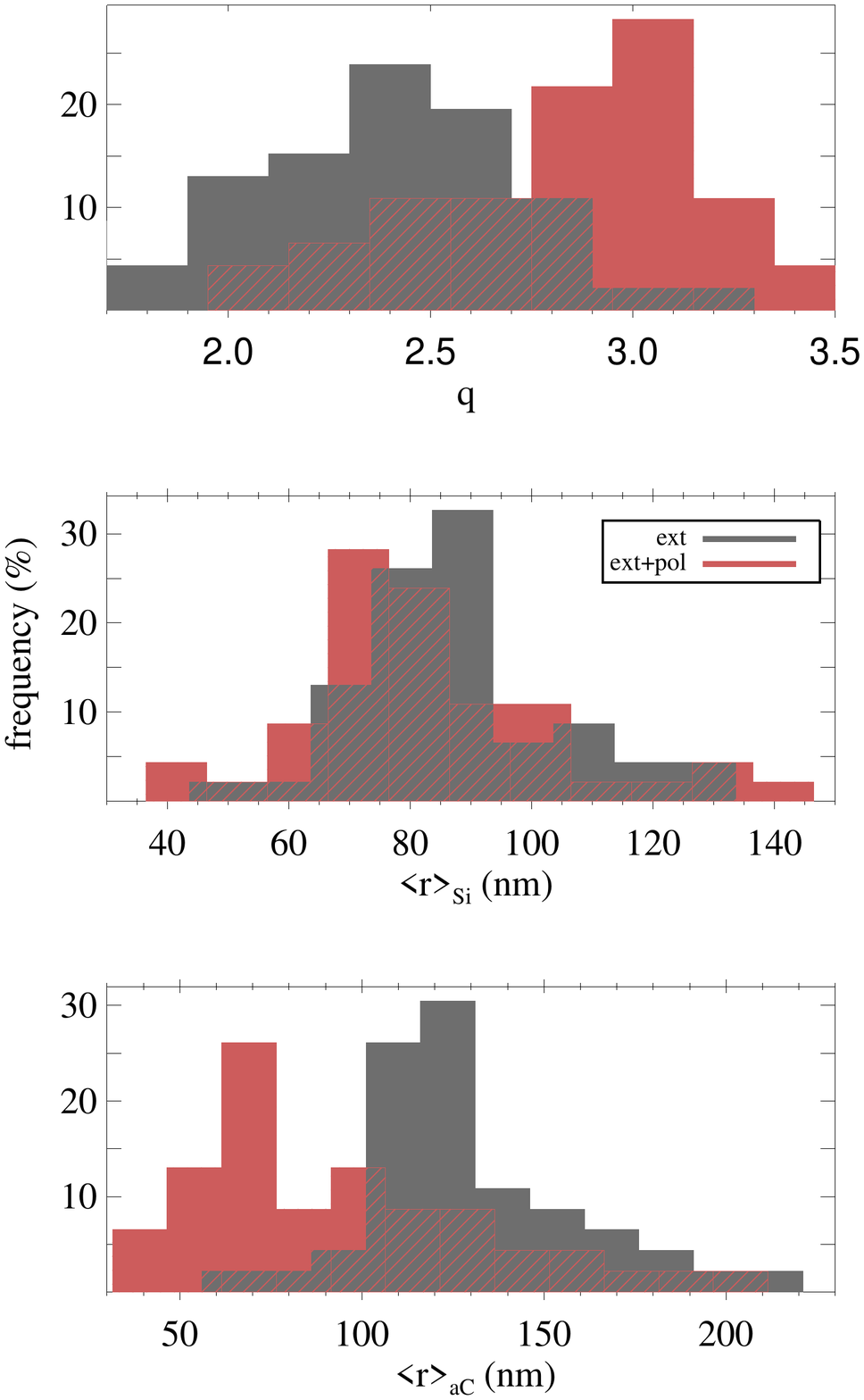}
\caption{Histograms of the exponent of the size distribution (top) and
  mean sizes of silicates (middle) and amorphous carbon (bottom) for
  the LIPS sample when fitting extinction and polarisation {
    {(shaded area in red) and extinction only (shaded area in
      grey). The intersection of both samples is shown as grey-red
      hatched area.}}\label{paraext.fig}}
\end{figure}

\section{Results \label{res.sec}}

We fit extinction and polarisation spectra of the LIPS sample applying
the procedure described by \citet{S17}, which is based on a {
  {minimum $\chi^2$}} technique utilizing the Levenberg--Marquardt
algorithm as implemented in MPFIT \footnote{http://purl.com/net/mpfit}
\citep{Markwardt09}. { {In the $\chi^2$-fit the extinction and
    polarisation data are treated with the same weight.}}

There are in total eight free parameters: four
size parameters $q$, $r_{+}^{\rm {Si}}$, $r_{+}^{\rm {aC}}$,
$r_{-}^{\rm {pol}}$, and one abundance for each dust component:
[C]/[H]$_{\rm {aC}}$, [C]/[H]$_{\rm {gr}}$, [C]/[H]$_{\rm {PAH}}$,
[Si]/[H]$_{\rm {sSi}}$, we remind that our normalisation is
[Si]/[H]$_{\rm {Si}} = 15$\, ppm.  The abundance is converted into
weight or specific mass $w_i = m_i$ of each dust component $i$ as in
Eq.~(15) by \citet{S14}. We included a slight improvement in the
fitting procedure by adjusting the center wavelength of the Drude
profile of the PAH cross section in the 217.5\,nm bump region to the
observed extinction peak.  The best-fit parameters with $1\,\sigma$
uncertainties are listed in
Table~\ref{para.tab}. Figure~\ref{1cl_ext.fig} shows the best-fit to
the extinction curves for the single-cloud sight-lines, together with
the contributions of the different dust populations, and
Fig.~\ref{1cl_pol.fig} shows the corresponding fits to the
polarisation curves. The best-fits of the other cases are shown in the
appendix. In the following we discuss the cloud-to-cloud variations of
the dust parameters and correlations of the dust parameters with
observing characteristics for extinction $c_1$, $c_2$, $c_3$, $c_4$,
$R_{\rm V}$, and polarisation $k_{\rm p}$, $\lambda_{\rm {max}}$, and
$p_{\rm {max}}$. We will see that generally the extinction and
polarisation curve of the LIPS sources are well fit by the dust model,
that the abundances in the dust model are consistent with present
cosmic abundance constraints, and that single-cloud sight-lines show
correlations that are { not present} in multiple-cloud cases.


\subsection{Parameter study}

The distribution of the best-fit parameters are displayed in
Fig.~\ref{param.fig} for the stars of the LIPS sample (grey) and the
single-cloud sight-lines (red).  Single-cloud
sight-lines have median values with 1\,$\sigma$ variations of the dust
abundances of [C]/[H]$_{\rm {tot}} = 94 \pm 10$\,ppm and
[Si]/[H]$_{\rm {tot}} = 20 \pm 2$\,ppm (Table~\ref{para.tab}). They
are in good agreement with the above estimates of the cosmic
abundances in dust. For some individual sources there are noticeable
outliers.  Examples are the translucent sight-line towards HD~147888
in the $\rho$~Orph complex and HD~37021 in the Orion nebulae.  For
HD~147888 \citet{VH10} finds a total [Si]/[H] $= 30 \pm 1.5$\,ppm in
dust. Applying this value increases [C]/[H]$_{\rm {tot}}$ to 169\,ppm,
in dust which would equal the gas phase abundance of [C]/[H]$_{\rm
  {gas}} = 169 \pm 38$\,ppm derived by \citet{Sofia04}. For HD~37021
we derive [C]/[H]$_{\rm {tot}} = 211 \pm 18$\,ppm and this value would
even double when applying [Si]/[H]$_{\rm {tot}} = 30 \pm 1.5$\,ppm
\citep{VH10}. It is certainly above the available C abundance as
estimated by \citet{Sofia04}. Strikingly the latter authors find
$R_{\rm V} = 4.6$ while we apply $R_{\rm V} = 5.84$ following
\citet{Valencic}. Extinction curves for both stars are displayed in
Fig.~\ref{1cl_ext.fig} and \ref{HD037021.fig}. For the [C]/[Si]
abundance ratio we find a lower bound of 3.9 as derived towards
HD~45314 (Table~\ref{para.tab}).

We find for single-cloud sight-lines typical size parameters of $q =
3.0 \pm 0.4$, $r_{+}^{\rm {Si}} = 303 \pm 50$\,nm, $r_{+}^{\rm {aC}} =
316 \pm 157$\,nm, and $r_{-}^{\rm {pol}} = 136 \pm 28$\,nm
(Table~\ref{para.tab}).  By considering the full sample one finds
parameter distributions that are wider but within 1\,$\sigma$ similar
median values, e.g.  $r_{+}^{\rm {aC}} = 374 \pm 162$\,nm
(Fig.~\ref{param.fig}). Sizes of large silicates are further
constrained by the polarisation spectra so that one finds a somewhat
larger scatter in the upper limit to the aC grain size $r_{+}^{\rm
  {aC}}$.

{ We remove from our LIPS sample all known single-clouds cases
  (eight), and consider in total 51 stars. This new subsample, will be
  called hereafter ``L-sample''. It includes all multiple-clouds
  cases, plus six cases which we cannot determine if they are single
  or multiple sight-lines.} For these 51 sight-lines we verify if
their average dust properties differ from the eight single-cloud
cases. To perform this check, we have taken $10^5$ random samples of 8
stars selected out of these 51 stars. For each of these elements we
have computed the median of the best-fit parameters.  We find within
1\,$\sigma$ the same median dust properties as computed for the LIPS
sample. By attempting to find average dust properties similar values
are derived for single- or multiple-cloud sight-lines.  Such average
dust properties will be always observed either by mixing of several
clouds in single-cloud sight-lines or by mixing of clouds in
multiple-cloud sight-lines. Exceptions are the abundances of the small
grains and the upper grain size of carbon dust $r_{+}^{\rm {aC}}$,
which are also the parameters affected by the largest errors. Note the
large cloud-to-cloud variations in each of the dust parameters when
comparing individual cases (Fig.~\ref{param.fig}).

Finally, we checked how the estimate of dust parameters change if we
neglect the constraint from polarimetric
measurements. Figure~\ref{paraext.fig} shows the histograms of $q$ and
mean sizes of large silicates $\langle r \rangle_{\rm {Si}}$ and amorphous carbon
$\langle r \rangle_{\rm {aC}}$. Mean sizes are computed by averaging over the dust
size distribution { {from 6\,nm to $r_{+}$}}. Histograms are shown
for these parameters when each target of the LIPS sample is either
simultaneously fitting the extinction and polarisation curve (red) or
the extinction curve only (grey). The median of the dust parameters
when fitting extinction and polarisation are $q = 3.0 \pm 0.4$,
$\langle r \rangle_{\rm {Si}} = 81\pm 21$\,nm, and $\langle r \rangle_{\rm {aC}}= 95 \pm 21$\,nm,
and when fitting extinction only $q= 2.4 \pm 0.4$, $\langle r \rangle_{\rm {Si}} =
85 \pm 16$\,nm, and $\langle r \rangle_{\rm {aC}} = 125 \pm 29$\,nm.  By deriving
dust properties only from extinction and ignoring polarisation
measurements, one retrieves a flatter dust size distribution and
larger mean grain sizes than when polarimetric measurements are
included in the analysis.


\subsection{Correlation study}

As a first step for a physical interpretation of the observations we
searched for relations between the dust model parameters and the
observing characteristics for extinction $c_1$, $c_2$, $c_3$, $c_4$,
$R_{\rm V}$, and polarisation $k_{\rm p}$, $\lambda_{\rm {max}}$.
{ {The linear Pearson correlation coefficient $-1 \leq \rho \leq 1$
    can be taken as a measure of the correlation strength. We assume
    that for $\vert{\rho}\vert \simgreat 0.8$ a correlation could
    exist that we further investigate by other means.  The existence
    of a trend in both data is tested by employing three different
    straight-line fits $y = a \cdot x + b$, where 1\,$\sigma$ errors
    $(\Delta x,\Delta y)$ in both coordinates $(x,y)$ are considered
    \citep[for a discussion of issues concerning straight-line fits see][]{Hogg10}.
    A first and most common applied treatment is the
    minimum $\chi^2$ technique called fitexy by \citet{Press}.  A
    second procedure is a principal component analysis (PCA). In PCA
    one computes eigenvectors and eigenvalues of the covariance
    matrix. The eigenvalues are renormalised so that their sum equals
    1. The eigenvector that belongs to the renormalised eigenvalue
    with the highest value is taken as the principal component and
    provides our fit parameters $a$, $b$. We include in PCA the
    measurement uncertainties by drawing 10,000 random samples that
    are consistent with the 1\,$\sigma$ error of the data and assuming
    Gaussian noise distribution.  For each of such samples we perform
    a PCA and derive the probability density function (PDF) of $a$ and
    $b$.  We are most interested in the slope of the potential
    correlation so that we take the value of $a$ where its
    distribution function peaks, and take the corresponding value of
    $b$. Finally, we apply the Bayesian maximum likelihood estimator
    (MLE) as provided by \citet{Kelly07}. The later provides also the
    PDF of the fitting parameters.  The range where 68\,\% of the
    parameter is scattered around its peak of the PDF is quoted as
    uncertainty.  MLE outperforms the $\chi^2$ and PCA estimators
    whenever the measurement uncertainties are not strictly Gaussian
    functions \citep{Hogg10}. For low values of $\vert{\rho}\vert$ the
    slope parameters $a$ derived by the $\chi^2$, PCA, and MLE might
    become arbitrary, while for $\vert{\rho}\vert \simgreat 0.8$ more
    consistent slopes are found. We consider that a correlation exists
    when all three regression estimators provide a similar trend in
    the data. It is then still important to consider the issue of
    outliers. Random samples that include a single extreme value in
    the abscissa result in high $\vert{\rho}\vert$ values. Such
    extreme data are better removed in the correlation study.

    We compare correlations of single-cloud sight-lines with different
    samples of multiple-cloud (dominated) sight-lines.  We recall that
    we have defined as S-sample the eight LIPS targets observed
    through single-cloud sight-lines, as identified through UVES
    data. The 24 multiple-cloud sight-lines as identified thanks to
    UVES data represent the M-sample.  The complete 59 LIPS objects
    excluding the eight stars of the S-sample (51 multiple-cloud
    dominated sight-lines) is called L-sample. For a proper comparison
    of the correlations, we will also consider the targets from
    M-sample that have $c_i$ and $R_V$ parameters in the same range as
    the stars of the S-sample. By so-doing, we have selected a
    sub-subsample of multiple-cloud sight-lines that we call M$_{\rm
      S}$. We do the same kind of selection for the L-sample, and we
    define the L$_{\rm S}$ sample as the list of LIPS targets that are
    not identified as single-clouds sight-lines, but that have
    parameters $c_i$ and $R_V$ in the same range as the S-sample
    stars.  In the observed extinction characteristics, the M-sample
    has a narrower range than the S-sample.  For example, the M-sample
    is observed between $0.87 \leq c_1 \leq 1.6$ and $2.8 \leq R_{\rm
      V} \leq 4.9$, whereas the S-sample is between $0.75 \leq c_1
    \leq 2.3$ and $2.3 \leq R_{\rm V} \leq 5$.  Finally, we consider
    further sub-samples of M$_{\rm S}$ and L$_{\rm S}$, by considering
    the stars for which the ratios $p_{\rm{max}}/ A_{\rm V}$ and
    either $k_p$ or $\lambda_{\rm max}$ are in the same range as the
    stars of the S-sample.  We will call these subsamples M$_{\rm
      {SP}}$ and L$_{\rm {SP}}$.

    Pearson's correlation coefficients $\rho$ between dust model and
    extinction or Serkowski parameters are given for the single-cloud
    and the other samples in Table~\ref{corrext.tab} and
    Table~\ref{corrpol.tab}. The high latitude star HD~210121 has
    peculiar far UV extinction \citep{WD01} with extreme values in the
    extinction parameters. Therefore it is excluded from the S-sample,
    except for the correlation study with $R_{\rm V}$, where HD~210121
    has a normal behaviour. For large $\vert \rho\vert$ we show data
    in {Figs.~\ref{pl_corC1.fig} -- \ref{pl_corPol.fig}}. Data
    considered in the correlation study are marked by filled
    symbols. Samples with the strongest $\vert \rho\vert$ are fit by
    the $\chi^2$, PCA, and MLE method. In Table~\ref{regress.tab} we
    report straight-line parameters derived by MLE with error
    estimates and confidence that the fit is not due to selection bias
    (Sect.~\ref{bias.sec}). }}

\begin{table}[!htbp]

\begin{center}
  \caption {Pearsons' correlation coefficient between dust
    model and the extinction curve parameters $c_1$, $c_2$, $c_3$, $c_4$ of Eq.~(2)
    and $R_{\rm {V}}$ for various LIPS sub-samples. Strong correlations
    { {or strong anti-correlations}} are marked in bold. \label{corrext.tab}}
 \begin{tabular}{clrrrrr}
\hline  \hline 
 &                & & &  & &\\                     
Para.           &Sample & $c_1$  &$c_2$  &$c_3$ &$c_4$  &$R_{\rm V}$ \\
  &               & & &  & &\\                     
\hline
$q$                   &  S  &  --0.33  & 0.74  &\bf{0.87}& 0.74   & \bf {--0.87} \\
              &M&  --0.61  & 0.65  &--0.51&0.05   &--0.45 \\   
              &L$_{\rm S}$&  --0.47  & 0.39  &0.26&0.11   & --0.42 \\   
                      &  L&  --0.47  & 0.48  &0.26&--0.06   & --0.37 \\   
  &               & & &  & &\\ 
$r^{+}_{\rm {Si}}$ & S&--0.31 &--0.12 &\bf{-0.83}& \bf{--0.79}  & \bf {0.89} \\
              &M& 0.46  &--0.64 &--0.57&--0.34   & 0.71 \\   
              &L$_{\rm S}$& 0.06 &--0.37  &--0.29&--0.47   & 0.79 \\   
                      &  L&0.06    &--0.50&--0.32&--0.39 & 0.69 \\  
  &               & & &  & &\\                     
$\langle r \rangle_{\rm {Si}}$      &  S  &   0.38  &\bf{--0.77}&--0.68& --0.38  & 0.77\\
              &M&\bf{0.84}  &\bf{--0.95} &--0.67&--0.20  & \bf{0.89} \\
              &L$_{\rm S}$& 0.58  & \bf{--0.87}  &-0.48&--0.30   & \bf{0.81} \\
                        &  L&  0.58   &\bf{--0.88}&--0.55& --0.17  &\bf{0.83} \\ 
  &               & & &  & &\\                     
$\langle r \rangle_{\rm {aC}}$      &  S  &  0.14   &--0.59     &\bf{-0.91}& \bf{--0.82} &\bf{0.92} \\
              &M& 0.67 &\bf{--0.81} &--0.64&--0.18   &0.68 \\   
              &L$_{\rm S}$& 0.41  &--0.54  &--0.39&--0.27   & 0.67 \\   
              &  L& 0.40    & --0.60    &-0.40& --0.13   & 0.58\\
  &               & & &  & &\\                     
$\left(\frac{m_{\rm {Si}}}{m_{\rm {C}}}\right)_{\rm {tot}}$ 
                      &  S  &\bf{--0.94} &\bf{0.87} &0.12&  0.09   & -0.32 \\
              &M&  0.13  &--0.09  &--0.36&0.44   & --0.04 \\   
              &L$_{\rm S}$&  --0.11  & 0.17  &--0.03&0.38   & 0.11 \\                         &  L&0.11     &  0.08       &0.09&  0.53  & --0.13 \\     &               & & &  & &\\                     
$\frac{m_{\rm {vsg}}}{m_{\rm {lgr}}}$ 
                      &  S  &--0.26   & 0.66    &\bf{0.84}&\bf{0.94}&  --0.72 \\
                     &M   &  --0.61  & 0.63  &0.46&0.76   &--0.68 \\   
              &L$_{\rm S}$&  --0.17  & 0.55  &0.41&\bf{0.85}   & --0.44 \\                         &  L&--0.17   & 0.50      &0.48&\bf{0.89}& --0.53 \\        
                  
\hline
\end{tabular}
\end{center}
\tablefoot{ $q$ is the exponent of the dust-grain size-distribution,
  $r^+$ is the upper radius of large silicates.  $\langle r
  \rangle_{\rm Si}$ and $\langle r \rangle_{\rm aC}$ are the mean
  radius of large silicates and carbon grains, respectively, $m_{\rm
    vsg}/m_{\rm lg}$ and $(m_{\rm Si}/m_{\rm C})_{\rm tot}$ are the
  mass ratios of the very small to large grains and of total silicate
  to carbon, respectively. Samples are defined as follows.\\
  \quad {\it {\/ -- \/ S-sample:}} Single-cloud sight-lines (Table~\ref{Cloud.tab};
  HD~210121 is excluded in the correlation study with extinction
  parameters but is considered in $R_V$).\\
  \quad   {\it { \/ -- \/ M-sample:}} Multiple-cloud sight-lines, including 24 stars of
  Table~\ref{Cloud.tab}. The M-sample has a narrower range in the
  observed extinction characteristics than the S-sample.\\
  \quad   {\it {\/ -- \/  L-sample:}} LIPS objects that are not single-cloud
  sight-lines, including 51 stars.\\
  \quad   {\it { \/ -- \/ L$_{\rm S}$-sample:}} Targets that are not seen through
  single-cloud sight-lines (i.e., that belong to the L-sample), but
  with extinction parameter range similar to the S-sample.  This
  sample includes between 47 and 51 stars.  }
\end{table}

\begin{table}[!htbp]

\begin{center}
  \caption {Pearsons' correlation coefficient $\rho$ between dust and
    Serkowski parameters $k_{\rm P}$, $\lambda_{\rm {max}}$ of various
    LIPS sub-samples.  Strong correlations { {or strong
        anti-correlations}} are marked in bold. \label{corrpol.tab}}
 \begin{tabular}{c l r r}
\hline \hline
                    &        &                   & \\                     
Parameter           & Sample & $k_{\rm P}$  &$\lambda_{\rm {max}}$  \\
                    &  &   & \\                     
\hline
$q$                 & S             &\bf{-0.81} &\bf{-0.84}  \\
                    & M$_{\rm {SP}}$  &-0.39      &-0.06  \\
                    & M$_{\rm S}$   &-0.70      & 0.11  \\
                    & M             & 0.36      & 0.10  \\
                    & L$_{\rm {SP}}$  &-0.25      &-0.04  \\
                    & L             &-0.25      &-0.01  \\
                    &   & &  \\                     
$r^{-}_{\rm {pol}}$ & S             &\bf{0.85}  & \bf{-0.94}   \\
                    & M$_{\rm {SP}}$  &$^{\dagger}$\bf{0.91} & 0.65  \\
                    & M$_{\rm S}$   & 0.51      & 0.52  \\
                    & M             & 0.24      & 0.03  \\
                    & L$_{\rm {SP}}$  & 0.74      & 0.53  \\
                    & L             & 0.26      & 0.01   \\
                    &   & &   \\                     
$\langle r \rangle ^{\rm {Si}}_{\rm {pol}}$    & S & 0.22      &\bf{0.95}  \\
                    & M$_{\rm {SP}}$  &-0.27      &$^{\dagger}$\bf{0.81}   \\
                    & M$_{\rm S}$   &-0.39      & 0.74  \\
                    & M             &-0.17      & 0.09      \\
                    & L$_{\rm {SP}}$  &-0.22      & 0.77      \\
                    & L             &-0.07      & 0.17      \\
\hline
\end{tabular}
\end{center} 
\tablefoot{$q$ is the exponent of the dust-grain size-distribution,
  $r^{-}_{\rm {pol}}$ is the minimum radius of aligned silicates,
  $\langle r \rangle ^{\rm {Si}}_{\rm {pol}}$ is the mean radius of
  aligned silicates.  Samples are defined as follows.\\
  \quad   {\it {\/ -- \/ S-sample:}} Single-cloud sight-lines (Table~\ref{Cloud.tab}). HD\,210121 is excluded.  For the correlation studies with $\lambda_{\rm {max}}$ we consider only 6 stars with  $\lambda_{\rm {max}} \leq 700$\,nm.\\
  \quad   {\it {M-sample:}}Multiple-cloud sight-lines with 24 stars as of Table~\ref{Cloud.tab}.\\
  \quad   {\it {\/ -- \/ M$_{\rm {SP}}$-sample: }}Stars belonging to the M-sample that
  have the ratio $p_{\rm{max}}/ A_{\rm V}$ in the observed range of
  the S-sample, and either $k_{\rm P}$ or $\lambda_{\rm {max}}$ also
  in the range of the S-sample.\\ $^{\dagger}$ The M$_{\rm
  \quad     {SP}}$-sample has a narrower range in the observing parameters
  than the S-sample. \\
  \quad   {\it {\/ -- \/ M$_{\rm S}$-sample:}} M-sample that are in the observed range of the S-sample either in $k_{\rm P}$ or $\lambda_{\rm {max}}$.\\
  \quad   {\it {\/ -- \/ L-sample:}} LIPS objects that are not single-cloud cases, including 51 stars.\\
  \quad    {\it {\/ -- \/ L$_{\rm {SP}}$-sample:}} L-sample that are in the observed range of the S sample
 in $p_{\rm{max}}/ A_{\rm V}$ and either $k_{\rm P}$ or $\lambda_{\rm {max}}$.
}
\end{table}

\subsubsection{Model versus extinction parameters}


\begin{figure} [!htbp]
\includegraphics[width=8.0cm,clip=true,trim=4.5cm 1.6cm 2.5cm 2.6cm]{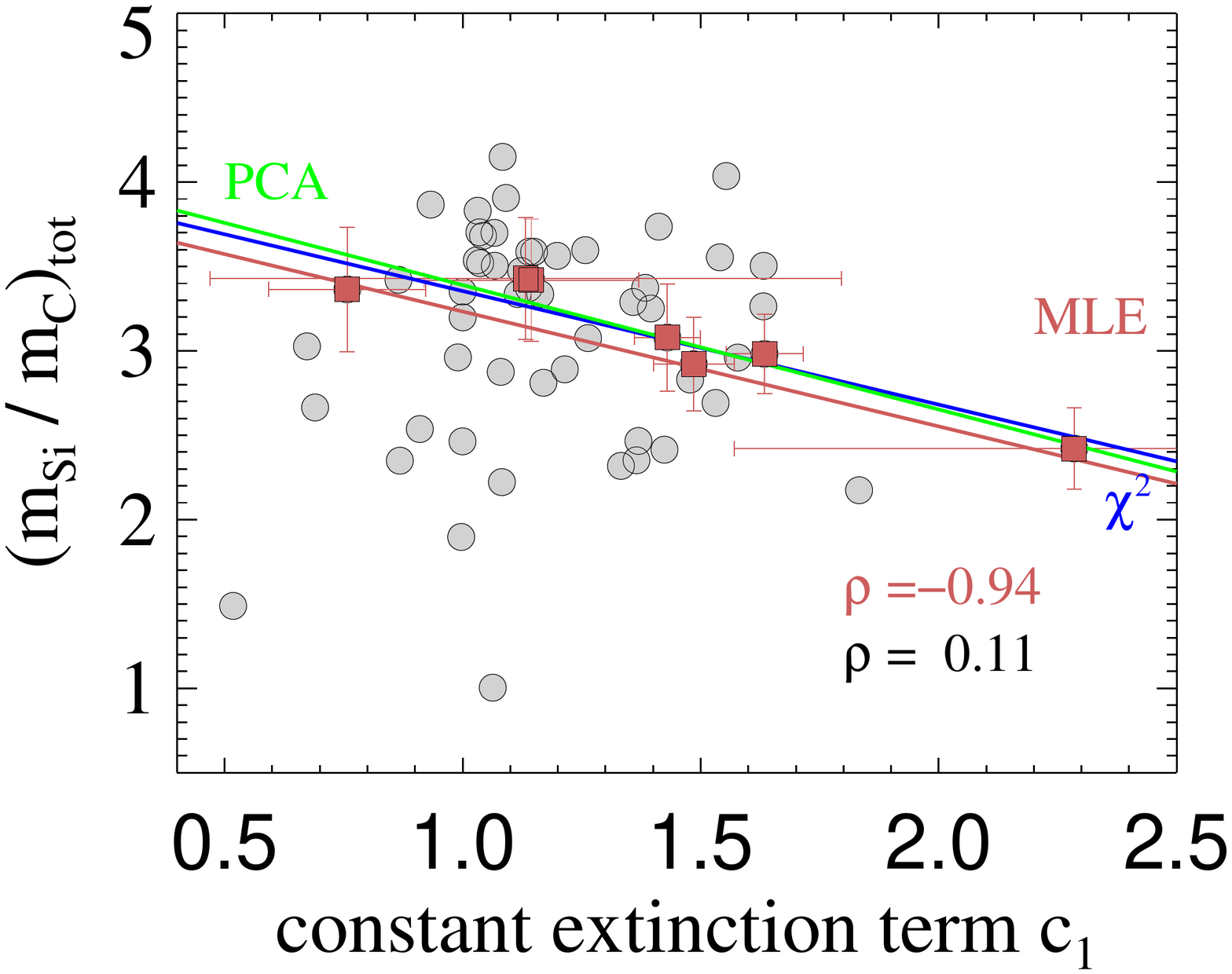}
\includegraphics[width=8.0cm,clip=true,trim=4.5cm 1.6cm 2.5cm 2.6cm]{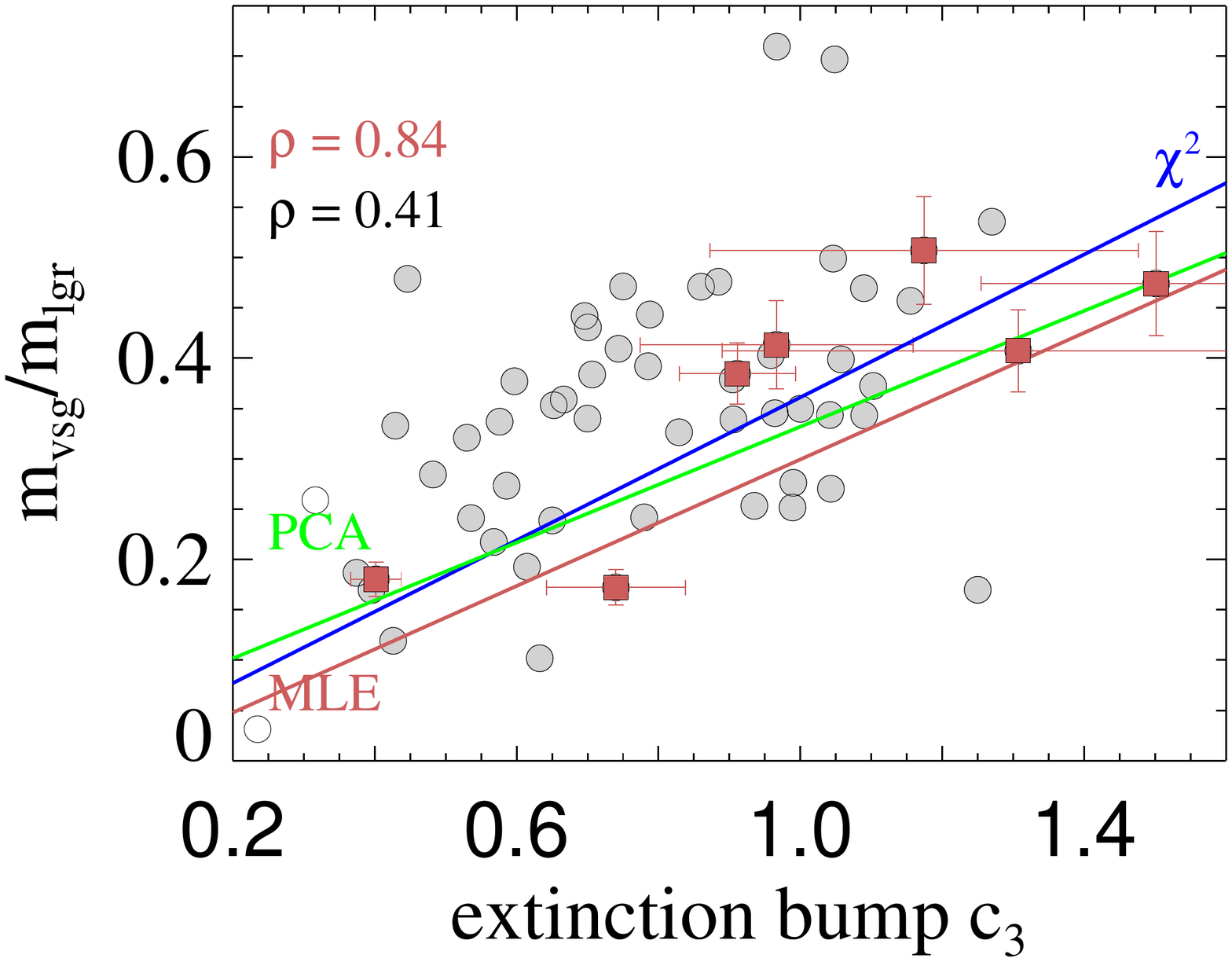}
\includegraphics[width=8.0cm,clip=true,trim=4.5cm 1.6cm 2.5cm 2.6cm]{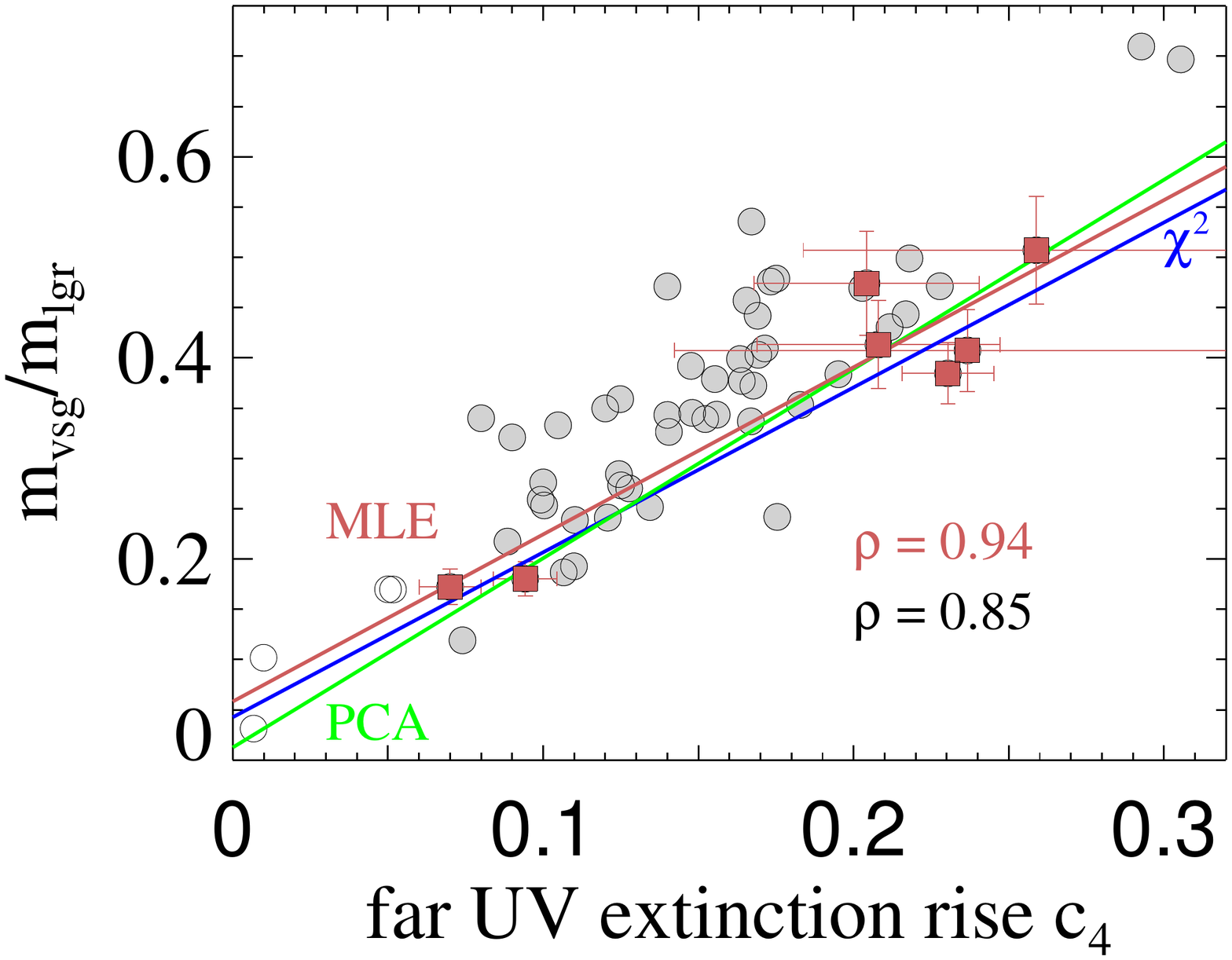}
\caption{\label{pl_corC1.fig} 
  { {\it Top panel:} Mass ratio of silicate and carbon dust versus
  the constant term $c_1$ of Eq.~(2) in observations of the extinction curve
  between 115 - 330\,nm \citep{Valencic,Gordon}.
  {\it Mid panel:} Mass ratio of very small ($r < 6\,{\rm {nm}}$) and
  large grains versus the parameter $c_3$ of Eq.~(2); the parameter $c_3$
  describes the strength of the extinction bump.
  {\it Bottom panel:}  Mass ratio of very small ($r < 6\,{\rm {nm}}$) and large
  grains versus the parameter $c_4$ of Eq.~(2), which describes the strength of the
  far UV rise.
   The L-sample is shown with open circles and the
      L$_{\rm S}$-sample with grey filled circles together with their
      Pearson coefficient (black).  Single-cloud sight-lines (red squares)
      with 1\,$\sigma$ error bars and Pearson coefficient are shown in
      red. They are fit by straight lines employing the MLE (red solid line), PCA
      (green solid line), and minimum $\chi^2$ (blue solid line)
      method.} 
  }
\end{figure}


\begin{figure} [!htbp]
\includegraphics[width=9cm,clip=true,trim=0.5cm 1.cm 2.7cm 2.5cm]{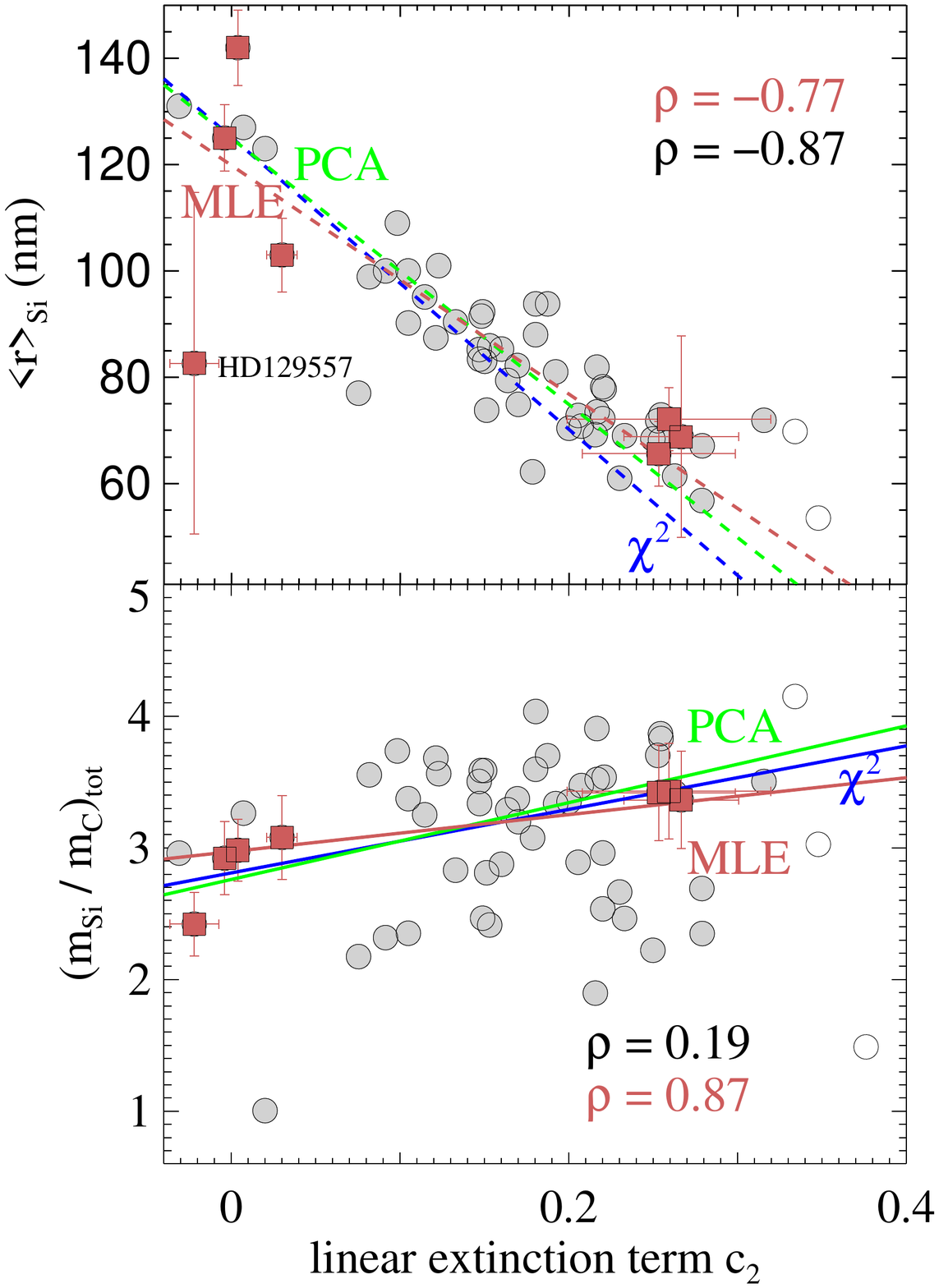}
\caption{{ {Mean radius of large silicate grains ({\it {top}}) and
      mass ratio of silicate and carbon}} ({\it {bottom}}) versus
  extinction fit parameter $c_2$. The $c_2$ parameter describes a
  linear term in observations of the extinction curve between 115 -
  330\,nm \citep{Valencic,Gordon}.  { {The dashed lines are
      straight-line fits of data belonging to the L$_{\rm S}$-sample
      (filled circles in grey) and full lines are fits}} to the
  single-cloud sight-lines. Symbols { same} as in
  Fig.~\ref{pl_corC1.fig}. \label{pl_corC2.fig}}
\end{figure}

\begin{figure} [!htbp]
\includegraphics[width=9cm,clip=true,trim=3.6cm 1.cm 5.5cm 2cm]{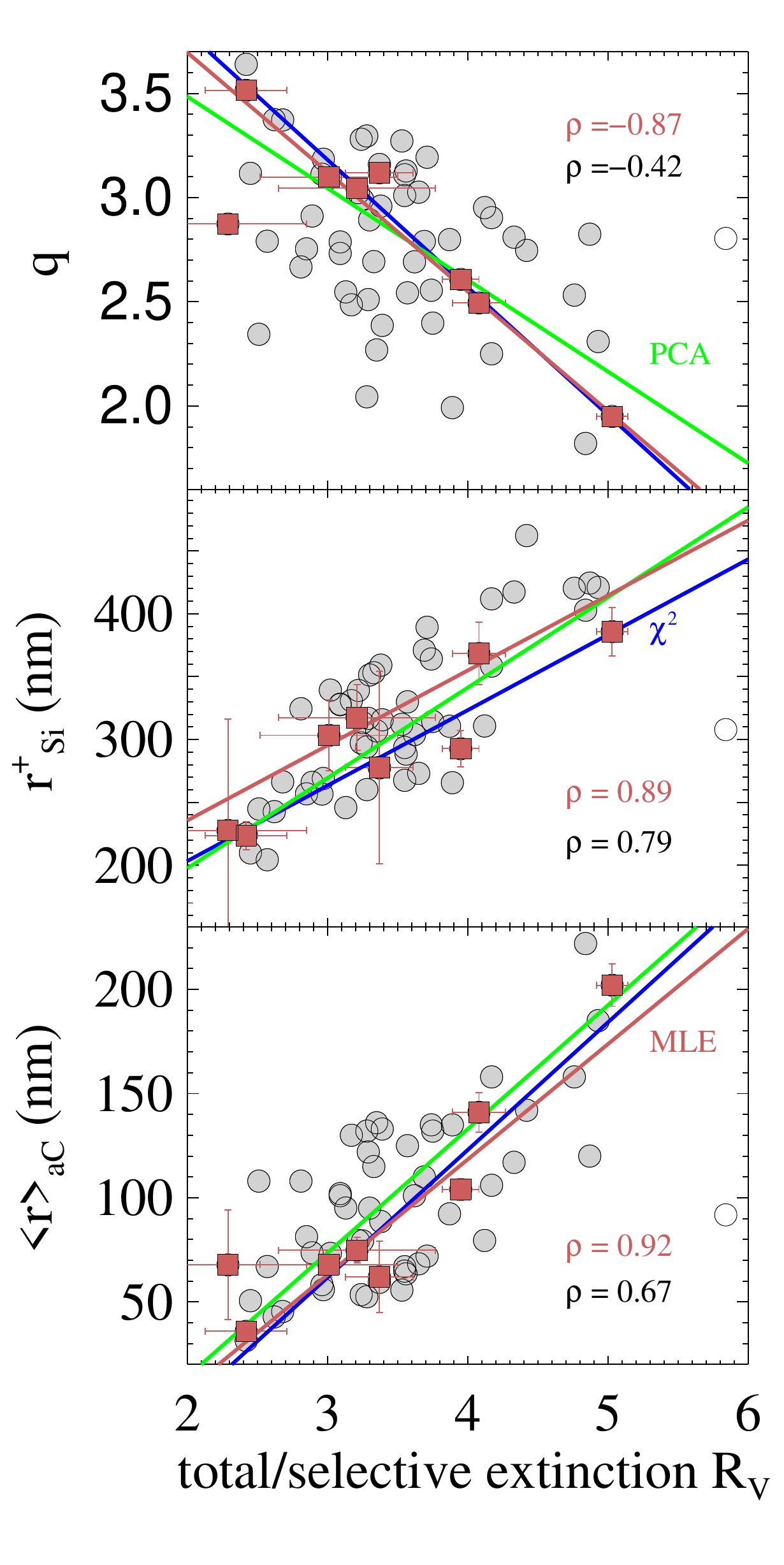}
\caption{Exponent of the dust size distribution ({\it {top}}), upper
  radius of large silicates ({\it {middle}}) and mean radius of large
  carbon grains ({\it {bottom}}) versus total-to-selective extinction
  $R_{\rm V}$. Symbols { same} as in
  Fig.~\ref{pl_corC1.fig} \label{pl_corRv.fig}.}
\end{figure}


In the range $x \simgreat 3 \ \mu \rm{m}^{-1}$, the extinction curve
is described by Eq.~(\ref{ext.eq}), which contains a constant term
$c_1$ and and a linear term with coefficient $c_2$. In our dust model,
this behaviour of the extinction curve { {depends}} on the mass
ratio of silicate to carbon grains $(m_{\rm {Si}}/m_{\rm
  {C}})_{\rm{tot}}$, i.e., on the chemical composition of the dust
cloud.  The higher the specific mass of silicate relative to
carbonaceous dust, the smaller the constant $c_1$, and the larger the
coefficient of the linear term $c_2$.  { {Typical examples for
    extinction curves that are flat in the spectral range $6 \leq x
    \leq 8 \ \mu \rm{m}^{-1}$ are HD~147889 and HD~164740
    (Fig.~\ref{1cl_ext.fig}). They show large values of $c_1$ of 1.6
    and 1.5, small values of $c_2 \leq 0.03$, and a mass ratio as low
    as $(m_{\rm {Si}}/m_{\rm {C}})_{\rm{tot}} \leq 3$. On the other
    hand, HD~79186 has a steep extinction in that spectral range
    (Fig.~\ref{1cl_ext.fig}) with $c_1=1.1$, $c_2=0.26$, and with
    $(m_{\rm{Si}}/m_{\rm {C}})_{\rm{tot}} = 3.4$. The star has about
    the largest of mass ratio in our sample.  We observe for
    single-cloud sight-lines a positive correlation of
    $(m_{\rm{Si}}/m_{\rm {C}})_{\rm{tot}}$ with $c_1$
    (Fig.~\ref{pl_corC1.fig}) and a negative for $c_2$
    (Fig.~\ref{pl_corC2.fig}, {\it {bottom}}). Both correlations
    vanish for the other samples of multiple-cloud (dominated)
    sight-lines. Apparently the correlations break down when observing
    through clouds with different chemical compositions. However, for
    single-cloud sight-lines there are only data available that are
    scattered near $c_2 \sim 0$ and $\sim 2.6$ so that these
    correlations need further proof.}}

Large carbon grains have a flat contribution to the extinction in the
range between $2 \leq x \leq 7 \mu \rm{m}^{-1}$ whereas large silicate
grains increase linearly with $x$. Examples are given in
Fig.~\ref{1cl_ext.fig}.  We find that by increasing the mean size of
silicates the linear extinction rise $c_2$ is reduced. We observe that
the anti-correlation of $ \langle r \rangle_{\rm {Si}}$ with $c_2$ is stronger for
the M, L samples with $\rho \sim -0.9$ than for the S sample having $\rho
\sim -0.8$ (Fig.~\ref{pl_corC2.fig}, {\it {top}}). The weaker
correlation is driven by the less well constrained $\langle r \rangle_{\rm {Si}}$
parameter of HD~129557.

We find for single-cloud cases that the strength of the extinction
bump $c_3$ is correlated with the mass ratio of small to large grains
$m_{\rm {vsg}}/m_{\rm{lrg}}$ (Fig.~\ref{pl_corC1.fig}). For example
(Table~\ref{para.tab}), the [C]/[H] abundance in PAH and small
graphite is 12\, ppm for HD~164740, which has $c_3 = 0.7$; whereas
much more (28\,ppm) C is locked in small grains for HD~79186 having a
larger $c_3$ of 1.2. By observing a mix of clouds the $m_{\rm
  {vsg}}/m_{\rm{lrg}}$ { {correlation with $c_3$}} breaks
(Fig.~\ref{pl_corC1.fig}).

Extinction curves with a large value of $c_4$ display a strong FUV
rise at $x \simgreat 5.9 \ \mu$m$^{-1}$ (Eq.~\ref{ext.eq}). We observe
a bi-modal distribution with two single-cloud cases with a FUV
extinction as flat as $c_4 \sim 0.1$ and five stars with a steep FUV
rise clustered near $c_4 = 0.25$. In the models the FUV extinction of
the later 5 stars have a significant contribution by small grains
(Fig.~\ref{1cl_ext.fig}, Table~\ref{para.tab}).  For example, the
abundance ratio of very small to large grains is for HD~129557 almost
a factor three larger than for the flat FUV extinction observed
towards HD~164740 (Table~\ref{para.tab}).  Indeed, we observe that the
$c_4$ parameter is strongly correlated with the mass ratio of very
small and large grains (Fig.~\ref{pl_corC1.fig}). The correlation is
very strong in the single-cloud cases ($\rho=0.94$), strong in the
{ {L- and L$_{\rm S}$-samples ($\rho=0.85-0.89$), and weak, if at
    all present, in the M-sample ($\rho <0.8$).}}

In the \citet{WD01} dust model, small grains coagulate onto large
grains in relatively dense environments. The ratio of visual
extinction to hydrogen column density $A_{\rm V}/N_{\rm H} \propto
R_{\rm V}$, and so one expects that an increase of $R_{\rm V}$ goes
hand in hand with an increase of the mean particle radius and a
flatter size distribution.  This was noted already by \citet{Kim} in
which for $R_{\rm V} = 5.3$ their size distribution has significantly
fewer grains with $r < 0.1 \mu$m than their $R_{\rm V} = 3.1$
distribution, as well as a modest increase at larger sizes. This
result is expected, at short wavelengths the extinction is provided by
small grains and for larger values of $R_{\rm V}$ there is relatively
less extinction at short wavelengths.  We prove these early
conclusions by our observations (Fig.~\ref{pl_corRv.fig}), however we
can do this only for single-cloud sight-lines and not for the 
multiple-cloud samples. { {There is some trend in a correlation
    of the mean size of large silicates with $R_{\rm V}$ for all
    sub-samples. For the S sample}} the mean size of large carbon
grains as well as the upper grain radius of silicates correlates
strongly with $R_{\rm V}$, and the exponent of the size distributions
is anti-correlated with $R_{\rm V}$ (Fig.~\ref{pl_corRv.fig}). Again,
this is because low values of $q$ provide a relative increase of large
grains.  For { {multiple-cloud samples}} the latter relation
becomes random with $\rho \sim -0.5$.  (Table~\ref{corrext.tab}).


\begin{table}[!htbp]
\begin{center}
  \caption{\label{regress.tab}
    Regression fits between dust and observing parameters.
    { Where no otherwise specified we use the single-cloud
        sight-lines with units as in Figs.~\ref{pl_corC1.fig} -
        \ref{pl_corPol.fig}. Columns 2 ad 3 gives the best-fit parameters
        of the relationship $y = ax+b$ derived by applying MLE.
        Column~4 gives the parameter $(1-\zeta)$ which relates to the
        confidence that no selection bias affects the statistics.}
    }
    
 \begin{tabular}{l|c|c|c}
\hline \hline
           &     & & \\                     
 Correlation  & $a_{-\Delta a}^{+\Delta a} $ & $ b_{-\Delta b}^{+\Delta b} $ & $1-\zeta$  \\
          &       & &(\%)  \\                     
\hline
          &     & & \\                     
  $c_1 \leftrightarrow   {m_{\rm {Si}}} / {m_{\rm C}}     $ & $-0.68_{-0.25}^{+0.22}$ & $3.9_{-0.3}^{+0.4} $ & 99.8 \\
           &      & & \\                     
  $c_2 \leftrightarrow   \langle r \rangle_{\rm {Si}} \/ ^{\dagger}     $ & $-215_{-5.2}^{+5.5}$ & $120_{-1.1}^{+0.4} $ & 98.5 \\
           &      & & \\                     
  $c_2 \leftrightarrow  { m_{\rm {Si}}} / {m_{\rm C}}     $ & $1.4_{-0.46}^{+0.54}$ & $3.0_{-0.16}^{+0.06} $ & 18.6 \\
           &      & & \\                     
  $c_3 \leftrightarrow  { m_{\rm {vsg}}} / {m_{\rm {lgr}}}$ & $0.31_{-0.04}^{+0.04}$ & $-0.02_{-0.02}^{+0.03}  $& 99.2 \\    
           &      & & \\                     
  $c_4 \leftrightarrow   {m_{\rm {vsg}}} / {m_{\rm {lgr}}}$ & $1.66_{-0.15}^{+0.17}$ & $0.06_{-0.01}^{+0.01}$& 70.0 \\    
           &      & & \\                     
  $R_{\rm V} \leftrightarrow   q                          $ & $-0.57_{-0.05}^{+0.05}$ & $4.84_{-0.18}^{+0.19} $& 95.5 \\
           &      & & \\                     
  $R_{\rm V} \leftrightarrow   r^{+}_{\rm {Si}}           $ & $60_{-7.3}^{+8.4}$ & $116_{-38}^{+22} $ & 83.0 \\
           &      & & \\                     
  $R_{\rm V} \leftrightarrow   \langle r \rangle_{\rm {aC}}             $ & $55.4_{-2.6}^{+2.3}$ & $103_{-7}^{+8} $& 93.1 \\
           &      & & \\                     
  $k_{\rm {p}} \leftrightarrow  r^{-}_{\rm {pol}}         $ & $117_{-11}^{+10}$ & $12_{-19}^{+14} $& 99.9 \\ 
          &     & & \\                     
$\lambda_{\rm {max}} \leftrightarrow \langle r \rangle ^{\rm {Si}}_{\rm {pol}} $ & $0.16_{-0.09}^{+0.09}$ & $104_{-54}^{+53} $& 98.5 \\
\hline
\end{tabular}
\end{center} 
 {\small {{\bf{Notes:}} $^\dagger$ Fit parameters are for the M sample.}}
\end{table}

\subsubsection{Model versus polarisation parameters}

\begin{figure} [!htbp]
  \includegraphics[width=9cm,clip=true,trim=3.cm 1.5cm 2.5cm 2cm]{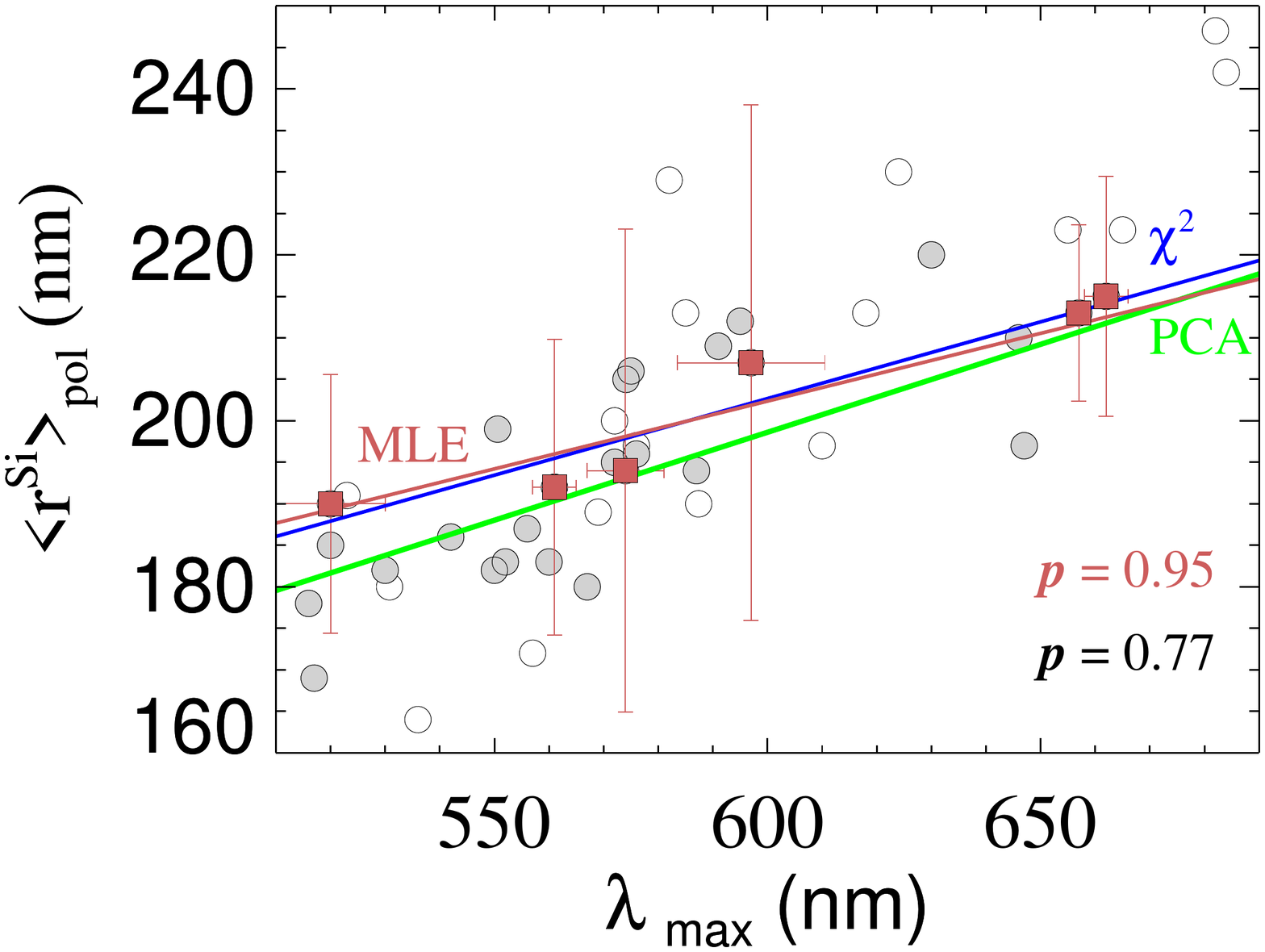}

\includegraphics[width=9cm,clip=true,trim=3.cm 1.5cm 2.5cm
2cm]{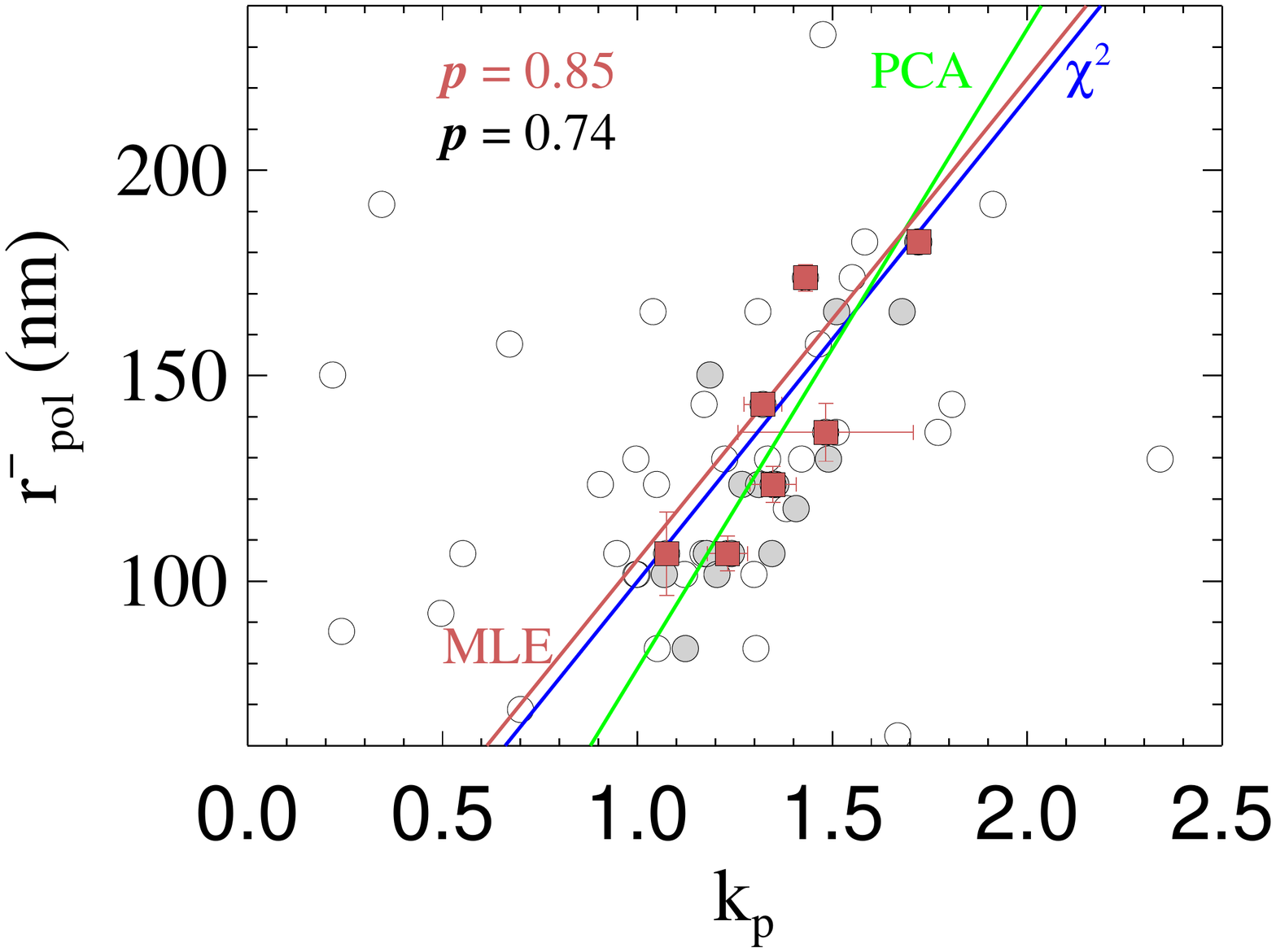}
\caption{Mean radius of large silicate grains versus wavelength at
  maximum polarisation ({\it {top}}). Minimum radius of aligned
  silicates versus width of the polarisation spectrum ({\it
    {bottom}}). Symbols as in
  Fig.~\ref{pl_corC1.fig} \label{pl_corPol.fig}.}
\end{figure}

We have investigated if the best-fit parameters of the Serkowski
curves exhibit differences between the case of single-cloud and
multiple-cloud { {sight-lines}}.  We found that the mean values of
$\lambda_{\rm{max}}$ and $k_{\rm p}$ are almost the same for the 8
single-cloud and the 51 multiple-cloud dominated sight-lines.
However, we noticed that single-cloud cases show on average higher
polarisation than the multiple-cloud sight-lines observed towards a
particular sight-line: the median polarisation of the single-cloud
cases is $p = 2.6$\,\% { {(S-sample)}}, whereas multiple-cloud 
{ {sight-lines (L-sample)}} show a median polarisation of $p =
1.3$\,\%. Similar, the median polarisation per visual extinction is
$p/A_{\rm V} = 1.8$ for single and $p/A_{\rm V} = 0.8$ (\%/mag) for
multiple-cloud sight-lines. The median polarisation per reddening is
$p/$E(B-V) = 5.8 for single and $p/$E(B-V) = 3\,(\%) for
multiple-cloud sight-lines. Both { {S and L}} samples have similar
median values for the visual extinction of $A_{\rm V} = 1.5$ and
$A_{\rm V} = 1.7$\,mag, { {and for}} the reddening of E(B-V)=0.50
and E(B-V)=0.45 for the single-cloud and multiple-cloud sight-lines
and have same median of $R_{\rm V} \sim 3.4$.  Apparently photons are
depolarised when they penetrate through different clouds, which is
explained by cloud-to-cloud variations of the magnetic field direction
and different grain alignment efficiency.

  The influence of grain sizes and the minimum alignment radius,
  $r^-_{\rm {pol}}$, on the polarisation spectrum is discussed by
  \citet{V13} and \citet{S14}. It is shown that the choice of
  $r^-_{\rm {pol}}$ and $r^+_{\rm {Si}}$ is sensitive to the
  polarisation spectrum.  Therefore one expects that these dust
  parameters are related to the Serkowski parameters $\lambda_{\rm
    {max}}$ and $k_p$. { {We also compute the mean size of aligned
      silicates $\langle r \rangle ^{\rm {Si}}_{\rm {pol}}$, which is given by
      averaging over the dust size distribution from $r^-_{\rm {pol}}$
      and $r^+_{\rm {Si}}$.}}

  { The parameters of the Serkowski curve and the dust parameters are
    strongly correlated in the single-cloud cases ($\rho\ga 0.9$),
    while no correlation is observed in the cases of multiple-cloud
    sight-lines, with two exceptions in the M$_{\rm {SP}}$-sample.
    However, the latter sample has a narrower observed parameter range
    than the S-sample so that this comparison needs to be taken with
    care.} In Fig.~\ref{pl_corPol.fig} we show that $k_p$ { {is
      correlated with}} $r^-_{\rm {pol}}$ and $\lambda_{\rm {max}}$ {
    {is correlated with}} $\langle r \rangle ^{\rm {Si}}_{\rm
    {pol}}$. The latter correlation has already been found by
  \citet{CK} and more recently by \citet{V16}. Apparently
  depolarisation is active when multiple-clouds are in the sight-line
  so that these correlations vanish.

\subsection{Observing bias \label{bias.sec}}

We wish to quantify a confidence level that the correlations are not
drawn by random selection effects of the sight-lines or a bias in the
observing sample.  We estimate a confidence level by means of Monte
Carlo and select $10^5$ random samples of 7 stars out of the total of
51 multiple-cloud sight-lines. For each of these elements we compute
the correlation strength as given by Pearsons' coefficient
$\rho'$. This allows computing the probability $\zeta$ of finding a
random correlation that is larger or equals the observed strength
$\rho' \geq \| \rho \|$, where $\rho$ is given in
Tables~\ref{corrext.tab} and \ref{corrpol.tab}. For example, we
observe that our single-cloud sight-lines $c_1$ is anti-correlated
with ${m_{\rm {Si}}} / {m_{\rm C}}$ at strength $\rho = -0.95$
(Table~\ref{corrext.tab}). We find that out of $10^5$ random samples
of 7 multiple-cloud dominated sight-lines selected from
Table~\ref{para.tab} there are only eight cases that show a
correlation strength $\rho' \leq -0.95$, hence $\zeta = 8/10000$. We
are confident at a $1- \zeta = 99.9$\,\% level that this correlation
is not drawn by a random selection and can only be found when
single-cloud sight-lines are observed. However, when a correlation
pre-exists in the LIPS sample such confidence level shall be
small. This is true for example in the $c_2 \leftrightarrow \langle r \rangle _{\rm Si}$
correlation, as given in Table~\ref{regress.tab}.  Excluding
the 8 known single-cloud cases from the 59 LIPS targets we find that
for any random selection of 7 out of the 51 such sight-lines there is
a fifty-fifty chance that $m_{\rm {vsg}}/m_{\rm{lrg}}$ { {is
    correlated with $c_4$ at $\rho \geq 0.9$.}}  This explains why it
was already observed before \citep{Desert90}, that a large amount of
very small relative to large grains steepens the FUV rise.


\section{Conclusion \label{concl.sec}}
For 59 diffuse ISM sight-lines of the LIPS targets observed in
spectro-polarimetric mode by \citet{Bagnulo17} we have retrieved from
previous literature the extinction curves in the range 2\,$\mu$m --
90\,nm. For this sample we have performed the simultaneous modelling
of normalised extinction and polarisation curves using an interstellar
dust model that includes a populations of carbon and silicate dust in
form of nano-sized particles and large ($\simgreat 6$\,nm) spheroidal
grains with a power law size distribution and imperfect rotational
orientation. Using archive spectroscopic observations of interstellar
absorption lines (Ca\,II, Na\,I, and K\,I), we have found that eight
sight-lines are crossing just a single absorbing interstellar
cloud. We have performed the analysis of our results
independently for single interstellar clouds and for multi-cloud
sight-lines. The main results of our analysis can be summarised as follows.

\begin{itemize}
\item[1.] For the eight single-cloud sight-lines, the ratio of
  total-to-selective extinction $R_{\rm V}$ correlates strongly with
  the mean size of large silicate and carbon grains and
  anti-correlates with the exponent $q$ of the dust-grain
  size-distributions (as expected, since the relative amount of large
  grains is larger for smaller values of $q$).
  
\item[2.] For single-cloud sight-lines we have revealed several strong
  correlations between the parameters $c_i$ of the UV extinction curve
  fitting (see Eq.~\ref{ext.eq}) and the dust model parameters.  For
  example, the mass ratio of total silicate to carbon $(m_{\rm
    {Si}}/m_{\rm C})_{\rm {tot}}$ is anti--correlated with $c_1$
      and correlated with $c_2$, and the mass ratio of very small to
  large grains $m_{\rm {vsg}}/m_{\rm {lgr}}$ is correlated
      with $c_3$.  These correlations disappear when instead of
      considering only the single-cloud sight-lines we include stars
      observed through multiple-clouds with different chemical
      compositions.  However, some relations are strong in both the
      single-cloud sample and the full LIPS sample. For instance, the
      $c_4$ parameter, which provides a measure of the strength of the
      FUV rise, is strongly correlated with the amount of small
          to large grains, $m_{\rm {vsg}}/m_{\rm {lgr}}$, as
      expected.

\item[3.] Polarimetry imposes additional constraints on the dust
  properties than those given by extinction data only; in particular,
  simultaneous modelling of extinction and polarisation gives a
  steeper dust-grain size-distributions and smaller mean grain sizes
  than what is deduced from extinction data only.

\item[4.] The interpretation of polarisation data from multiple-clouds
  sight-lines is more complicated After the radiation is polarised in
  a dust cloud, the intersection of a second or more clouds
  towards the sight-line leads to depolarisation. This fact can be
  understood by cloud-to-cloud variations of the grain properties,
  dust alignment efficiency and the direction of the magnetic field.
  
\item[5.] In the single-cloud cases we have found strong correlations
  between the parameters of the Serkowski curve and the silicate dust
  parameters.  For example, the wavelength at which the polarisation
  reaches maximum $\lambda_{\max}$, correlates with the mean radius of
  aligned silicates $\langle r\rangle ^{\rm {Si}}_{\rm
    {pol}}$, and the width of the polarisation spectrum $k_{\rm p}$
  with the minimum radius of aligned silicates $r^{-}_{\rm
    {pol}}$. No dependencies of the Serkowski parameters on the
      mean radius of large carbon particles $\langle r
      \rangle_{\rm{aC}}$ are found. This confirms conclusions of the
  previous modelling by \cite{V14}.

\item[6.] The strong correlationships between parameters found for
  single-cloud cases that are not seen in the full sample are
  statistically not likely to be due to selection bias in the LIPS
  sample. This is demonstrated by the fact that when one arbitrarily
  selects seven out of 51 stars of the, L-sample the dependencies are still not detected.

\item[7.]  Any mixing of clouds results in similar
  average dust model parameters, however there are strong variations
  of the dust properties and from cloud-to-cloud.
\end{itemize}

\noindent
We demonstrated that only the framework of single-cloud analysis
provides an unambiguous view of relations between dust properties and
observables such as extinction and polarisation.  The framework will
help understanding the dust evolution from cloud-to-cloud and within
their particular physical environments.  We conclude that a most
urgent task is to find more such single-cloud sight-lines and develop
dust models for them.

\begin{acknowledgements}
  This research is based on data obtained from the ESO Science Archive
  Facility and in particula on observations collected under ESO
  programme 095.C-0855 and 096.C-0159 (PI=N.Cox).  This research has
  also made use of the SIMBAD database, operated at the CDS,
  Strasbourg, France.  We thank E. Kr{\"u}gel and J. Kre{\l}owski for
  interesting discussions and the referee V.\ Guillet for helping us
  to improve our work.  NVV was partly supported by the RFBR grant
  16-02-00194 and RFBR--DST grant 16-52-45005.

\end{acknowledgements}


\bibliographystyle{aa} 
\bibliography{References}


\clearpage
\begin{appendix} 
\section{Extinction and polarisation fits }

\begin{figure} [h!tb]
\includegraphics[width=8.0cm,clip=true,trim=2.1cm 2.7cm 1cm 3.0cm]{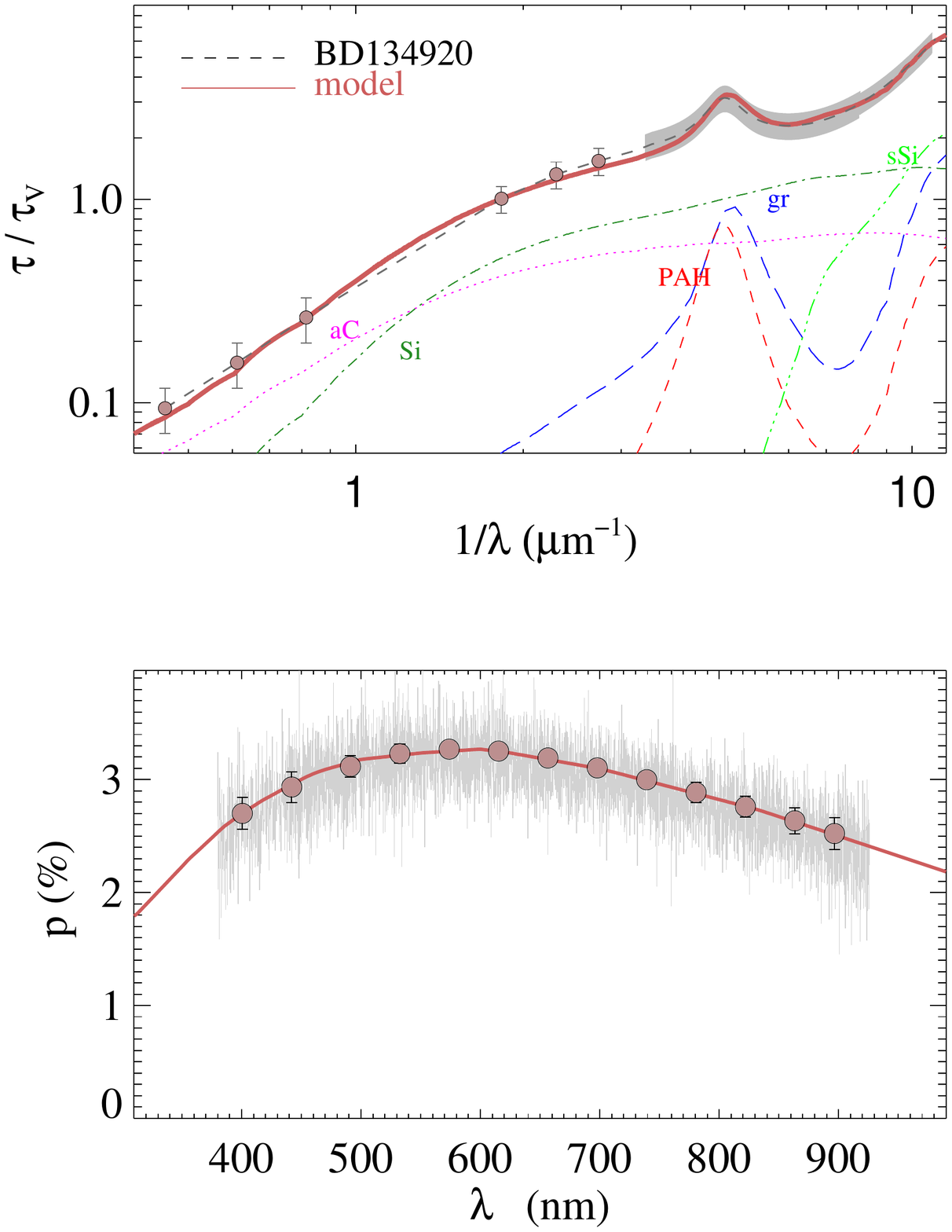}
\caption{{\it Top:} Observed and modeled extinction curve of
  BD~134920.  { {The extinction curve by
      \citet{Fitzpatrick04}, \citet{Valencic} and \citet{Gordon} is shown with a dashed line,
      $UBVJHK$ photometry with filled circles, IUE/FUSE spectra are
      represented by the grey shaded area, and uncertainties are
      1\,$\sigma$.}}  The model is shown (brown solid line) as well as the
  contribution of the different dust populations to the extinction
  (various colored and labeled lines). {\it Bottom:} 
  polarisation spectra observed with FORS2 are shown with grey lines, and the
  and best-fit model with brown solid lines;
  the filled circles (with 1\,$\sigma$ error bars) are the same data
  rebinned to a spectral resolution of $\lambda/\Delta \lambda \sim
  50$.\label{appstart.fig}}
\end{figure}

\begin{figure} [h!tb]
\includegraphics[width=9.0cm,clip=true,trim=2.1cm 2.7cm 1cm 3.0cm]{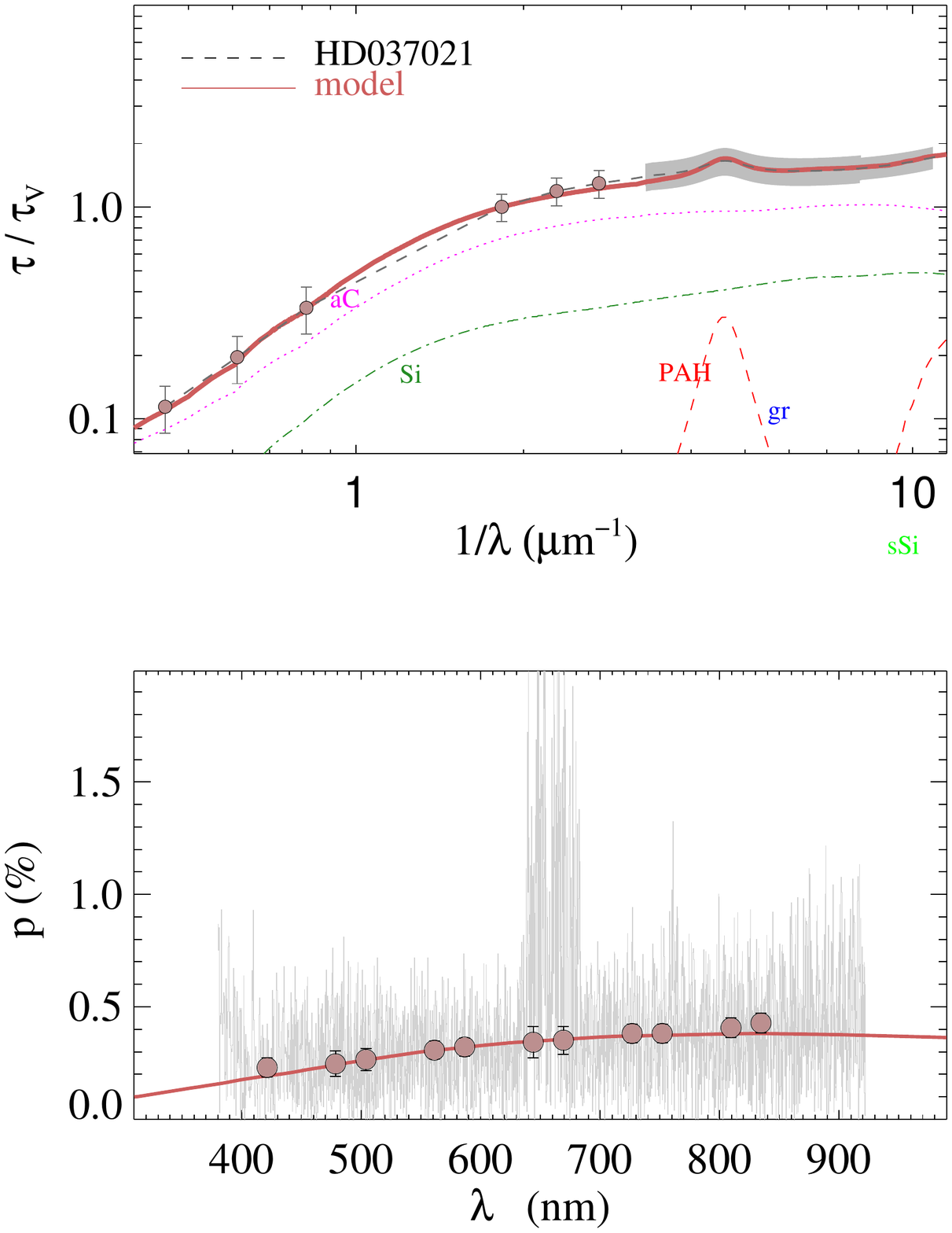}
\caption{Notation same as in Fig.~\ref{appstart.fig}  for
  HD~37021 \label{HD037021.fig}.}
\end{figure}

\begin{figure} [h!tb]
\includegraphics[width=9.0cm,clip=true,trim=2.1cm 2.7cm 1cm 3.0cm]{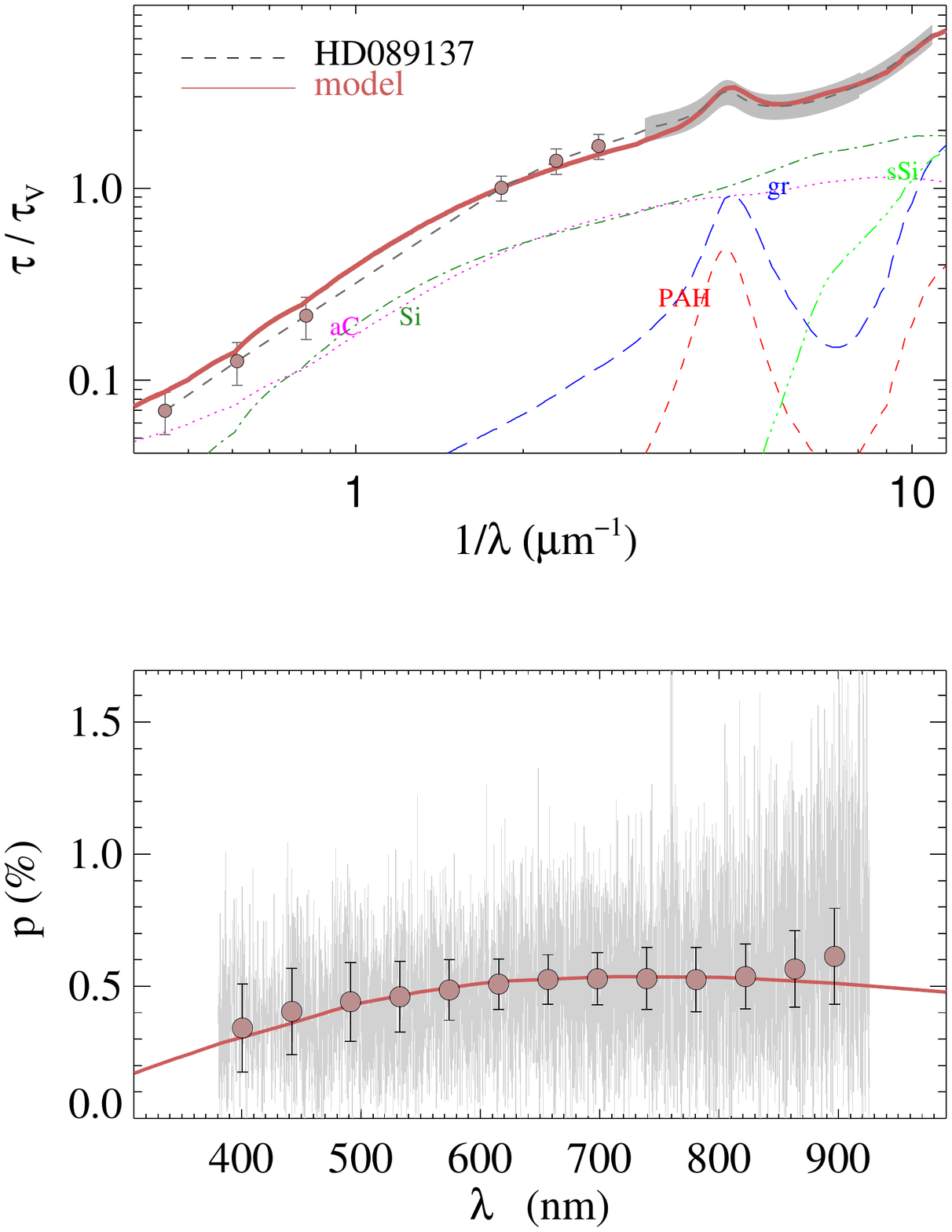}
\caption{Notation same as in Fig.~\ref{appstart.fig}  for HD~89137.  \label{applast.fig}}
\end{figure}

Remaining extinction and polarisation fits of individual targets are
displayed in Fig.~\ref{appstart.fig} to Fig.~\ref{applast.fig}.

\begin{figure} [h!tb]
\includegraphics[width=9.0cm,clip=true,trim=2.1cm 2.7cm 1cm 3.0cm]{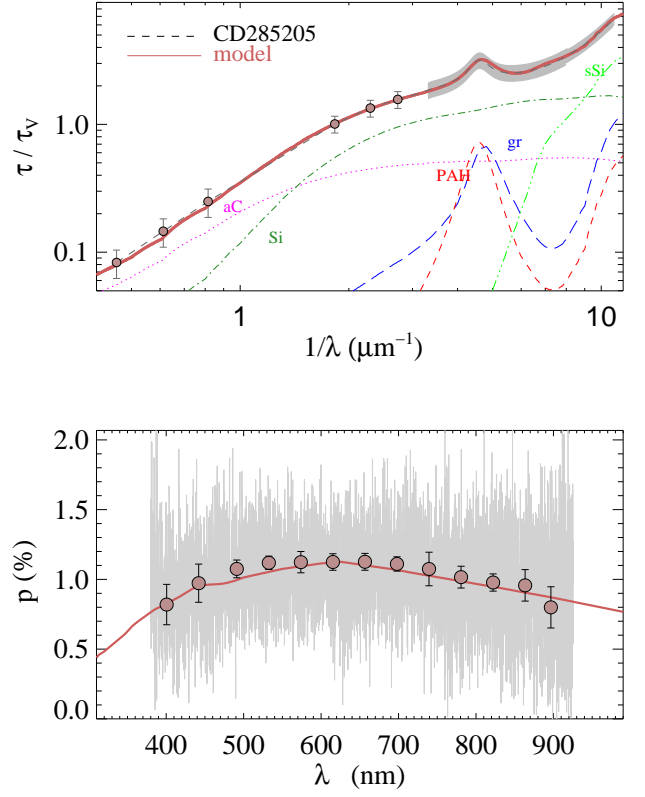}
\caption{Notation same as in Fig.~\ref{appstart.fig}  for CD~285205.}
\end{figure}

\begin{figure} [h!tb]
\includegraphics[width=9.0cm,clip=true,trim=2.1cm 2.7cm 1cm 3.0cm]{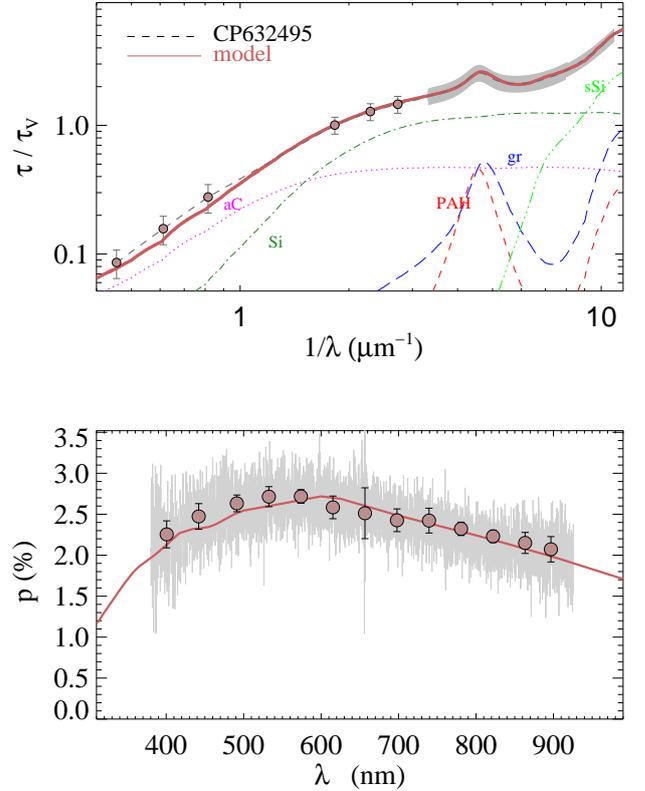}
\caption{Notation same as in Fig.~\ref{appstart.fig}  for CP~632495.}
\end{figure}

\begin{figure} [h!tb]
\includegraphics[width=9.0cm,clip=true,trim=2.1cm 2.7cm 1cm 3.0cm]{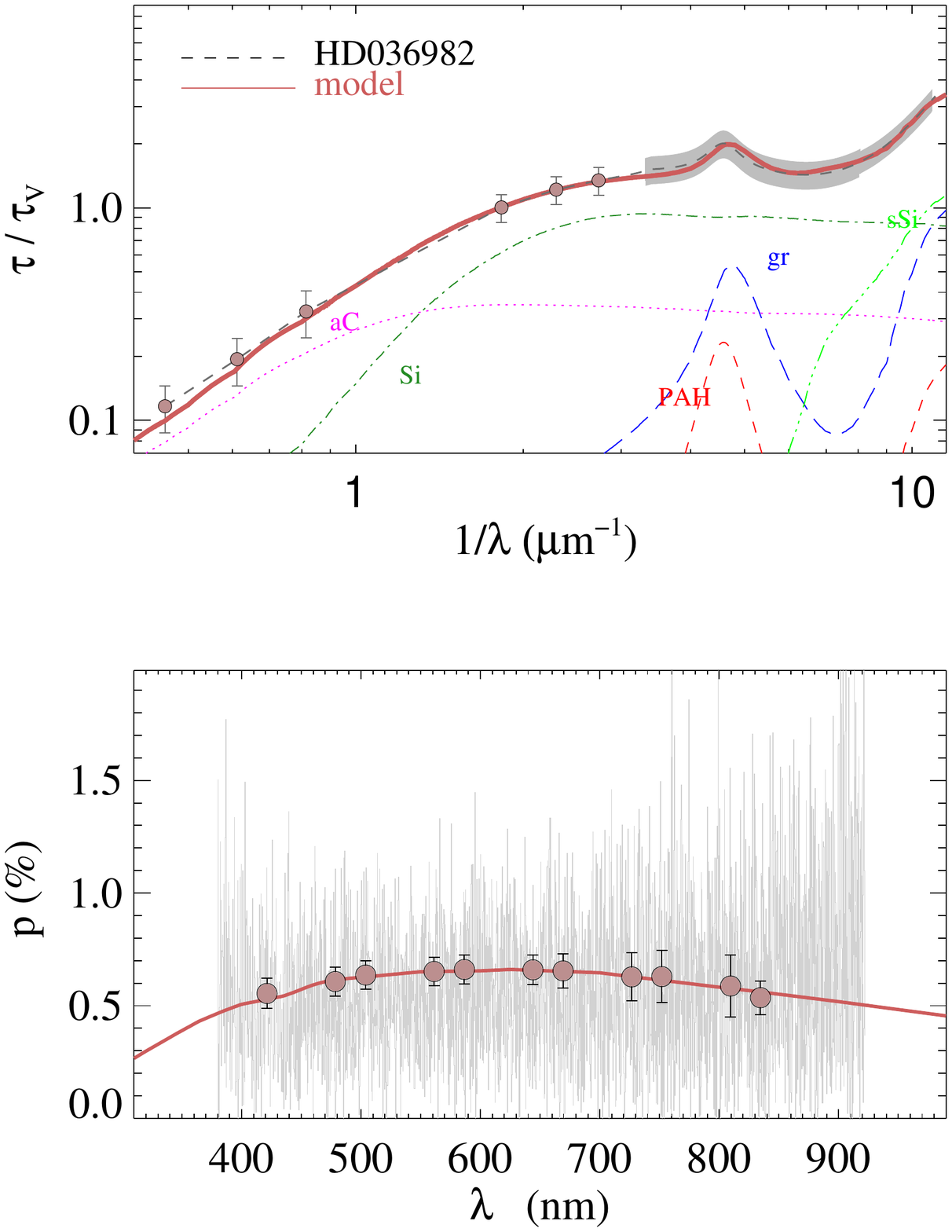}
\caption{Notation same as in Fig.~\ref{appstart.fig}  for HD~36982.}
\end{figure}

\begin{figure} [h!tb]
\includegraphics[width=9.0cm,clip=true,trim=2.1cm 2.7cm 1cm 3.0cm]{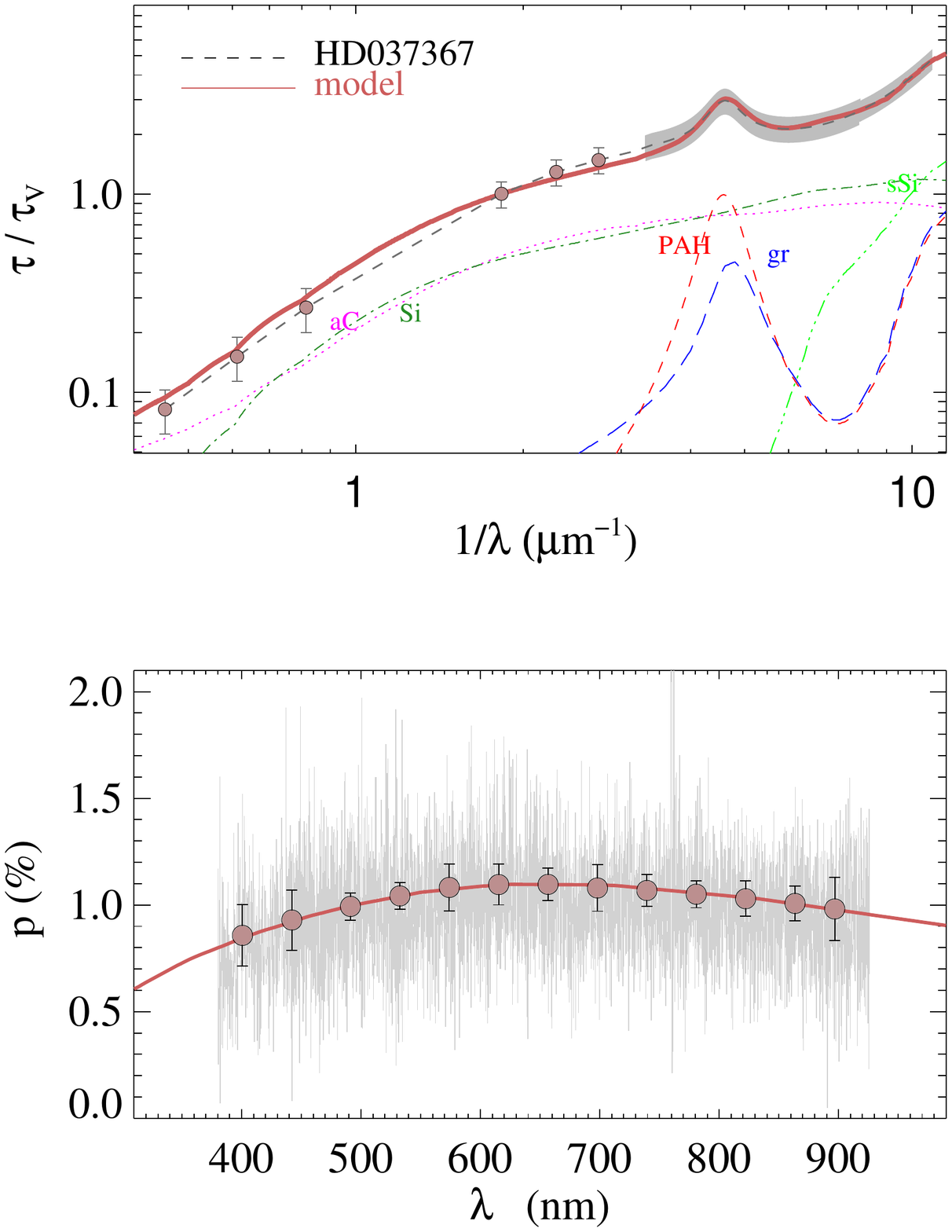}
\caption{Notation same as in Fig.~\ref{appstart.fig}  for HD~37367.}
\end{figure}

\begin{figure} [h!tb]
\includegraphics[width=9.0cm,clip=true,trim=2.1cm 2.7cm 1cm 3.0cm]{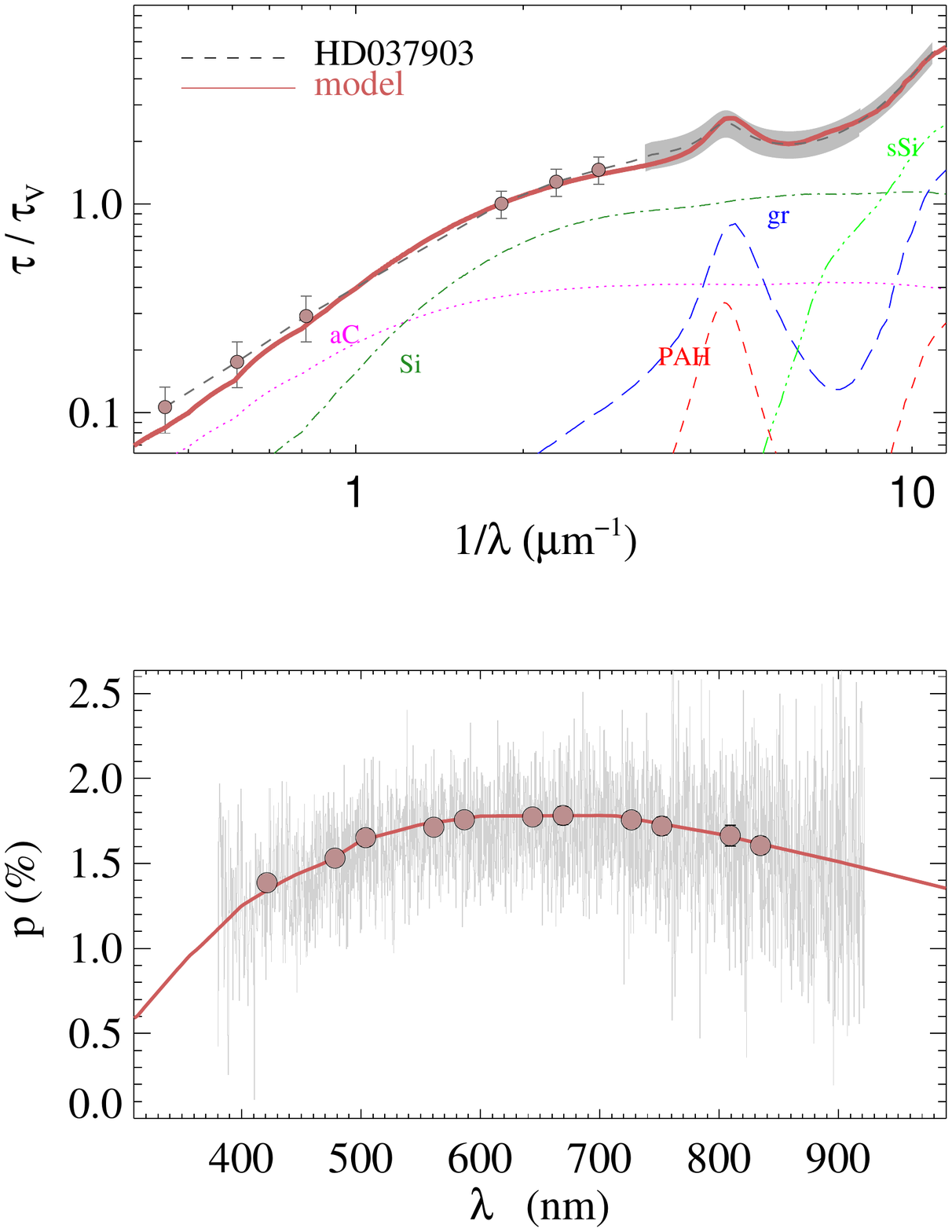}
\caption{Notation same as in Fig.~\ref{appstart.fig}  for HD~37903.}
\end{figure}

\begin{figure} [h!tb]
\includegraphics[width=9.0cm,clip=true,trim=2.1cm 2.7cm 1cm 3.0cm]{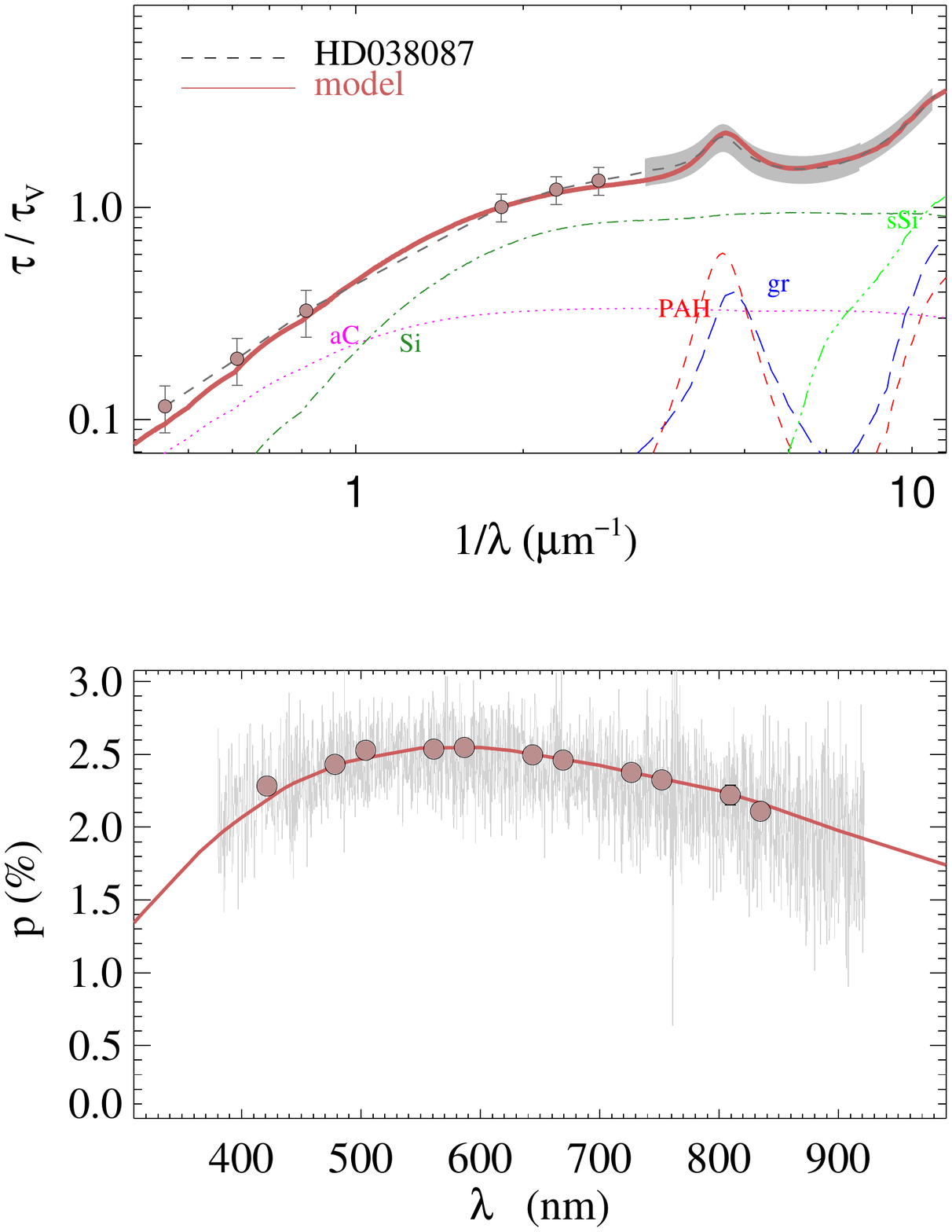}
\caption{Notation same as in Fig.~\ref{appstart.fig}  for HD~38087.}
\end{figure}

\begin{figure} [h!tb]
\includegraphics[width=9.0cm,clip=true,trim=2.1cm 2.7cm 1cm 3.0cm]{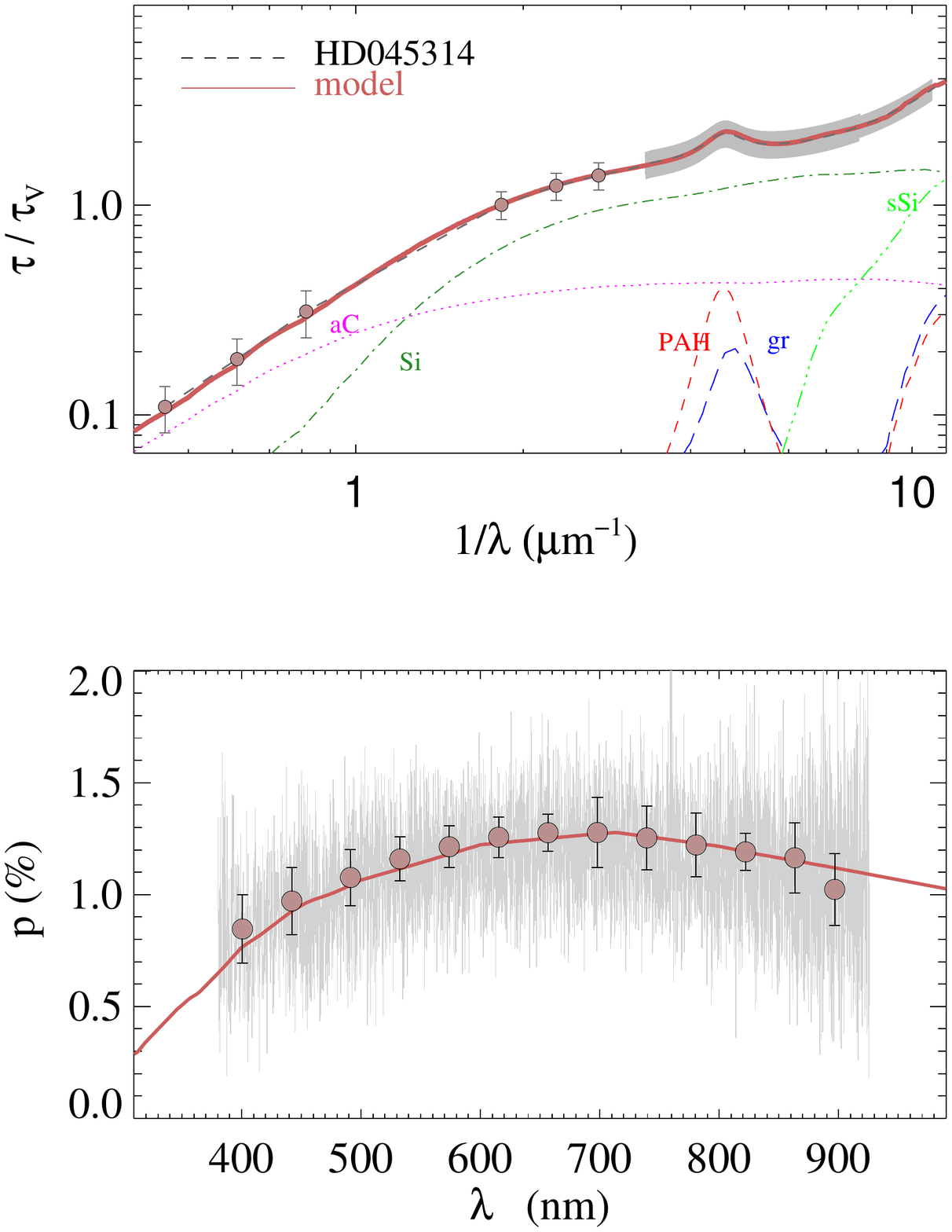}
\caption{Notation same as in Fig.~\ref{appstart.fig}  for HD~45314.}
\end{figure}

\begin{figure} [h!tb]
\includegraphics[width=9.0cm,clip=true,trim=2.1cm 2.7cm 1cm 3.0cm]{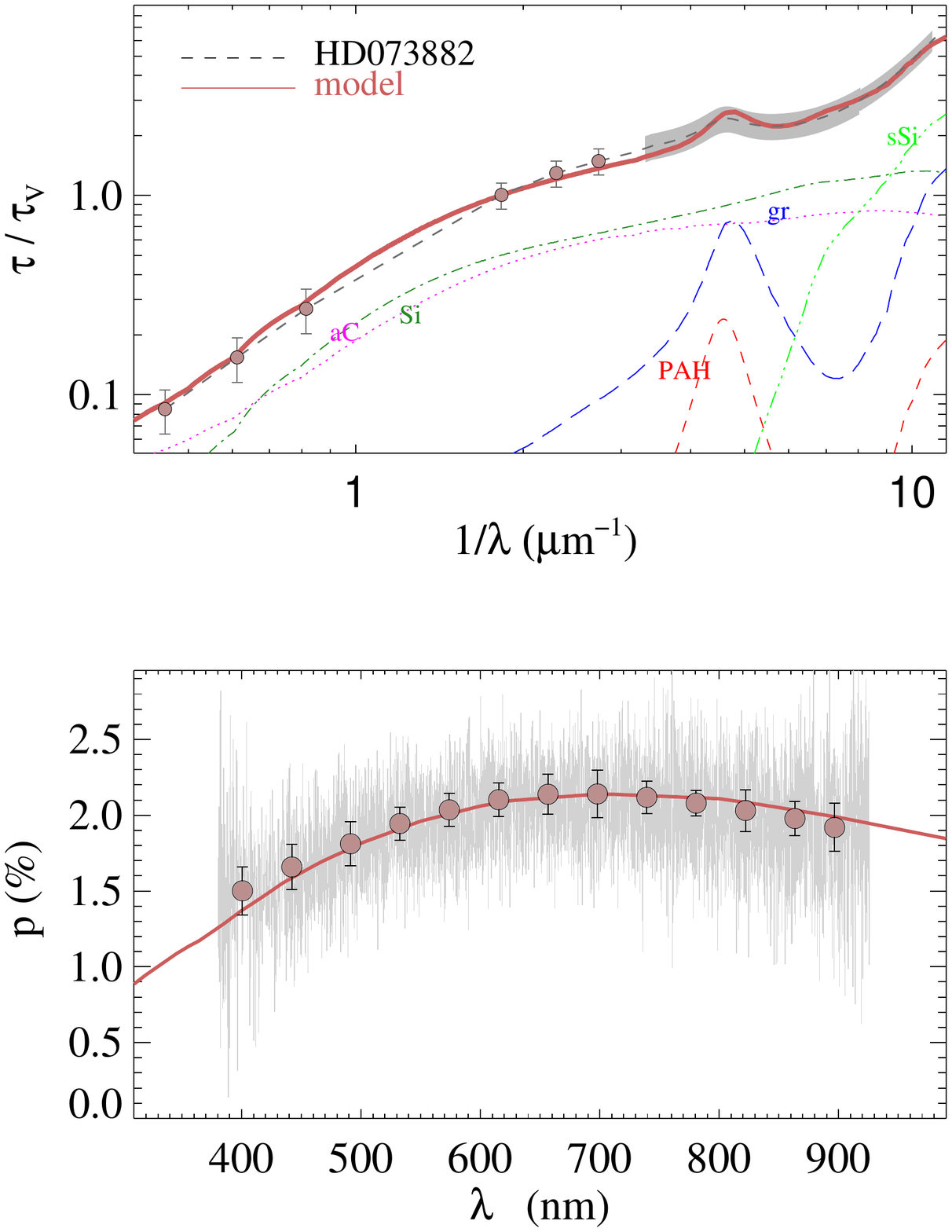}
\caption{Notation same as in Fig.~\ref{appstart.fig}  for HD~73882.}
\end{figure}

\begin{figure} [h!tb]
\includegraphics[width=9.0cm,clip=true,trim=2.1cm 2.7cm 1cm 3.0cm]{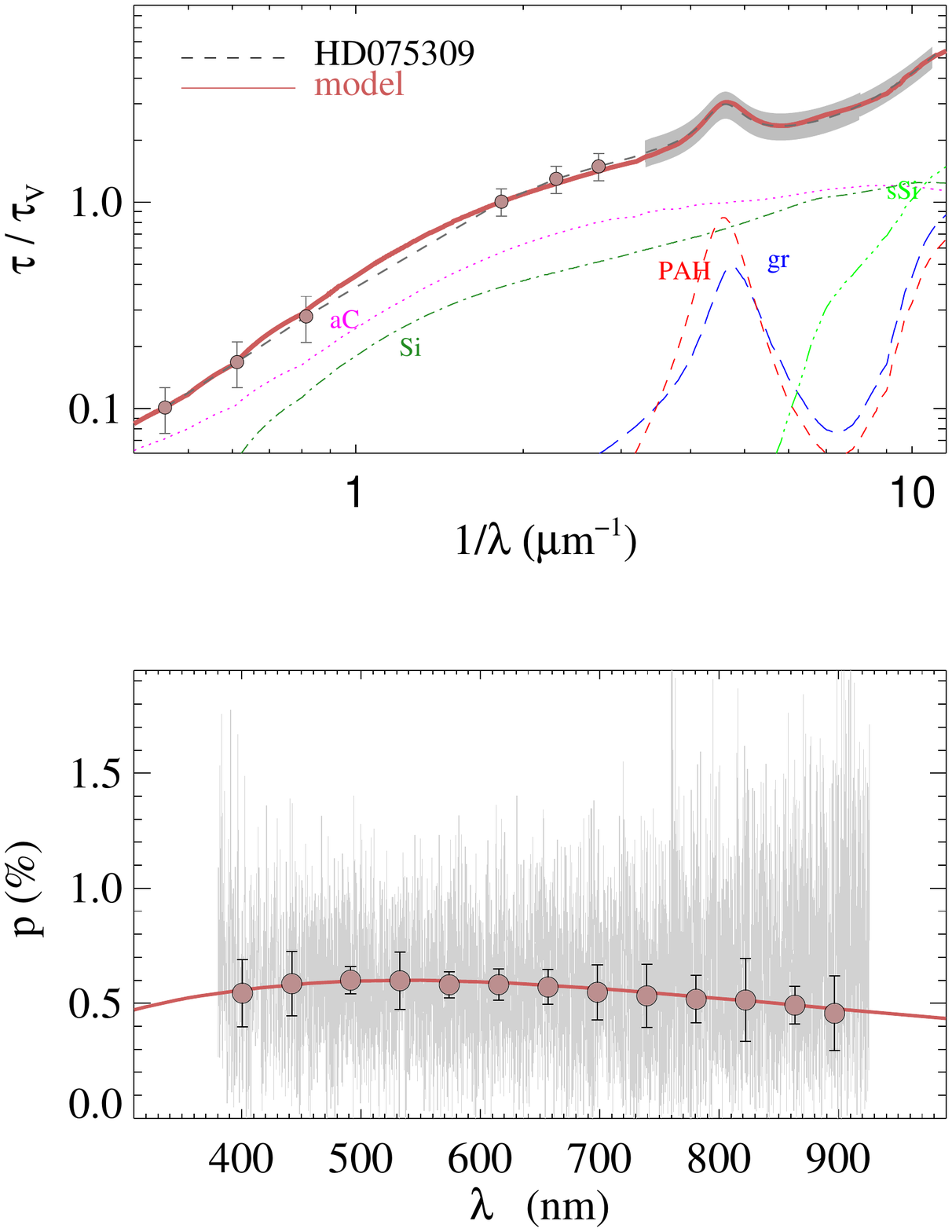}
\caption{Notation same as in Fig.~\ref{appstart.fig}  for HD~75309.}
\end{figure}

\begin{figure} [h!tb]
\includegraphics[width=9.0cm,clip=true,trim=2.1cm 2.7cm 1cm 3.0cm]{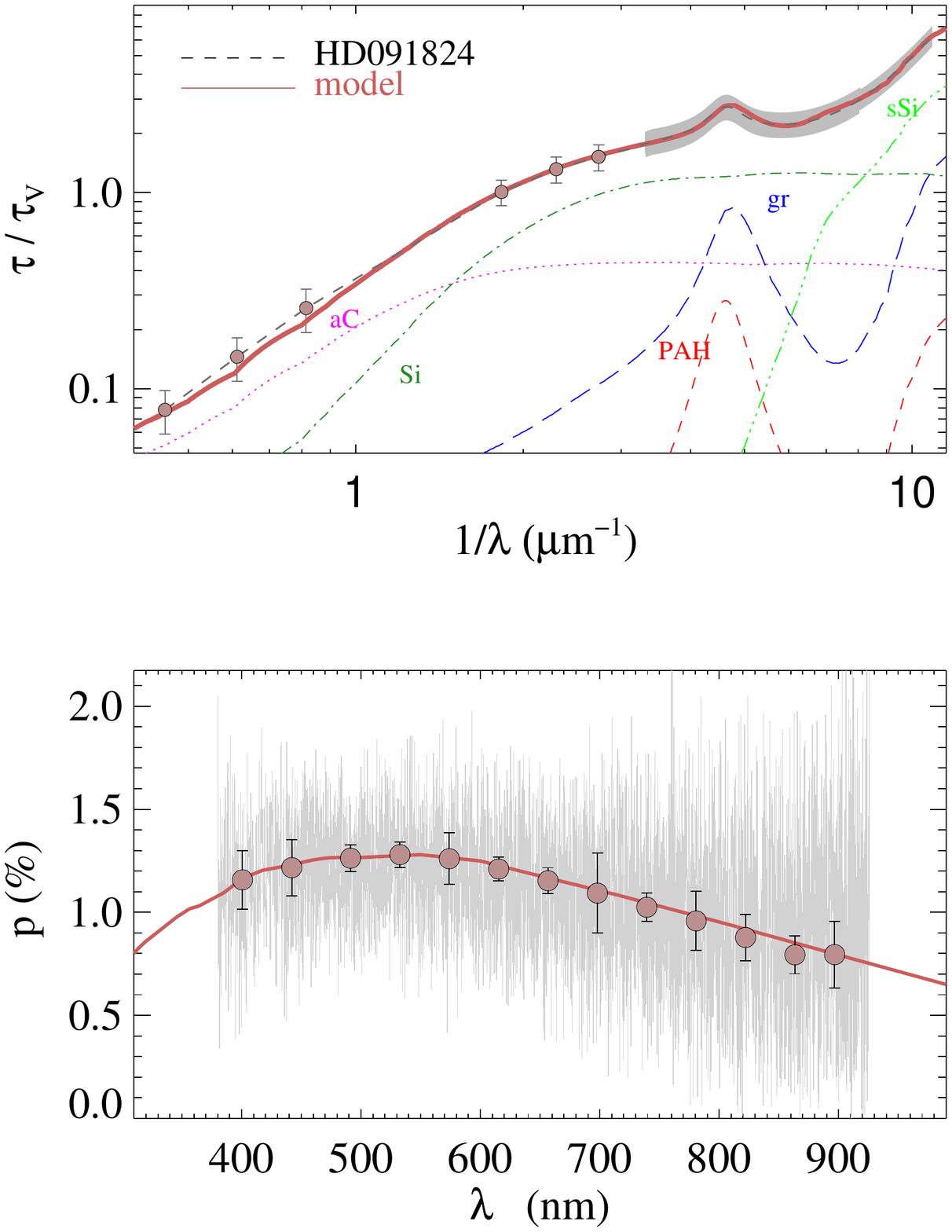}
\caption{Notation same as in Fig.~\ref{appstart.fig}  for HD~91824.}
\end{figure}

\begin{figure} [h!tb]
\includegraphics[width=9.0cm,clip=true,trim=2.1cm 2.7cm 1cm 3.0cm]{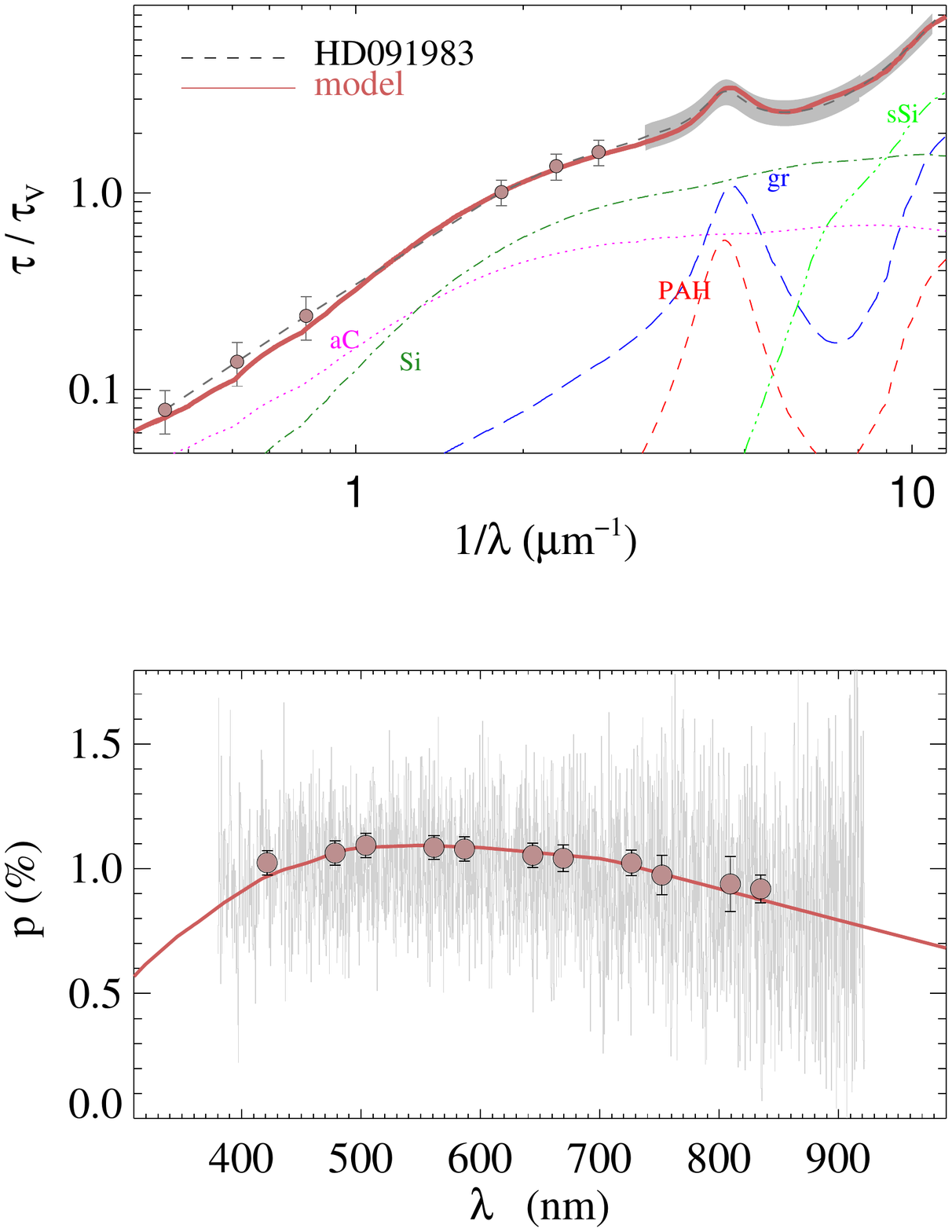}
\caption{Notation same as in Fig.~\ref{appstart.fig}  for HD~91983.}
\end{figure}

\begin{figure} [h!tb]
\includegraphics[width=9.0cm,clip=true,trim=2.1cm 2.7cm 1cm 3.0cm]{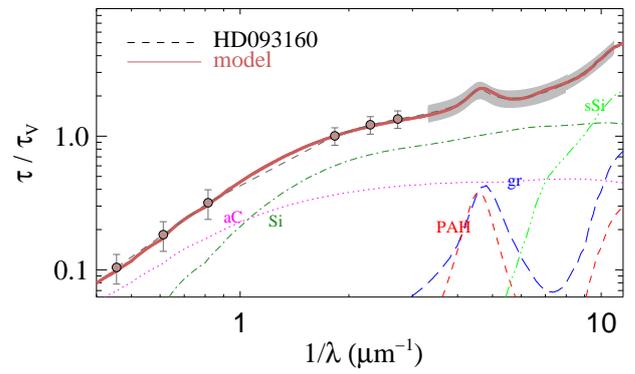}
\caption{Notation same as in Fig.~\ref{appstart.fig}  for HD~93160.}
\end{figure}

\clearpage
\begin{figure} [h!tb]
\includegraphics[width=9.0cm,clip=true,trim=2.1cm 2.7cm 1cm 3.0cm]{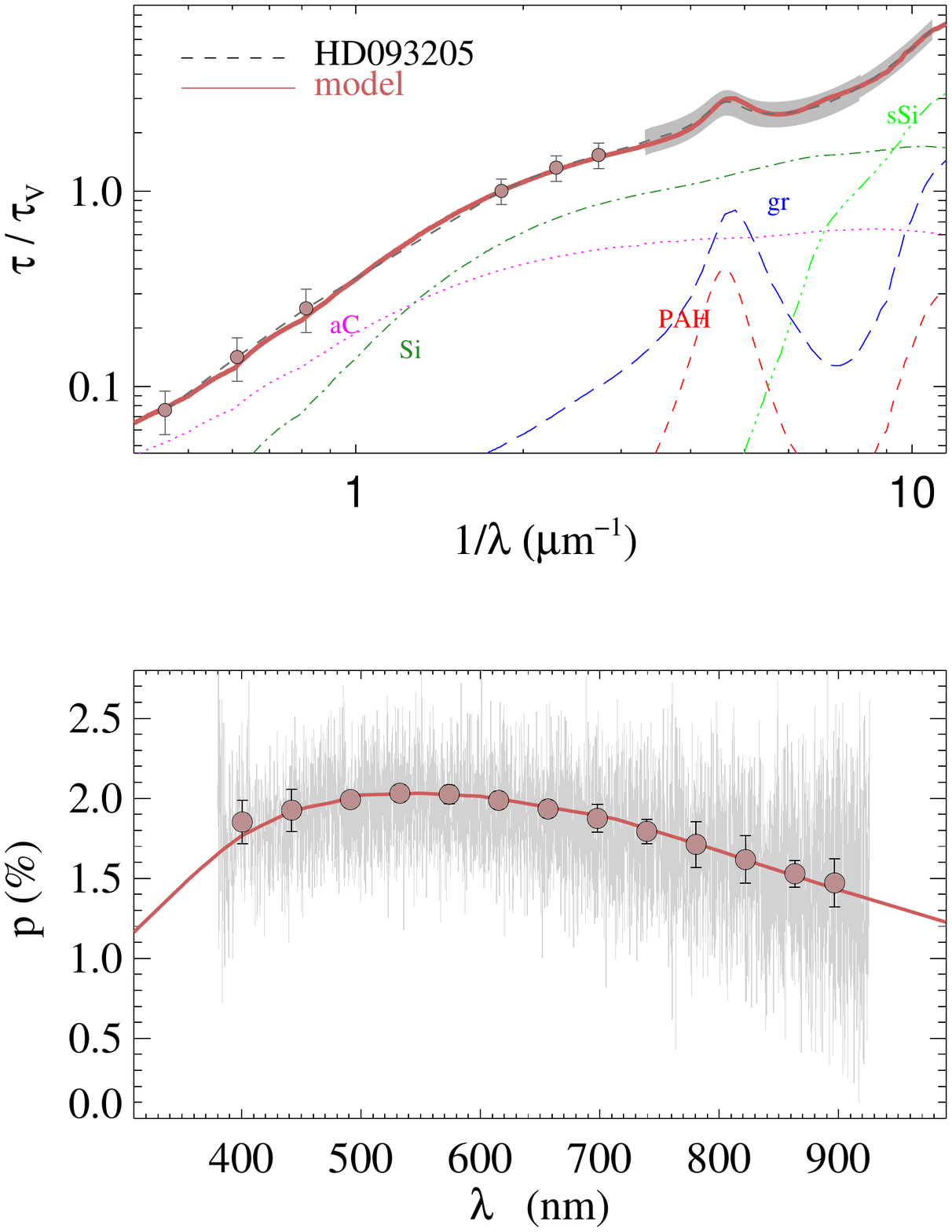}
\caption{Notation same as in Fig.~\ref{appstart.fig}  for HD~93205.}
\end{figure}

\begin{figure} [h!tb]
\includegraphics[width=9.0cm,clip=true,trim=2.1cm 2.7cm 1cm 3.0cm]{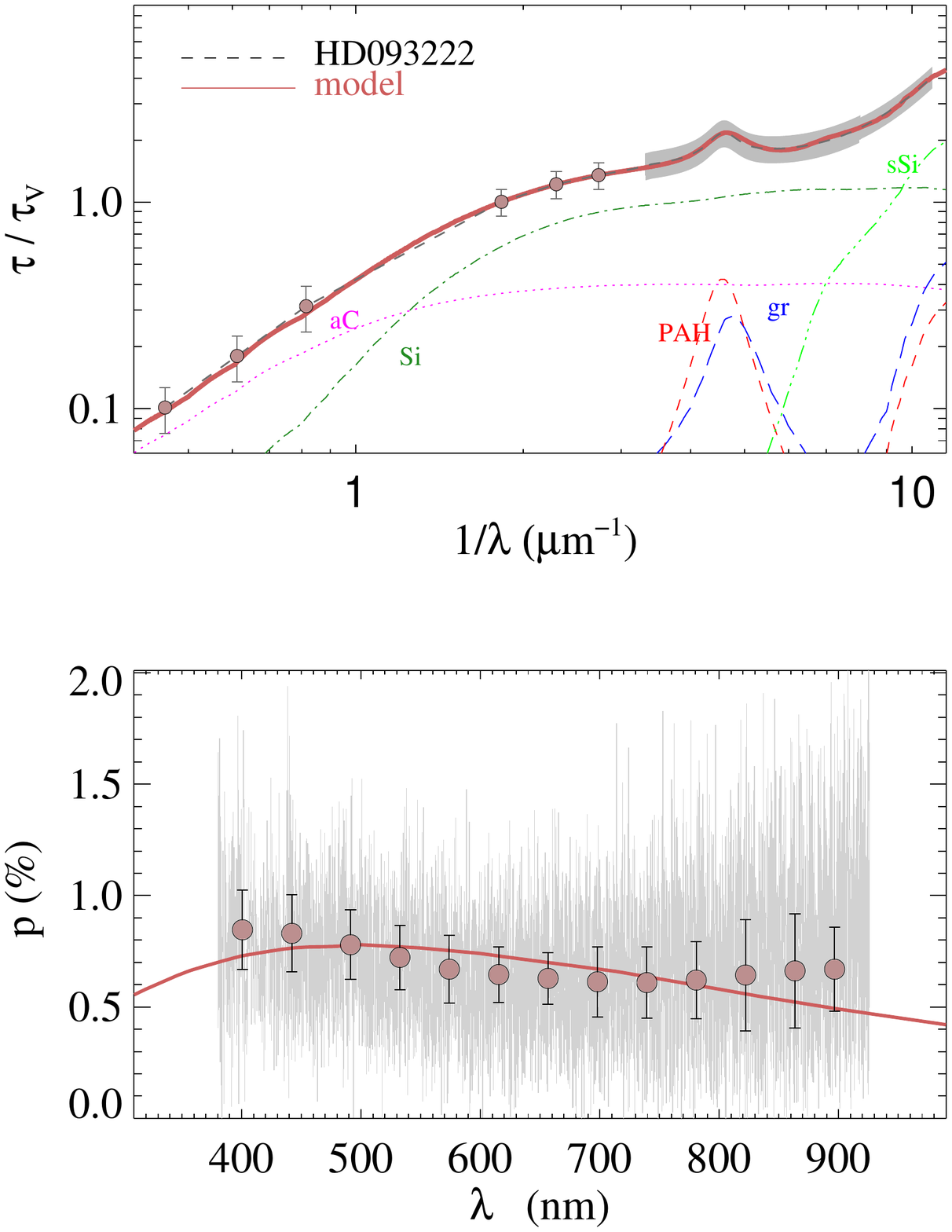}
\caption{Notation same as in Fig.~\ref{appstart.fig}  for HD~93222.}
\end{figure}

\begin{figure} [h!tb]
\includegraphics[width=9.0cm,clip=true,trim=2.1cm 2.7cm 1cm 3.0cm]{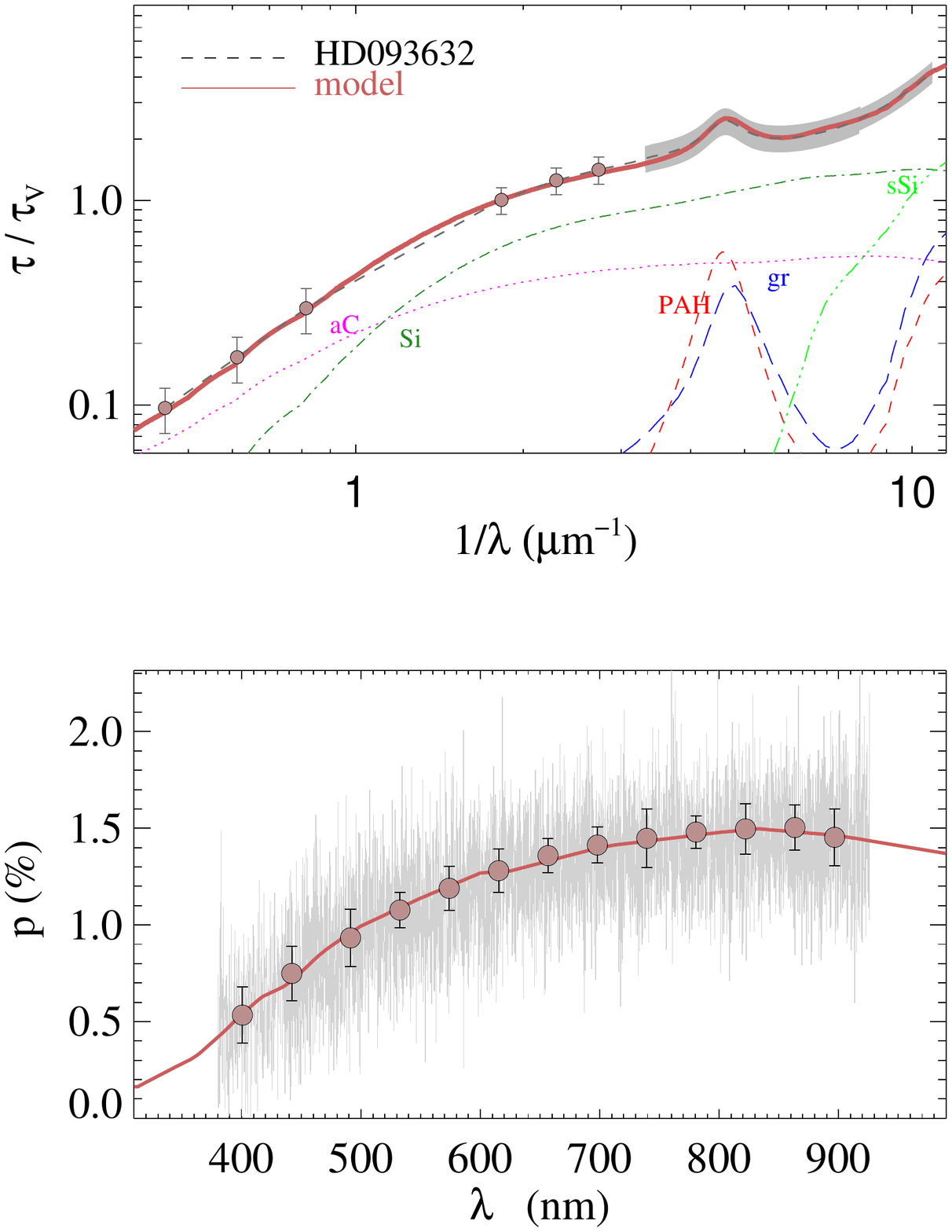}
\caption{Notation same as in Fig.~\ref{appstart.fig}  for HD~93632.}
\end{figure}

\begin{figure} [h!tb]
\includegraphics[width=9.0cm,clip=true,trim=2.1cm 2.7cm 1cm 3.0cm]{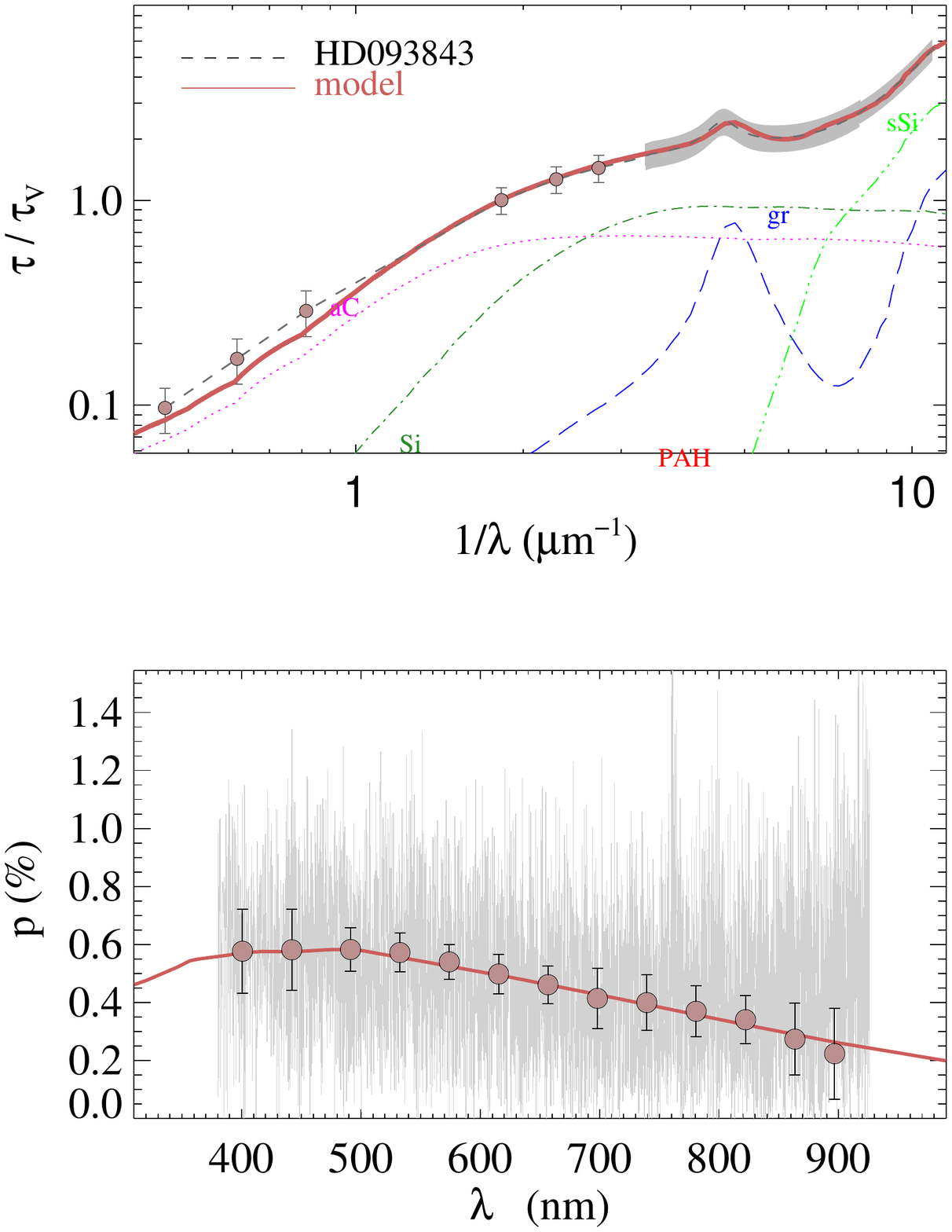}
\caption{Notation same as in Fig.~\ref{appstart.fig}  for HD~93843.}
\end{figure}

\begin{figure} [h!tb]
\includegraphics[width=9.0cm,clip=true,trim=2.1cm 2.7cm 1cm 3.0cm]{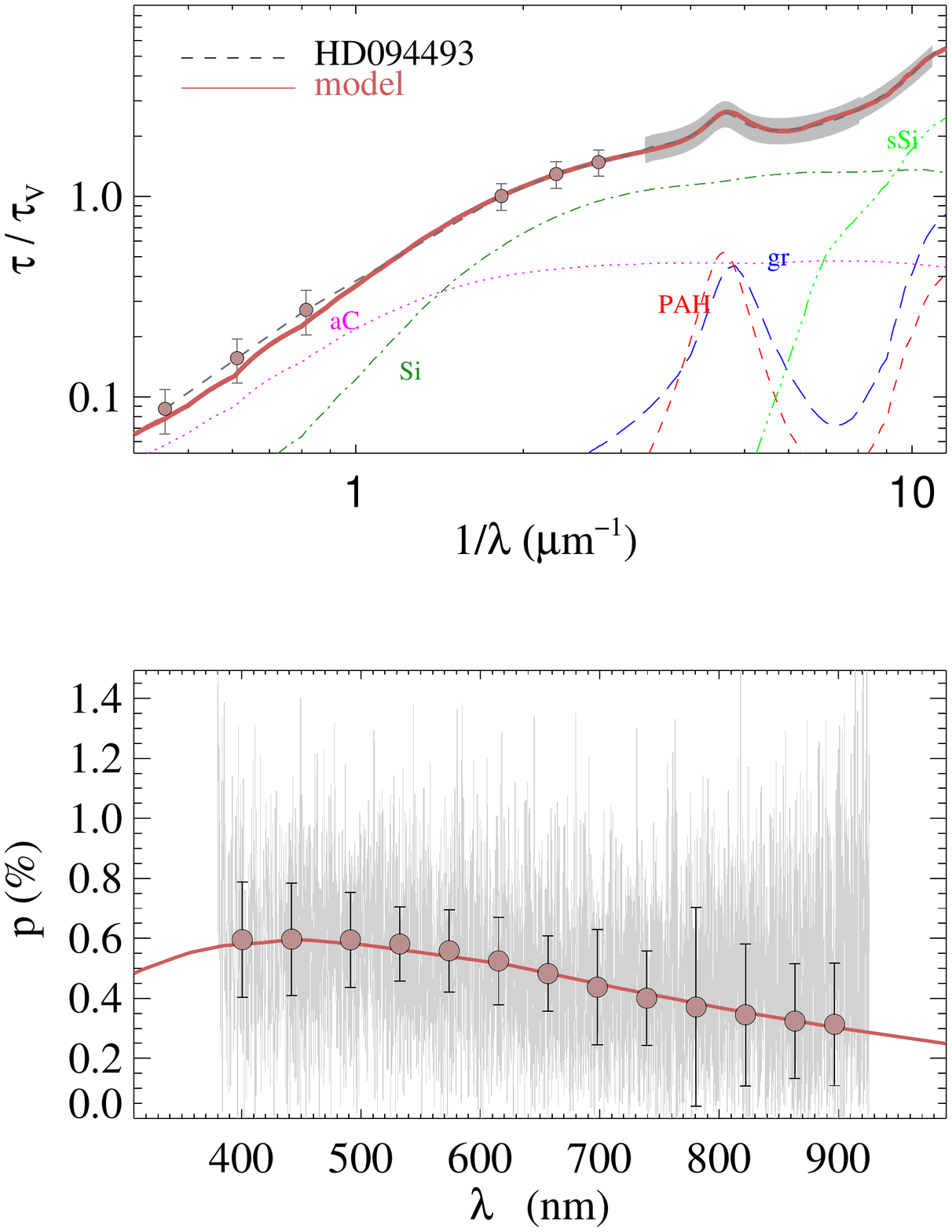}
\caption{Notation same as in Fig.~\ref{appstart.fig}  for HD~94493.}
\end{figure}

\begin{figure} [h!tb]
\includegraphics[width=9.0cm,clip=true,trim=2.1cm 2.7cm 1cm 3.0cm]{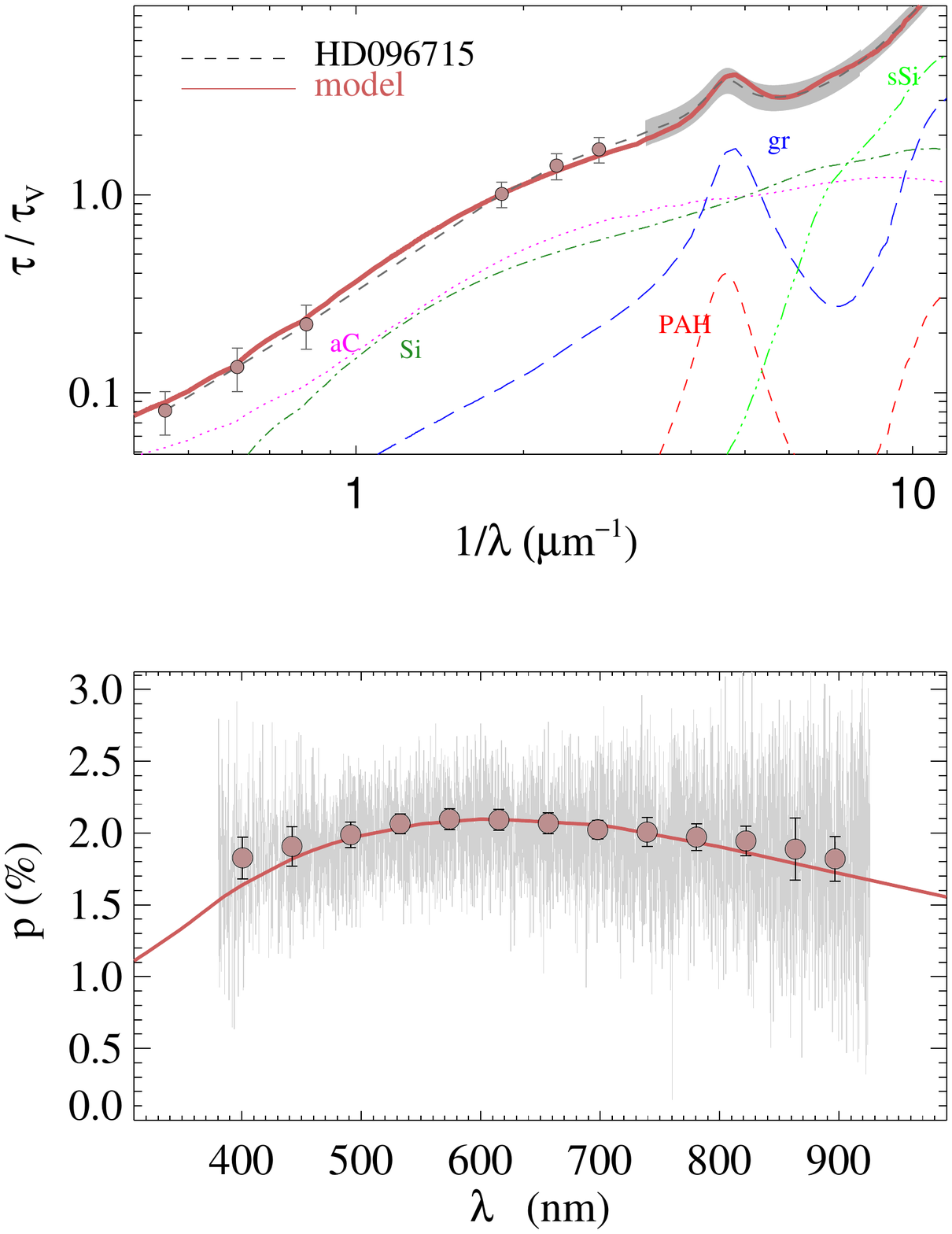}
\caption{Notation same as in Fig.~\ref{appstart.fig}  for HD~96715.}
\end{figure}

\begin{figure} [h!tb]
\includegraphics[width=9.0cm,clip=true,trim=2.1cm 2.7cm 1cm 3.0cm]{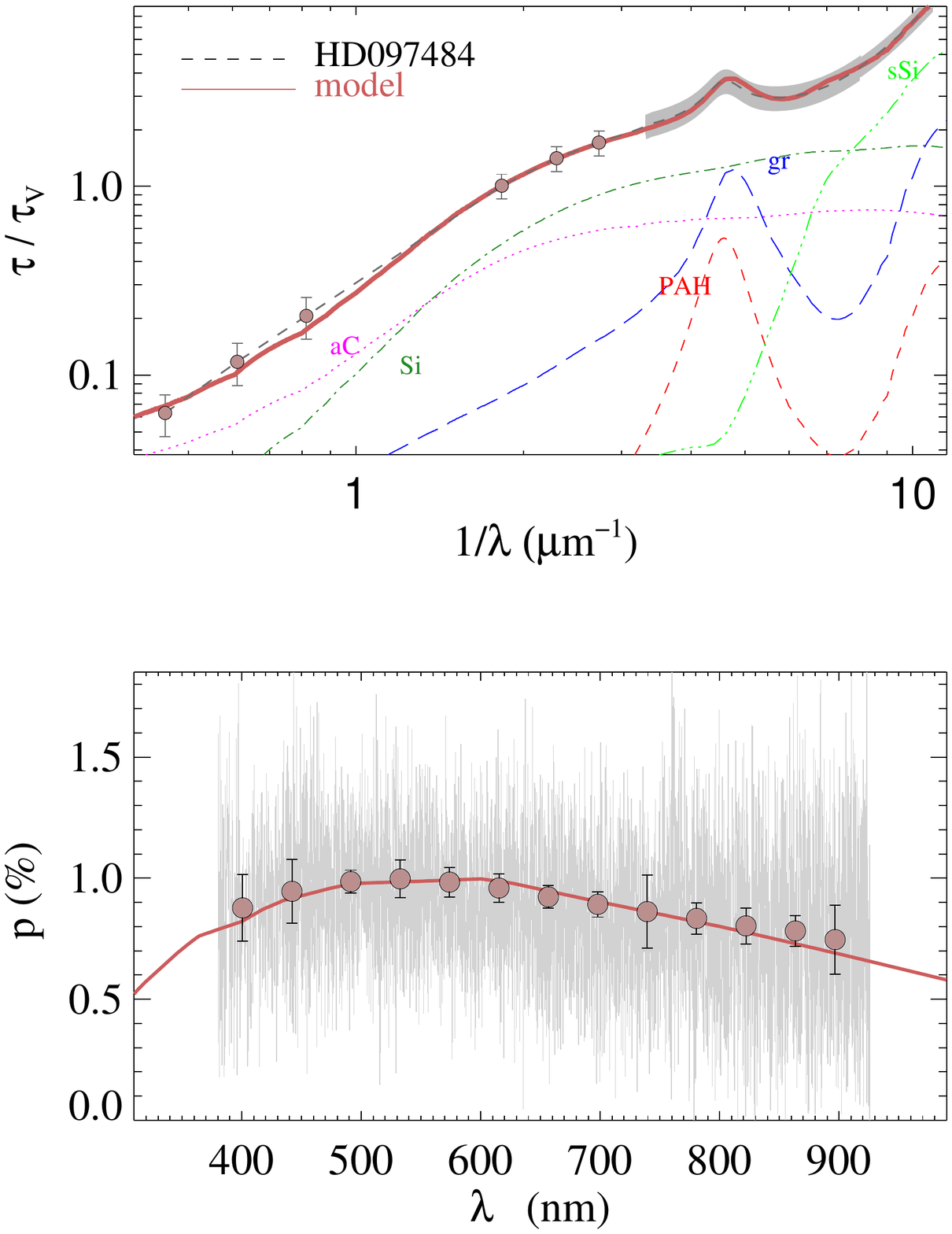}
\caption{Notation same as in Fig.~\ref{appstart.fig}  for HD~97484.}
\end{figure}

\begin{figure} [h!tb]
\includegraphics[width=9.0cm,clip=true,trim=2.1cm 2.7cm 1cm 3.0cm]{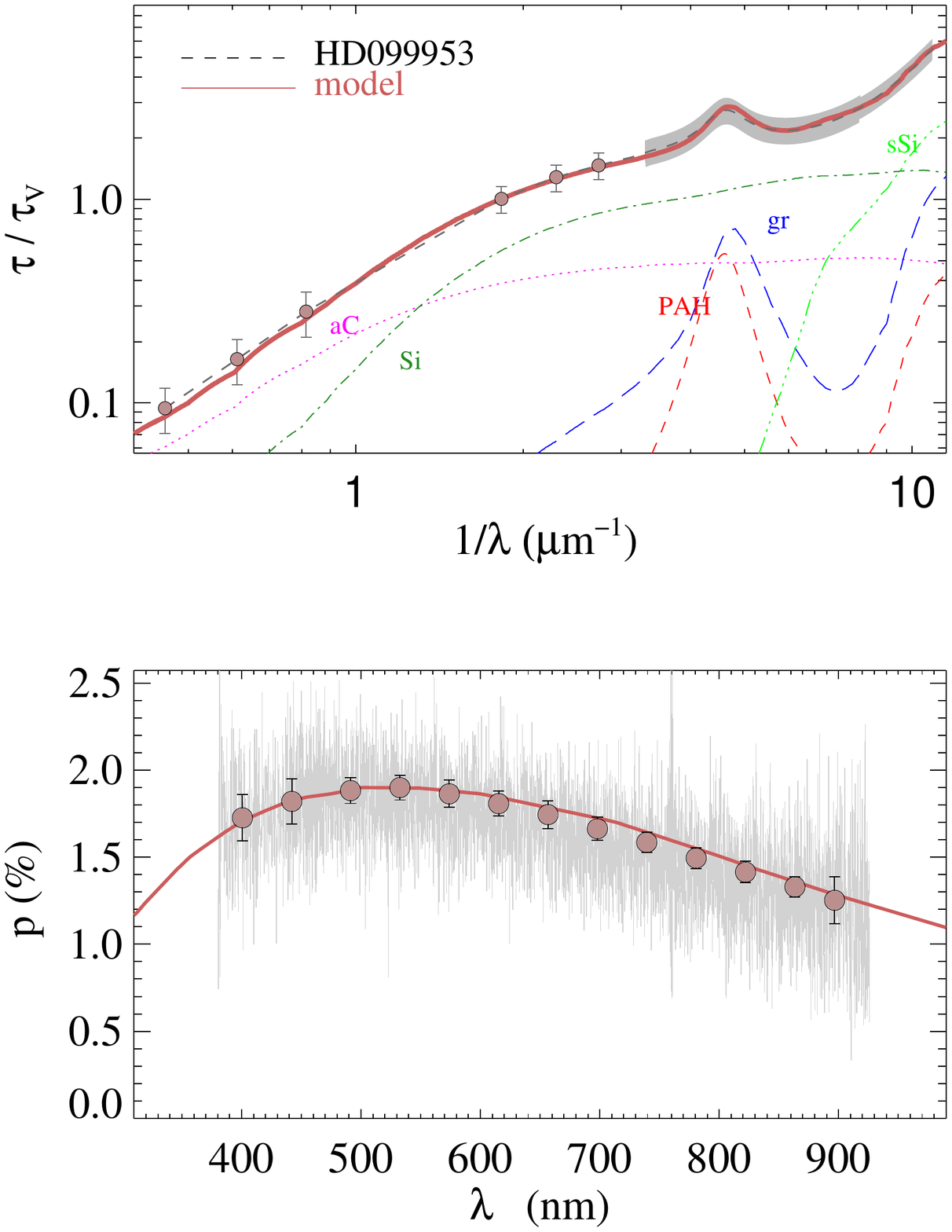}
\caption{Notation same as in Fig.~\ref{appstart.fig}  for HD~99953.}
\end{figure}

\begin{figure} [h!tb]
\includegraphics[width=9.0cm,clip=true,trim=2.1cm 2.7cm 1cm 3.0cm]{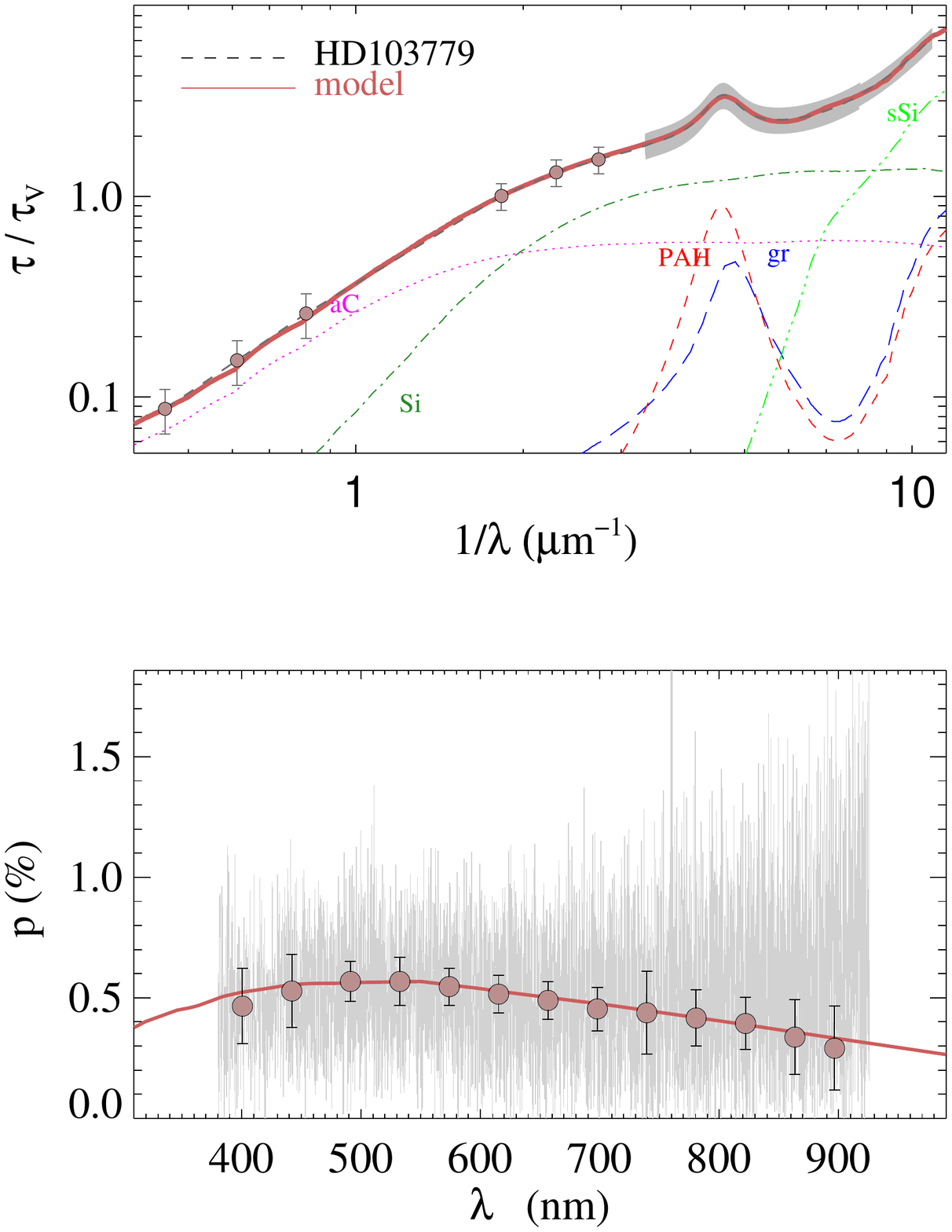}
\caption{Notation same as in Fig.~\ref{appstart.fig}  for HD~103779.}
\end{figure}

\begin{figure} [h!tb]
\includegraphics[width=9.0cm,clip=true,trim=2.1cm 2.7cm 1cm 3.0cm]{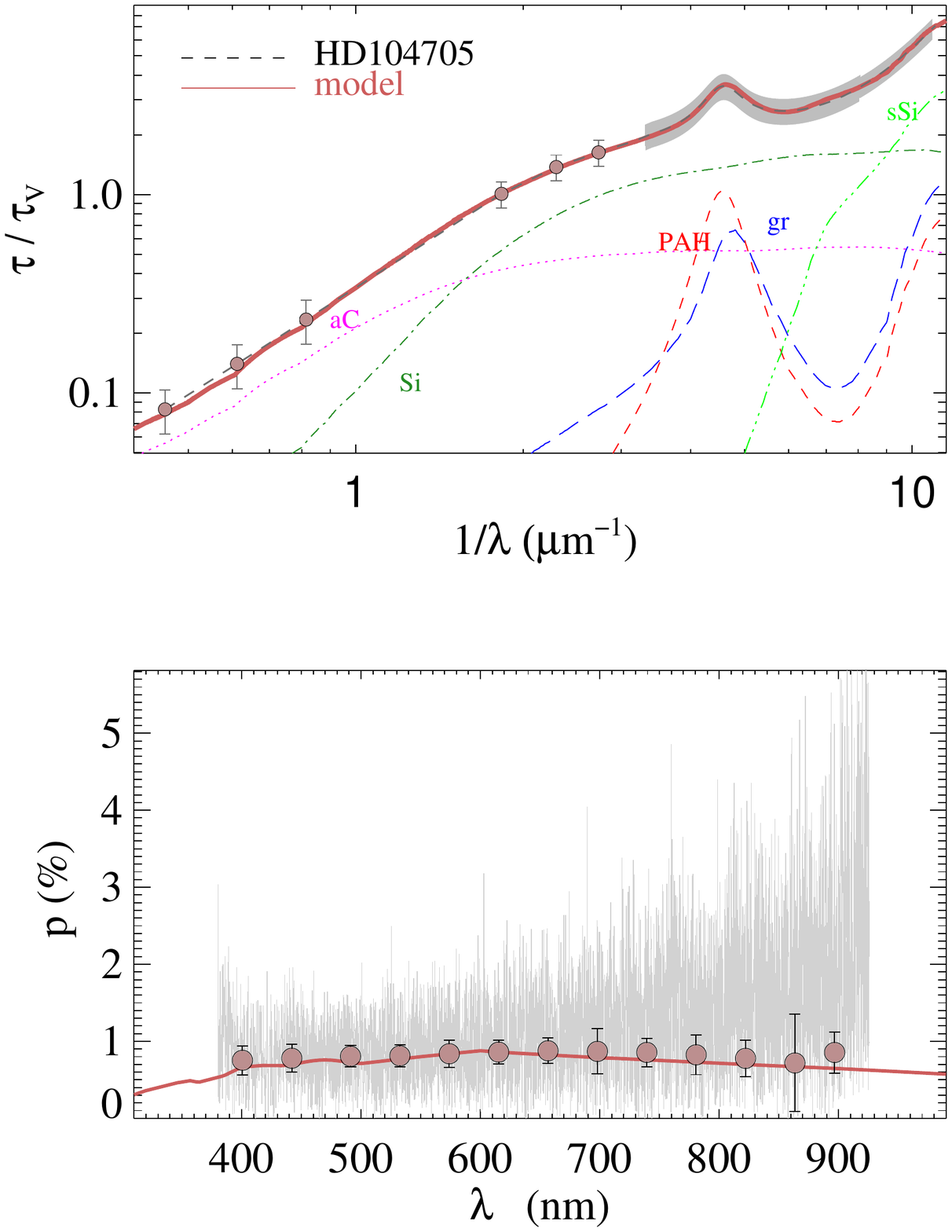}
\caption{Notation same as in Fig.~\ref{appstart.fig}  for HD~104705.}
\end{figure}

\begin{figure} [h!tb]
\includegraphics[width=9.0cm,clip=true,trim=2.1cm 2.7cm 1cm 3.0cm]{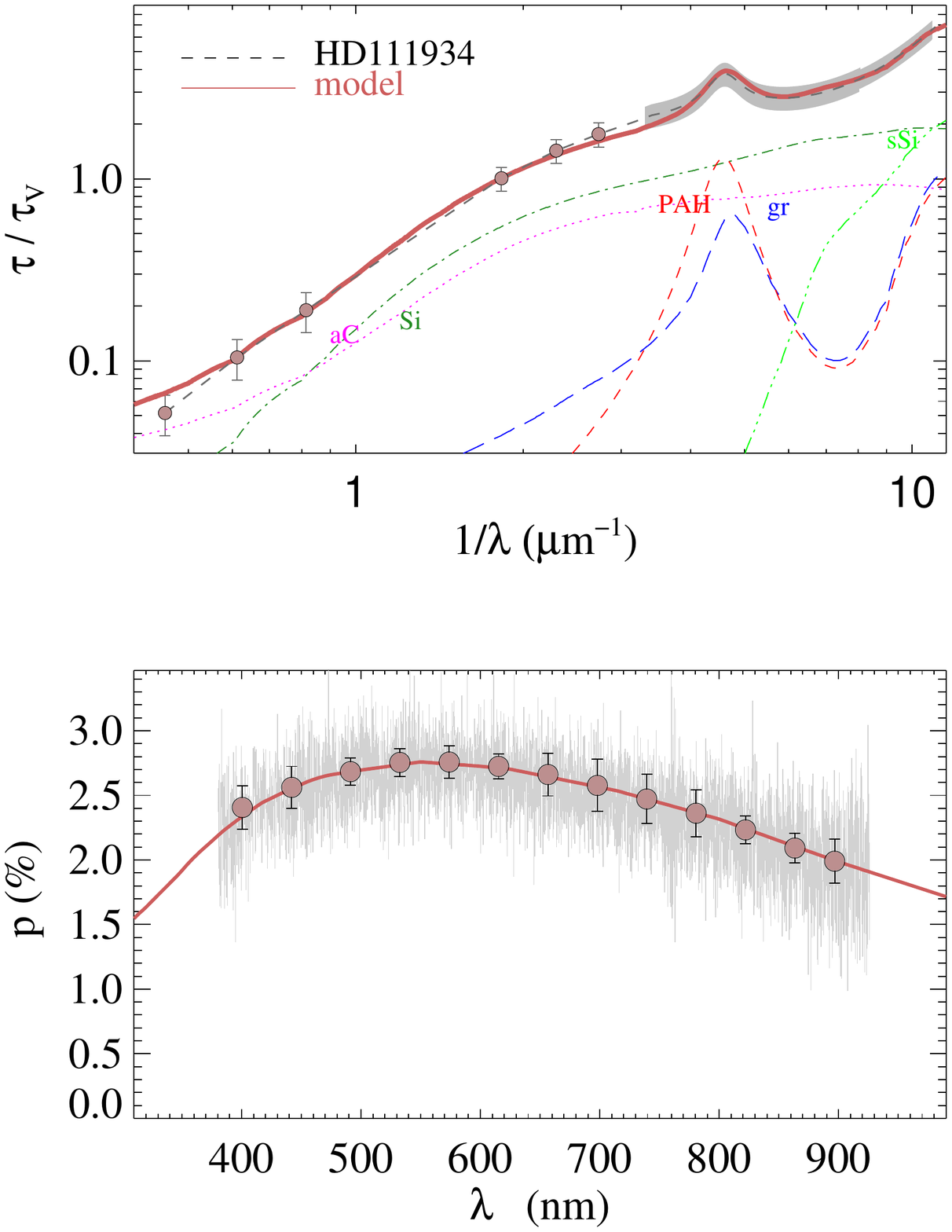}
\caption{Notation same as in Fig.~\ref{appstart.fig}  for HD~111934.}
\end{figure}

\begin{figure} [h!tb]
\includegraphics[width=9.0cm,clip=true,trim=2.1cm 2.7cm 1cm 3.0cm]{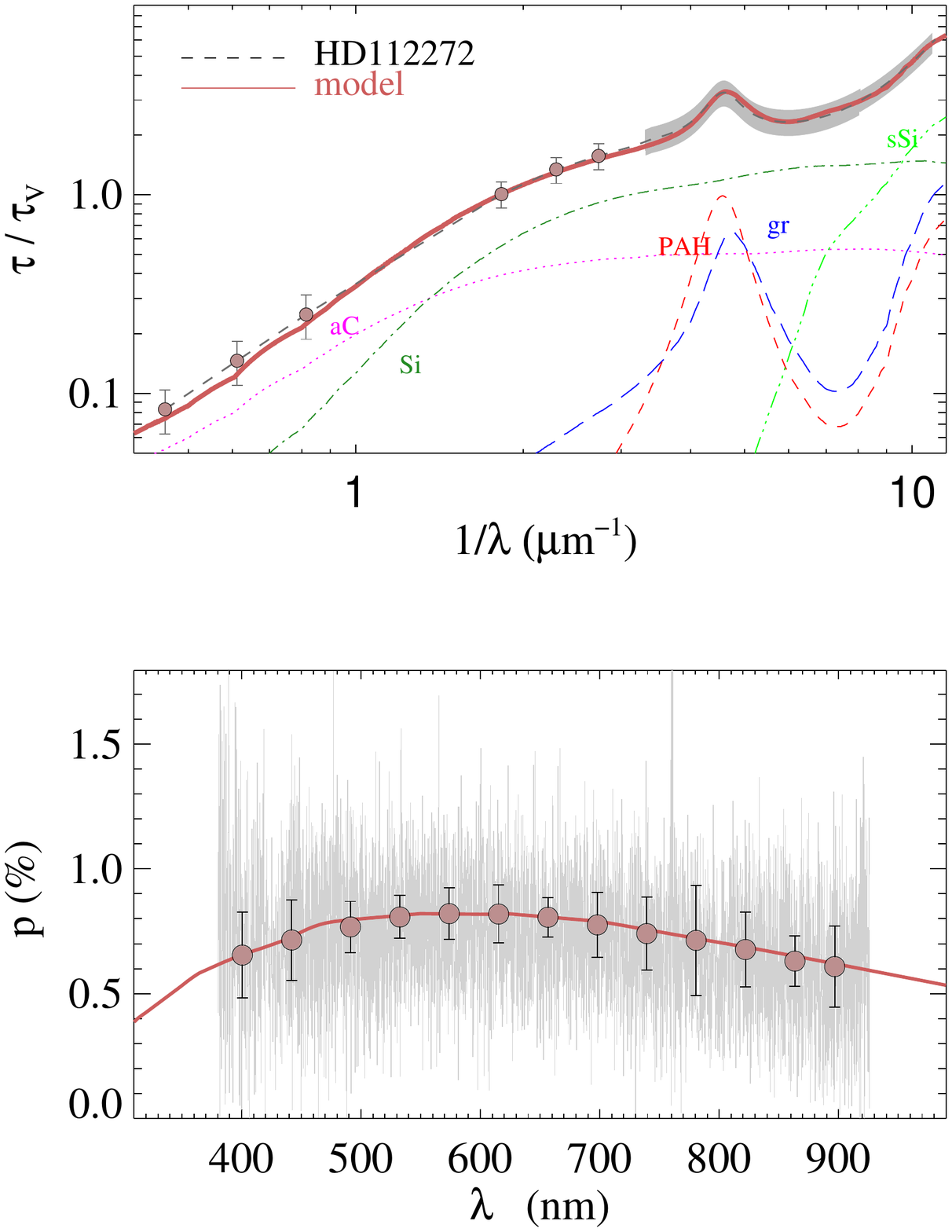}
\caption{Notation same as in Fig.~\ref{appstart.fig}  for HD~112272.}
\end{figure}

\begin{figure} [h!tb]
\includegraphics[width=9.0cm,clip=true,trim=2.1cm 2.7cm 1cm 3.0cm]{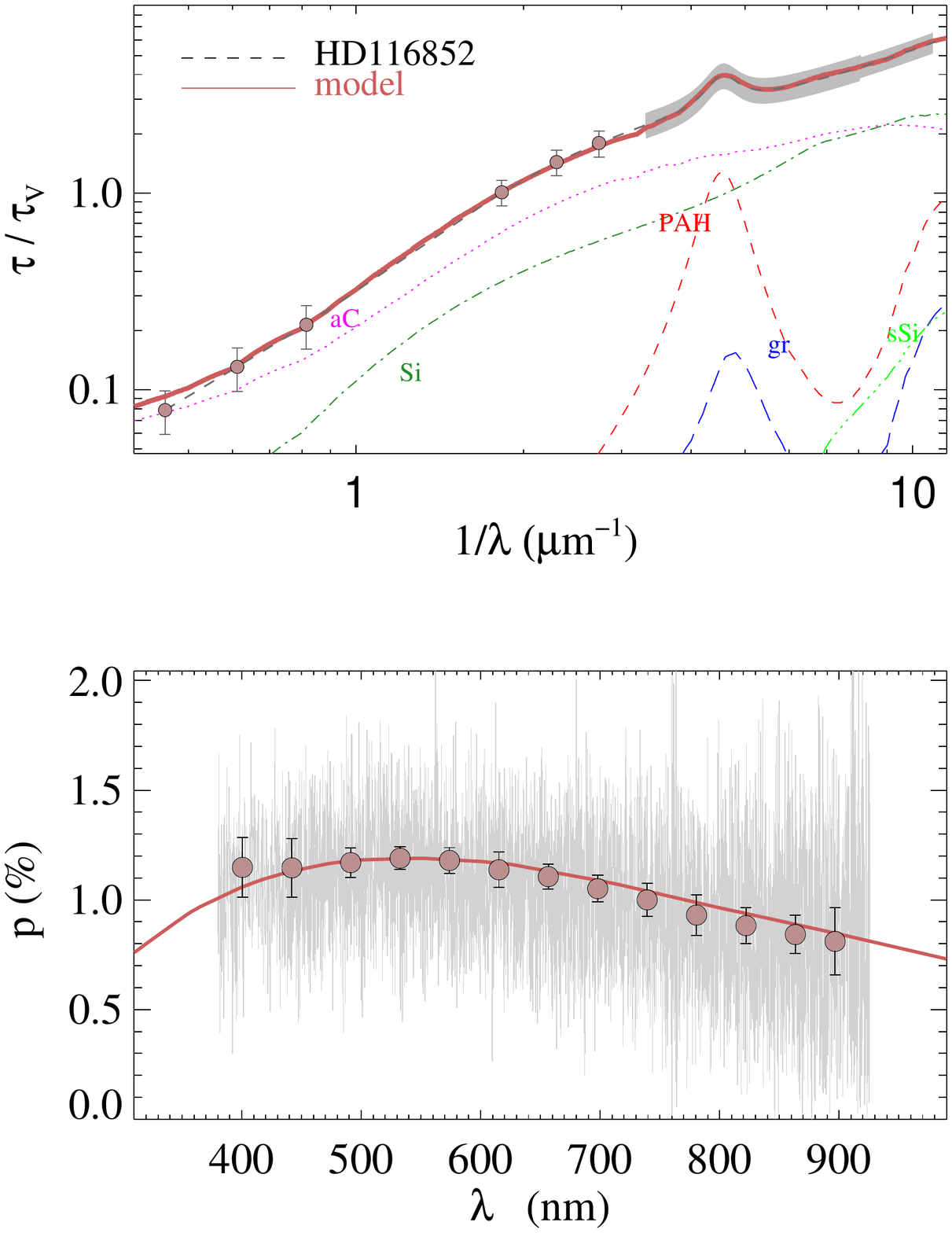}
\caption{Notation same as in Fig.~\ref{appstart.fig}  for HD~116852.}
\end{figure}

\begin{figure} [h!tb]
\includegraphics[width=9.0cm,clip=true,trim=2.1cm 2.7cm 1cm 3.0cm]{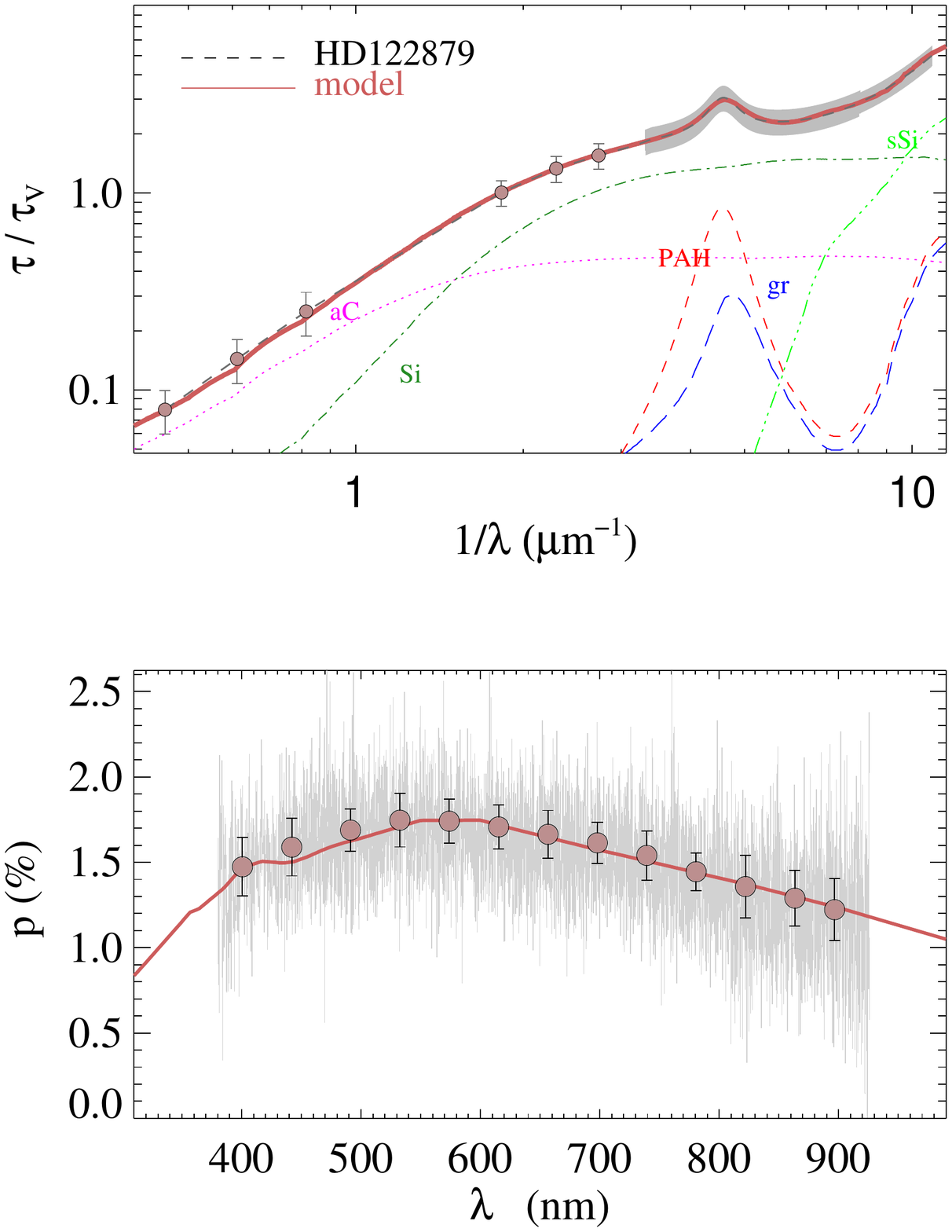}
\caption{Notation same as in Fig.~\ref{appstart.fig}  for HD~122879.}
\end{figure}

\begin{figure} [h!tb]
\includegraphics[width=9.0cm,clip=true,trim=2.1cm 2.7cm 1cm 3.0cm]{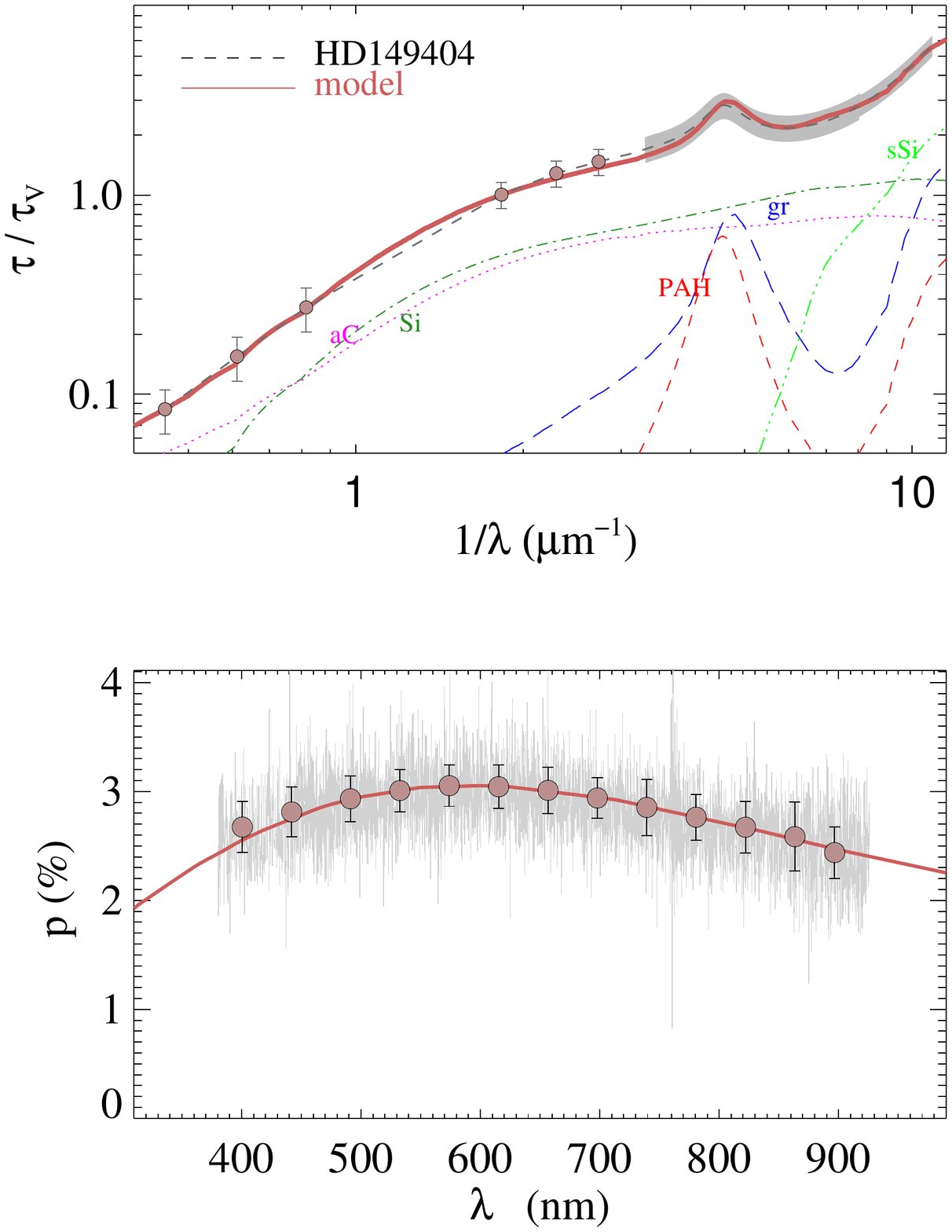}
\caption{Notation same as in Fig.~\ref{appstart.fig}  for HD~149404.}
\end{figure}

\begin{figure} [h!tb]
\includegraphics[width=9.0cm,clip=true,trim=2.1cm 2.7cm 1cm 3.0cm]{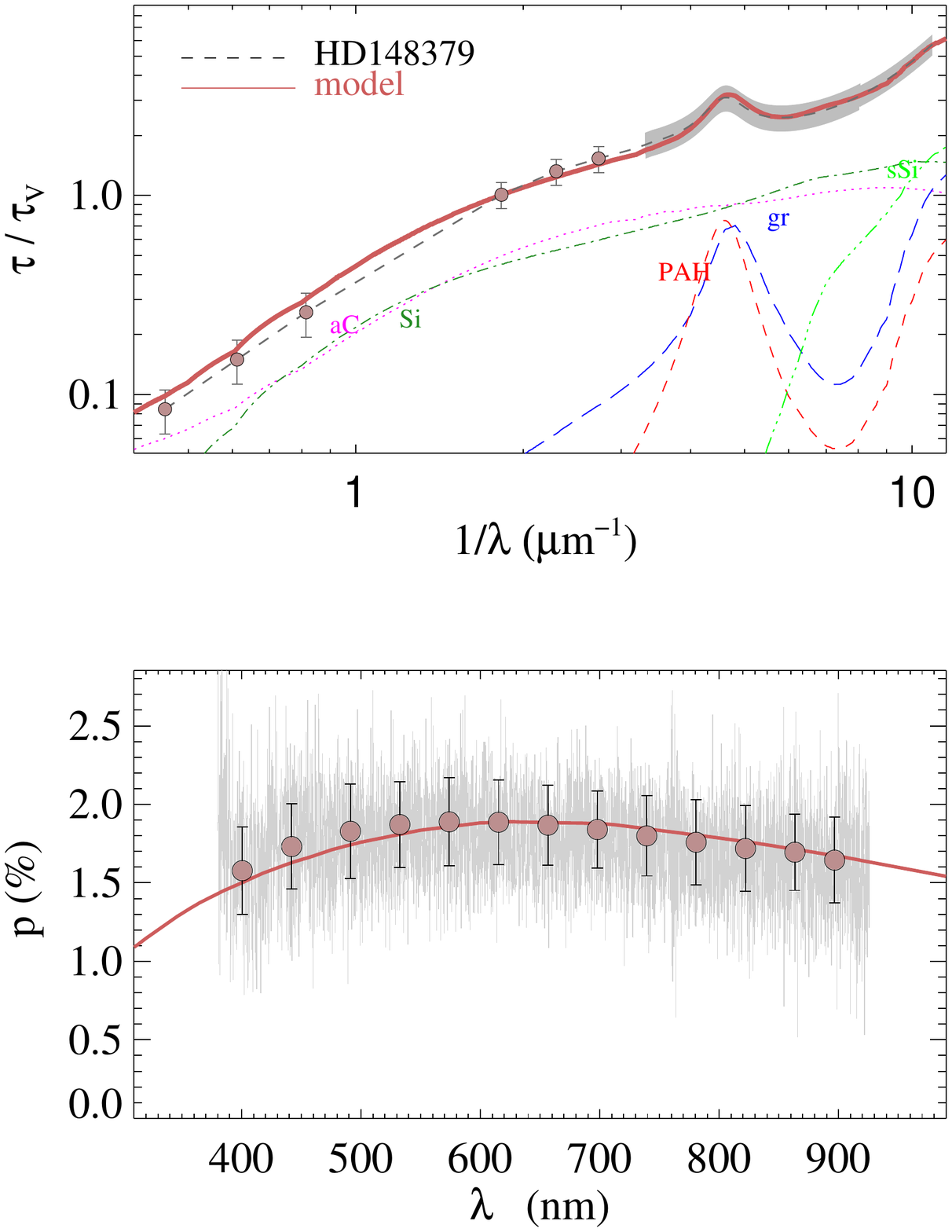}
\caption{Notation same as in Fig.~\ref{appstart.fig}  for HD~148379.}
\end{figure}

\clearpage
\begin{figure} [h!tb]
\includegraphics[width=9.0cm,clip=true,trim=2.1cm 2.7cm 1cm 3.0cm]{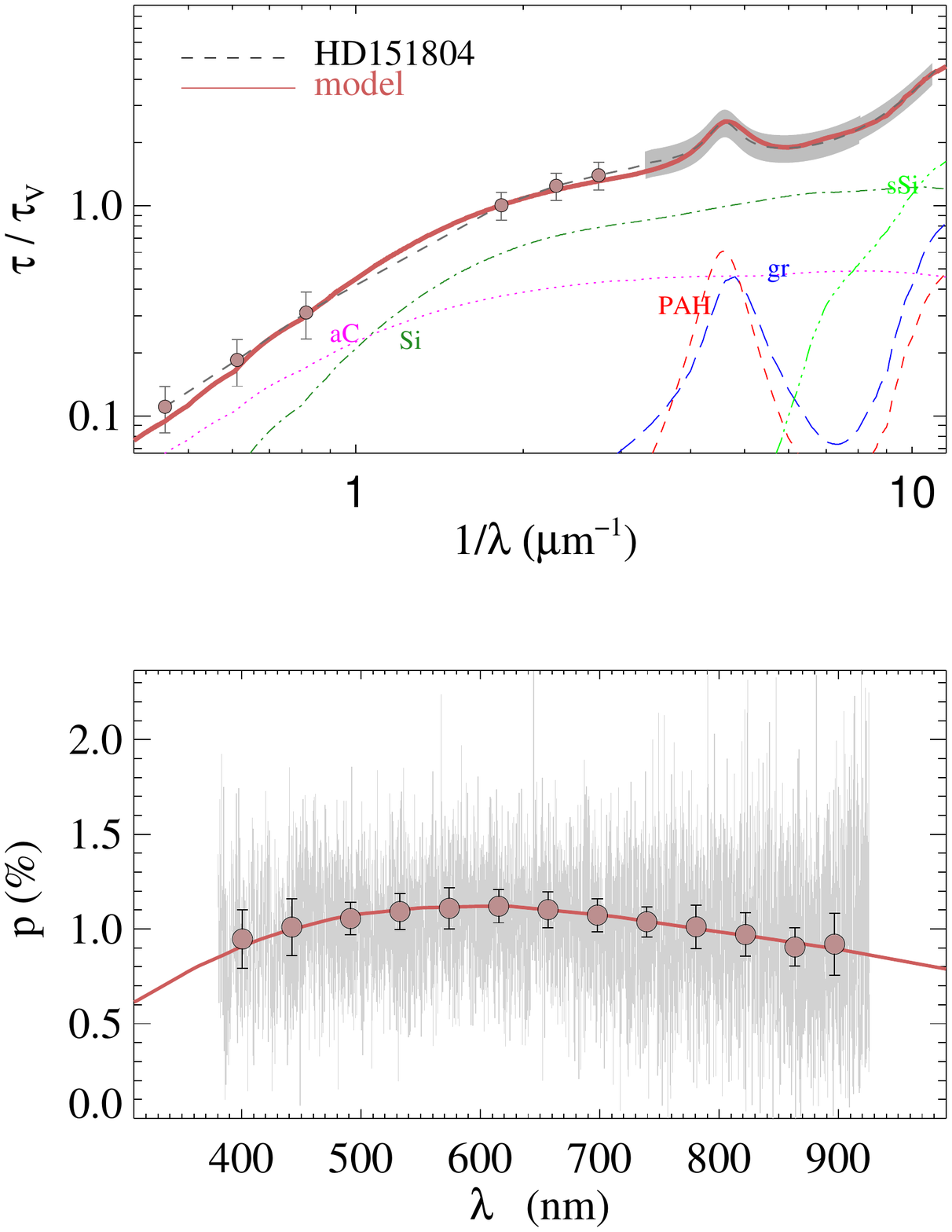}
\caption{Notation same as in Fig.~\ref{appstart.fig}  for HD~151804.}
\end{figure}

\begin{figure} [h!tb]
\includegraphics[width=9.0cm,clip=true,trim=2.1cm 2.7cm 1cm 3.0cm]{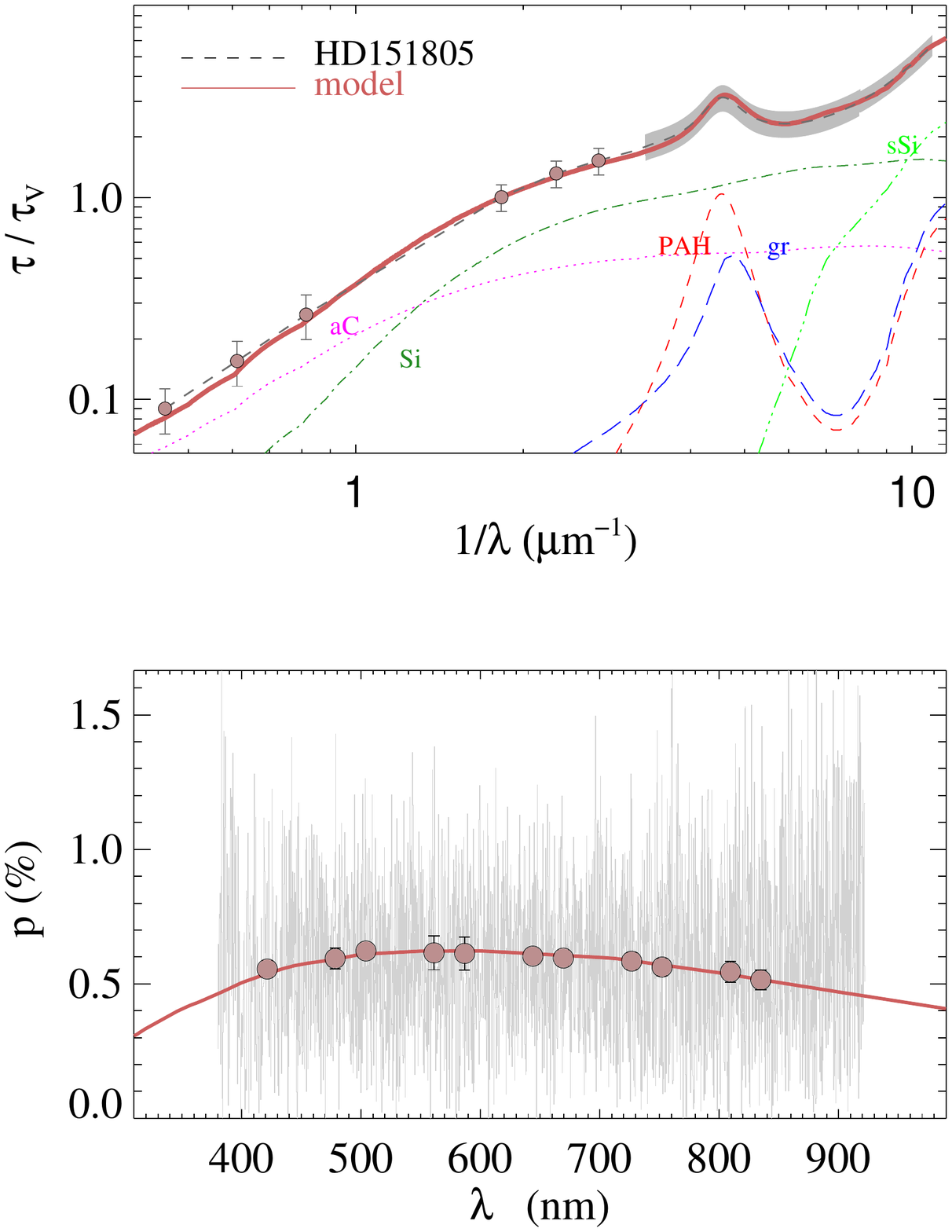}
\caption{Notation same as in Fig.~\ref{appstart.fig}  for HD~151805.}
\end{figure}

\begin{figure} [h!tb]
\includegraphics[width=9.0cm,clip=true,trim=2.1cm 2.7cm 1cm 3.0cm]{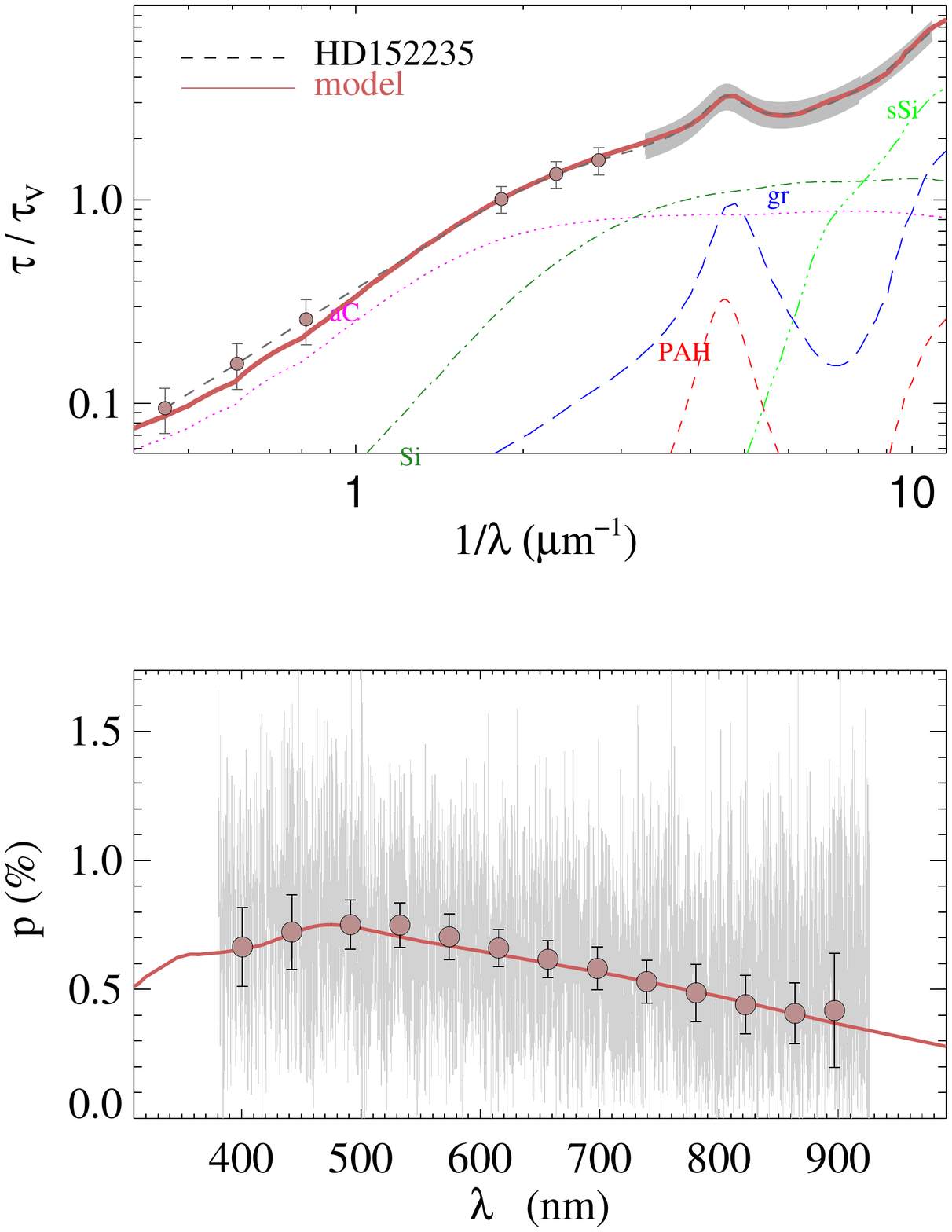}
\caption{Notation same as in Fig.~\ref{appstart.fig}  for HD~152235.}
\end{figure}

\begin{figure} [h!tb]
\includegraphics[width=9.0cm,clip=true,trim=2.1cm 2.7cm 1cm 3.0cm]{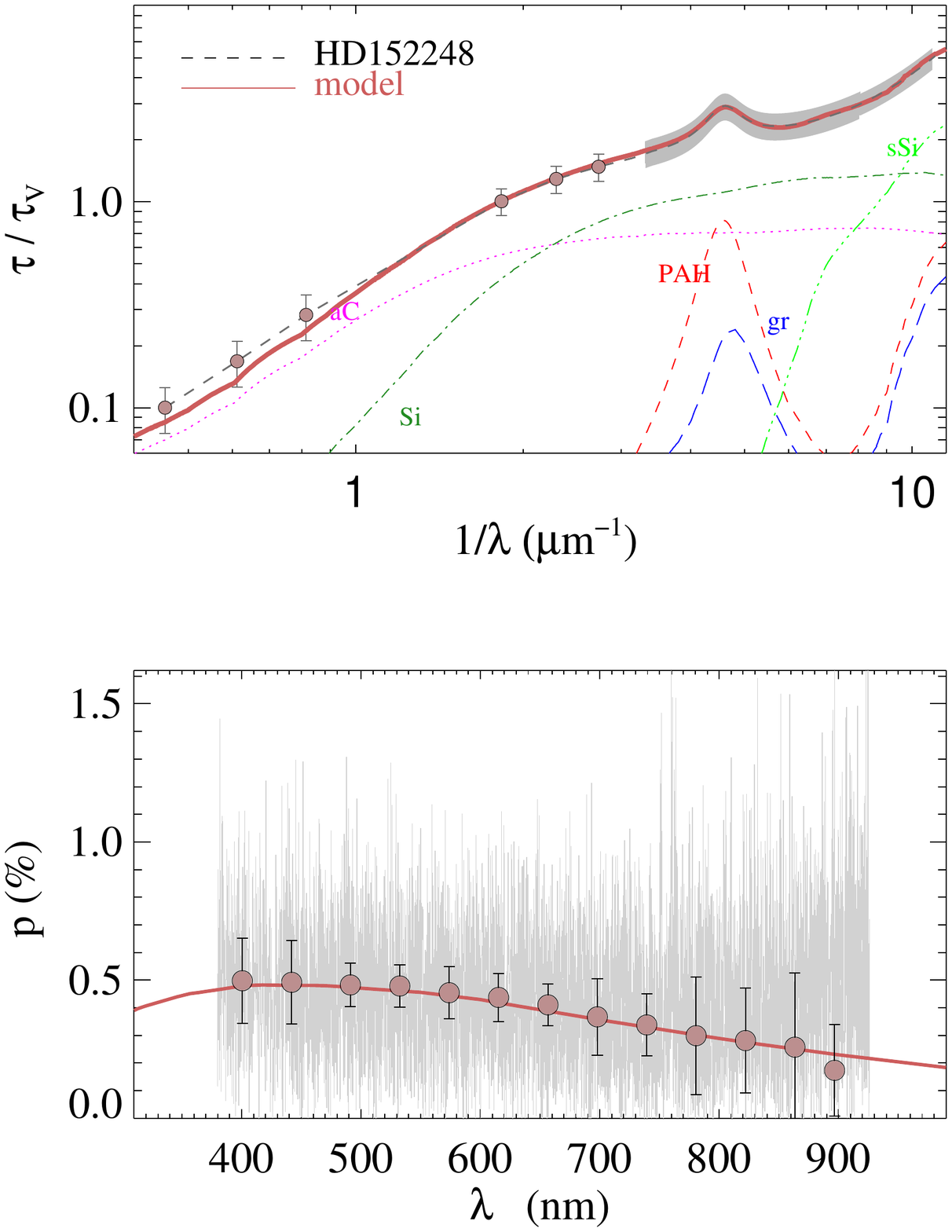}
\caption{Notation same as in Fig.~\ref{appstart.fig}  for HD~152248.}
\end{figure}

\begin{figure} [h!tb]
\includegraphics[width=9.0cm,clip=true,trim=2.1cm 2.7cm 1cm 3.0cm]{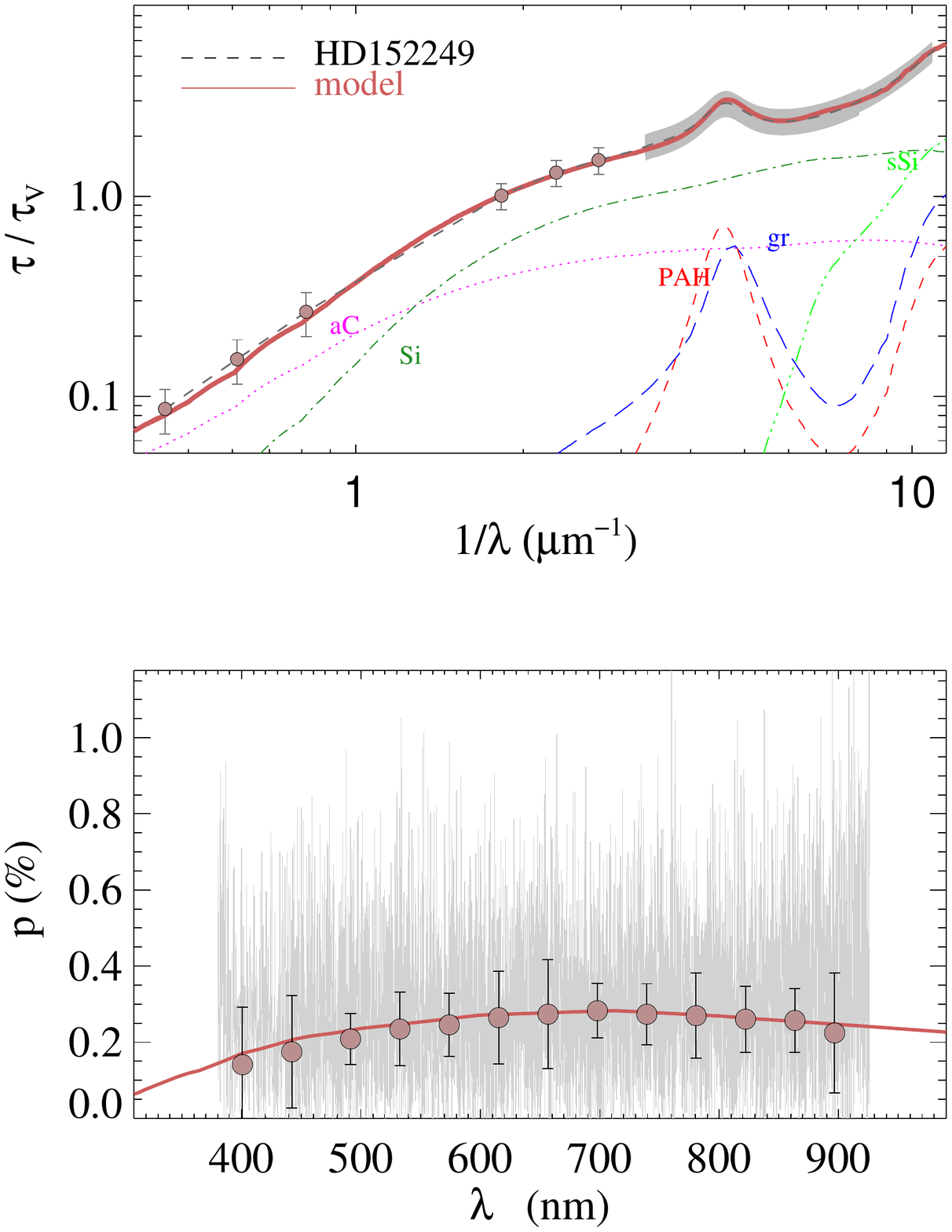}
\caption{Notation same as in Fig.~\ref{appstart.fig}  for HD~152249.}
\end{figure}

\begin{figure} [h!tb]
\includegraphics[width=9.0cm,clip=true,trim=2.1cm 2.7cm 1cm 3.0cm]{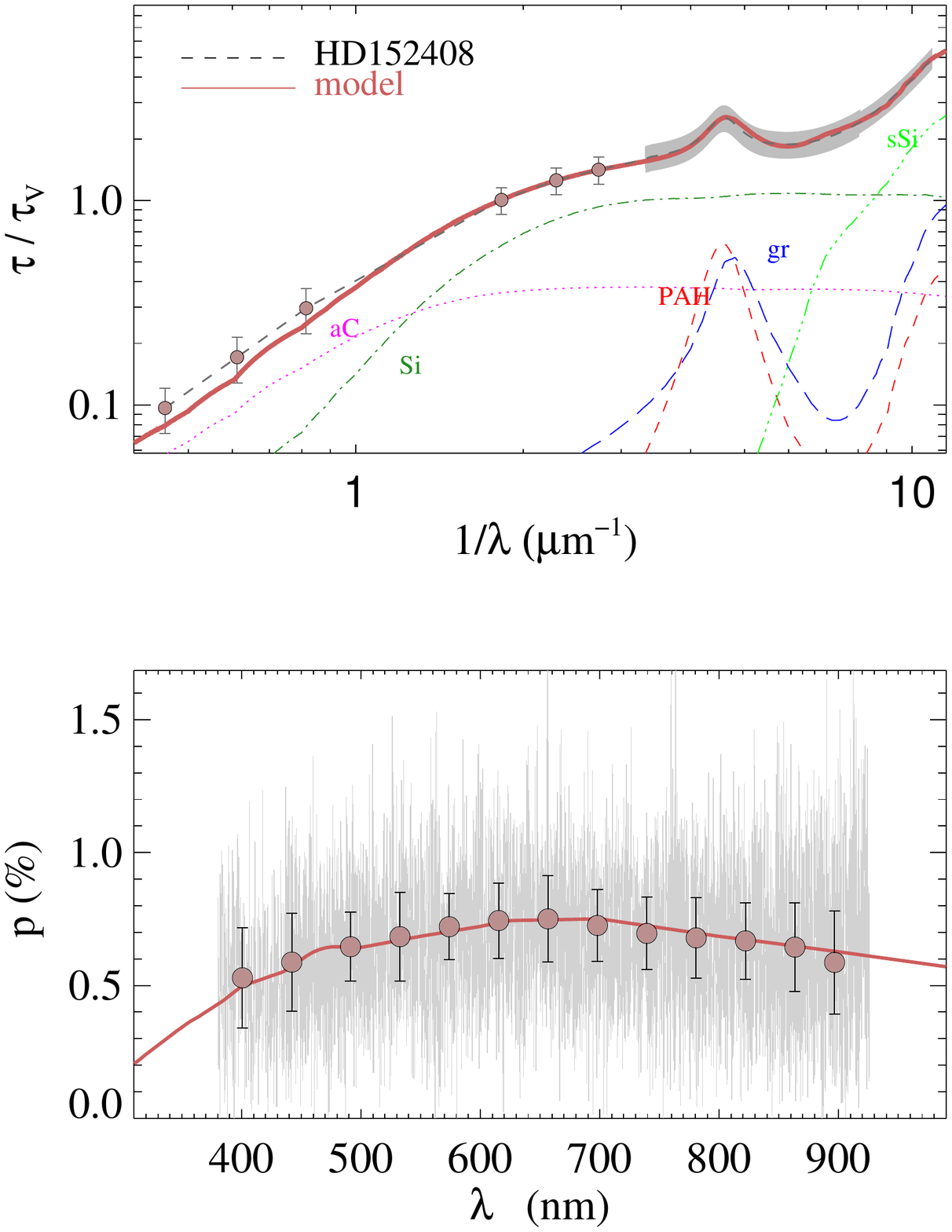}
\caption{Notation same as in Fig.~\ref{appstart.fig}  for HD~152408.}
\end{figure}

\begin{figure} [h!tb]
\includegraphics[width=9.0cm,clip=true,trim=2.1cm 2.7cm 1cm 3.0cm]{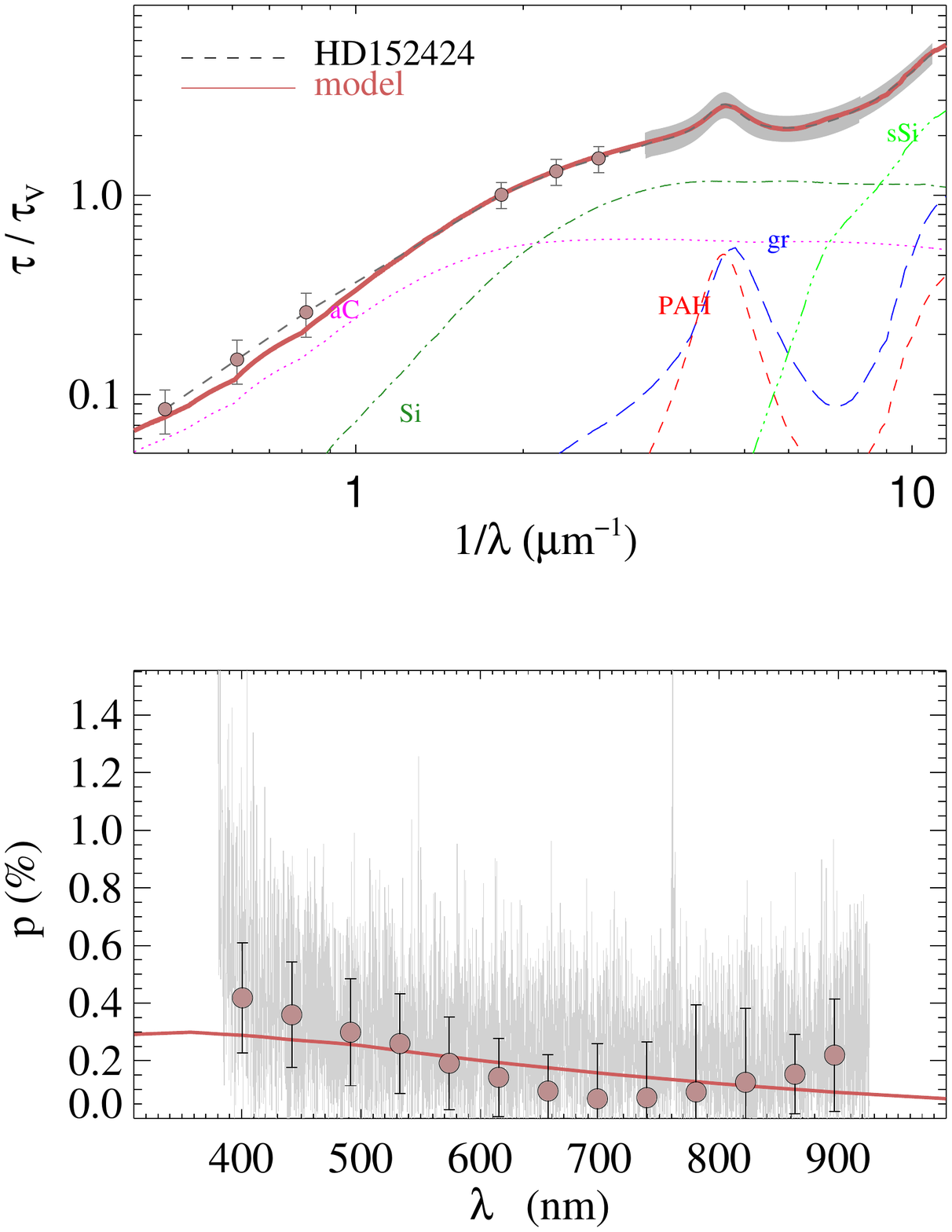}
\caption{Notation same as in Fig.~\ref{appstart.fig}  for HD~152424.}
\end{figure}

\begin{figure} [h!tb]
\includegraphics[width=9.0cm,clip=true,trim=2.1cm 2.7cm 1cm 3.0cm]{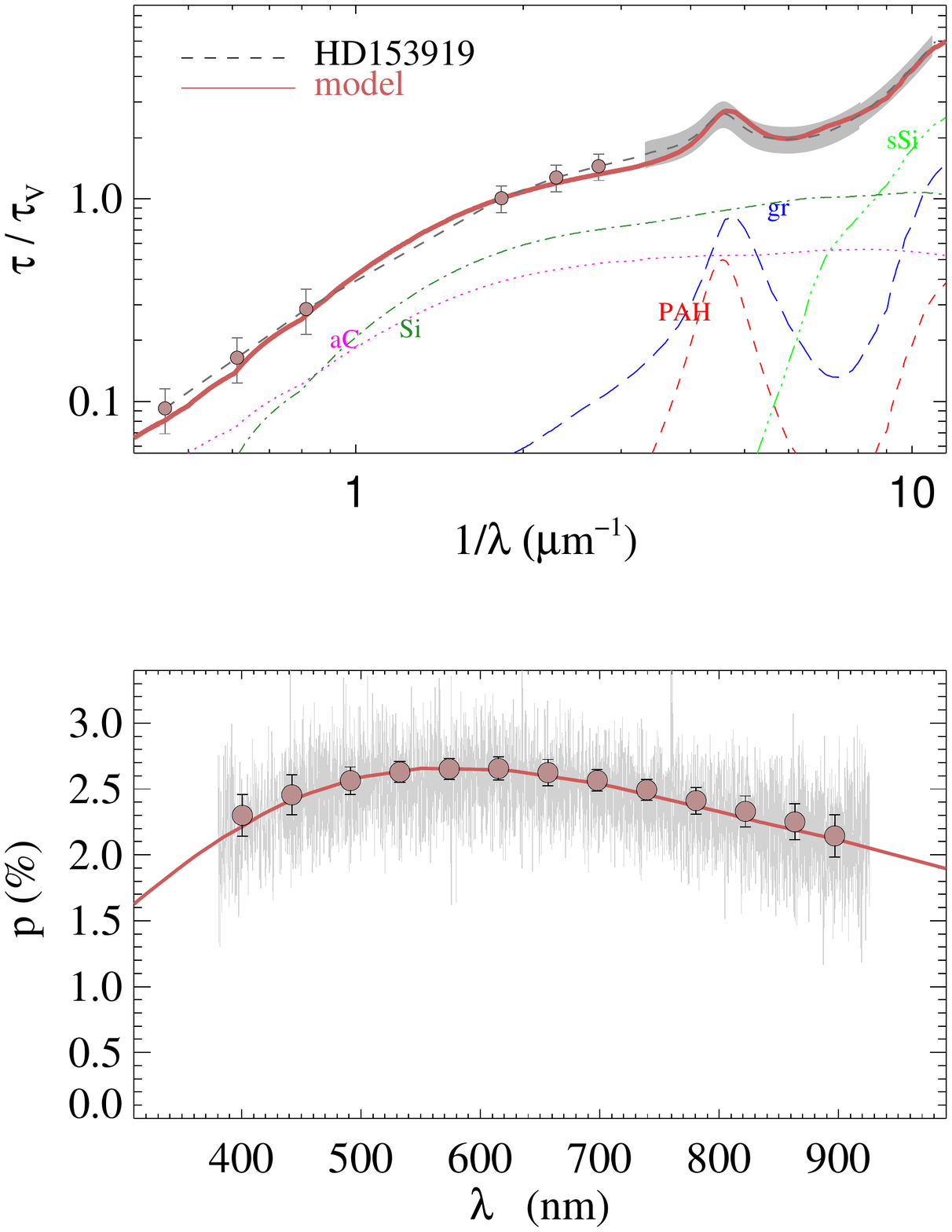}
\caption{Notation same as in Fig.~\ref{appstart.fig}  for HD~153919.}
\end{figure}

\begin{figure} [h!tb]
\includegraphics[width=9.0cm,clip=true,trim=2.1cm 2.7cm 1cm 3.0cm]{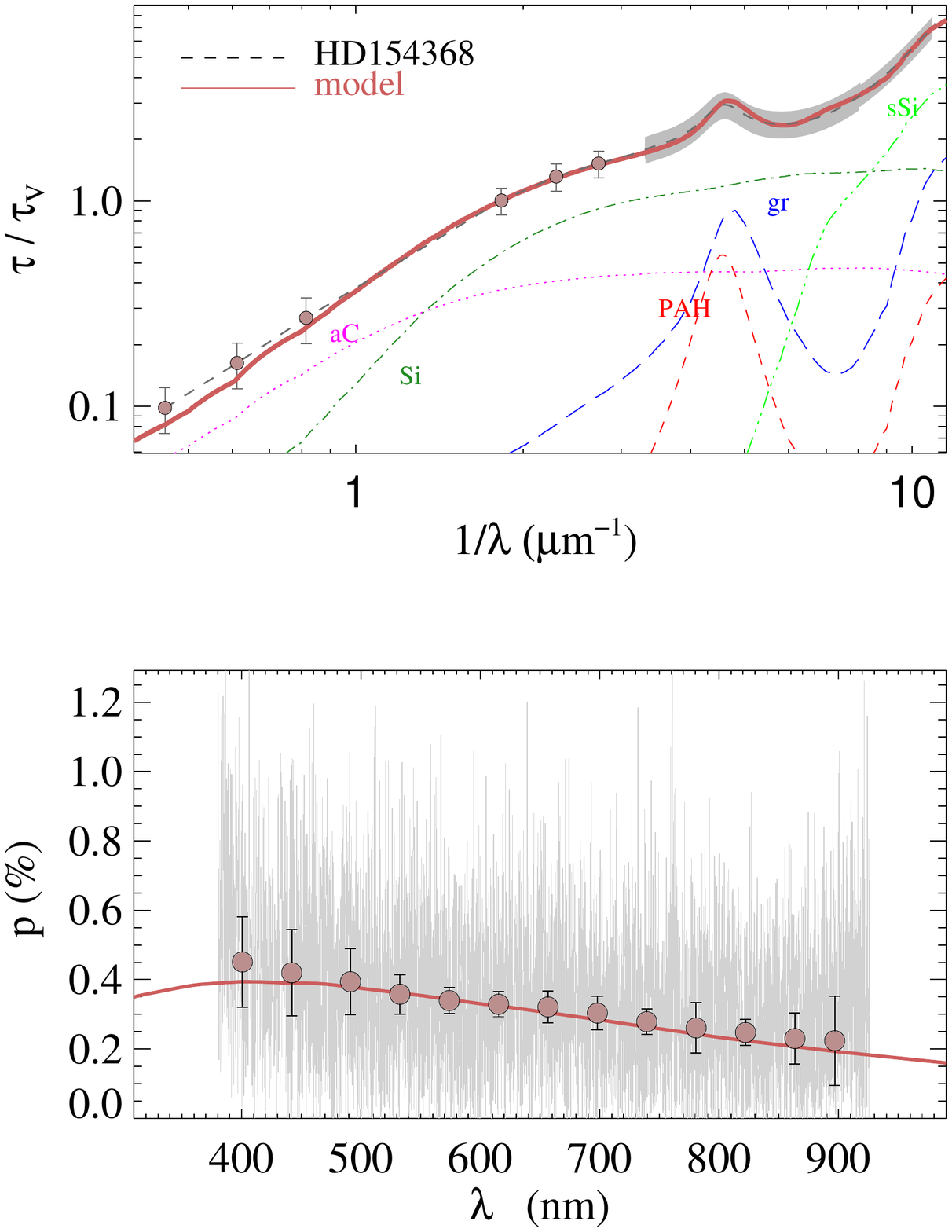}
\caption{Notation same as in Fig.~\ref{appstart.fig}  for HD~154368.}
\end{figure}

\begin{figure} [h!tb]
\includegraphics[width=9.0cm,clip=true,trim=2.1cm 2.7cm 1cm 3.0cm]{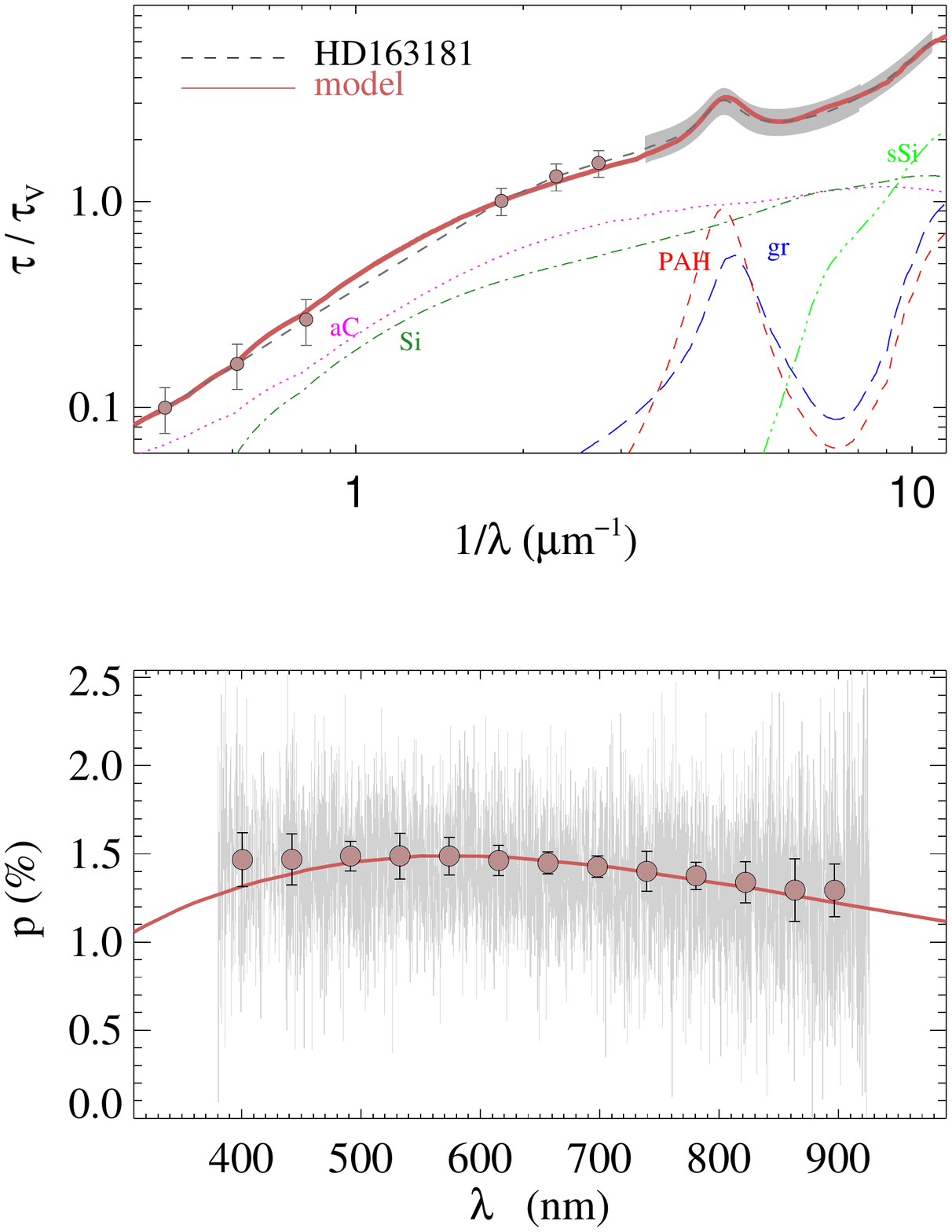}
\caption{Notation same as in Fig.~\ref{appstart.fig}  for HD~163181.}
\end{figure}

\begin{figure} [h!tb]
\includegraphics[width=9.0cm,clip=true,trim=2.1cm 2.7cm 1cm 3.0cm]{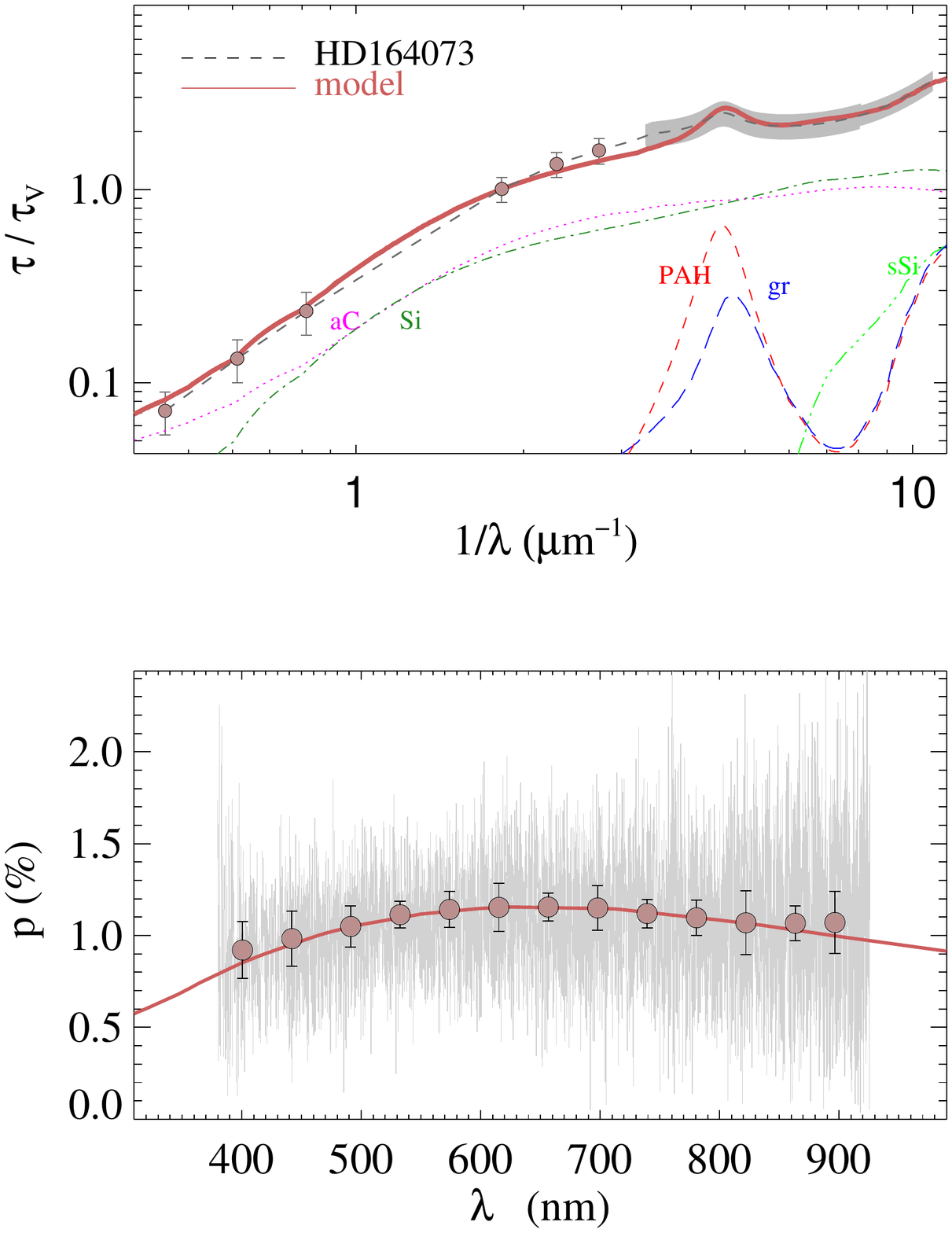}
\caption{Notation same as in Fig.~\ref{appstart.fig}  for HD~164073.}
\end{figure}

\begin{figure} [h!tb]
\includegraphics[width=9.0cm,clip=true,trim=2.1cm 2.7cm 1cm 3.0cm]{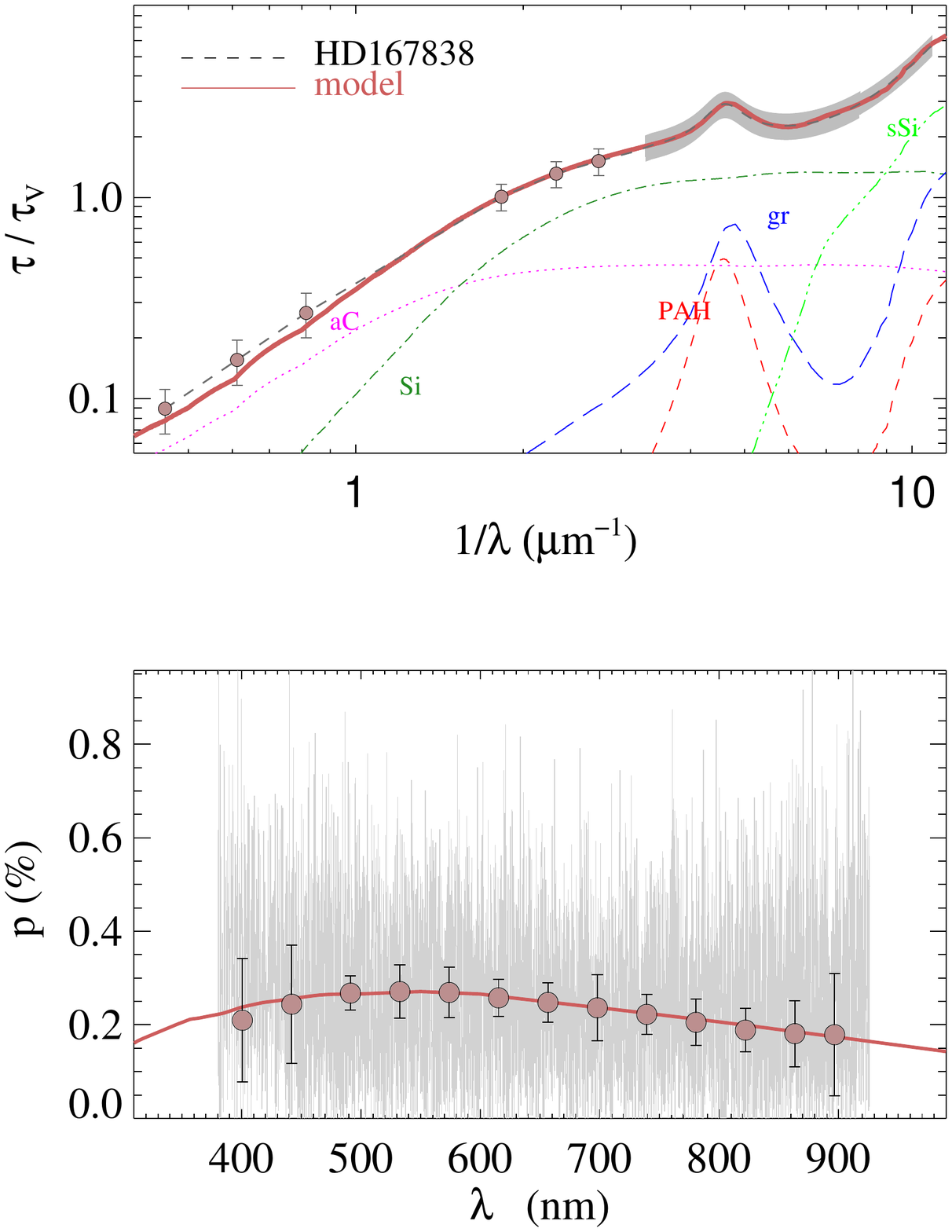}
\caption{Notation same as in Fig.~\ref{appstart.fig}  for HD~167838.}
\end{figure}

\begin{figure} [h!tb]
\includegraphics[width=9.0cm,clip=true,trim=2.1cm 2.7cm 1cm 3.0cm]{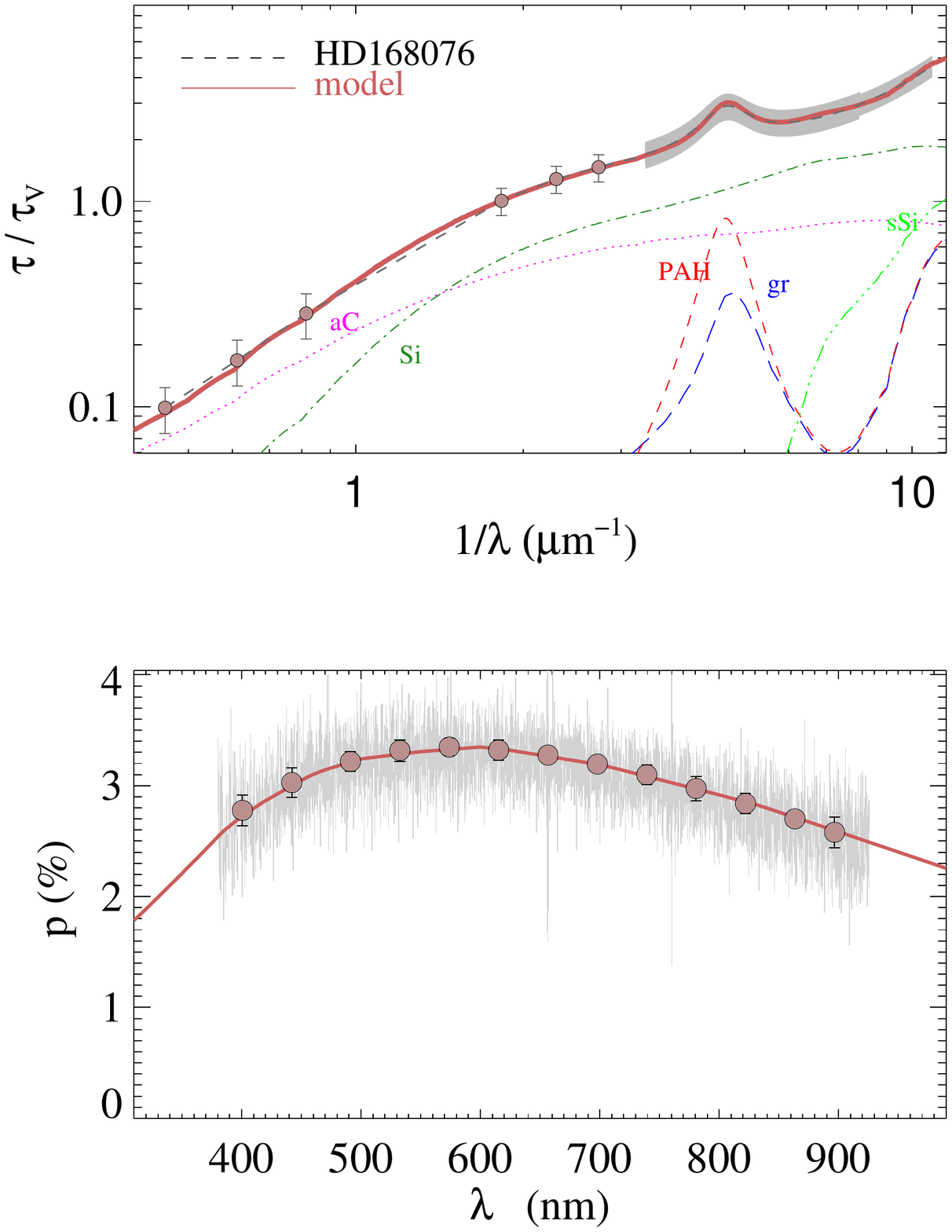}
\caption{Notation same as in Fig.~\ref{appstart.fig}  for HD~168076.}
\end{figure}

\begin{figure} [h!tb]
\includegraphics[width=9.0cm,clip=true,trim=2.1cm 2.7cm 1cm 3.0cm]{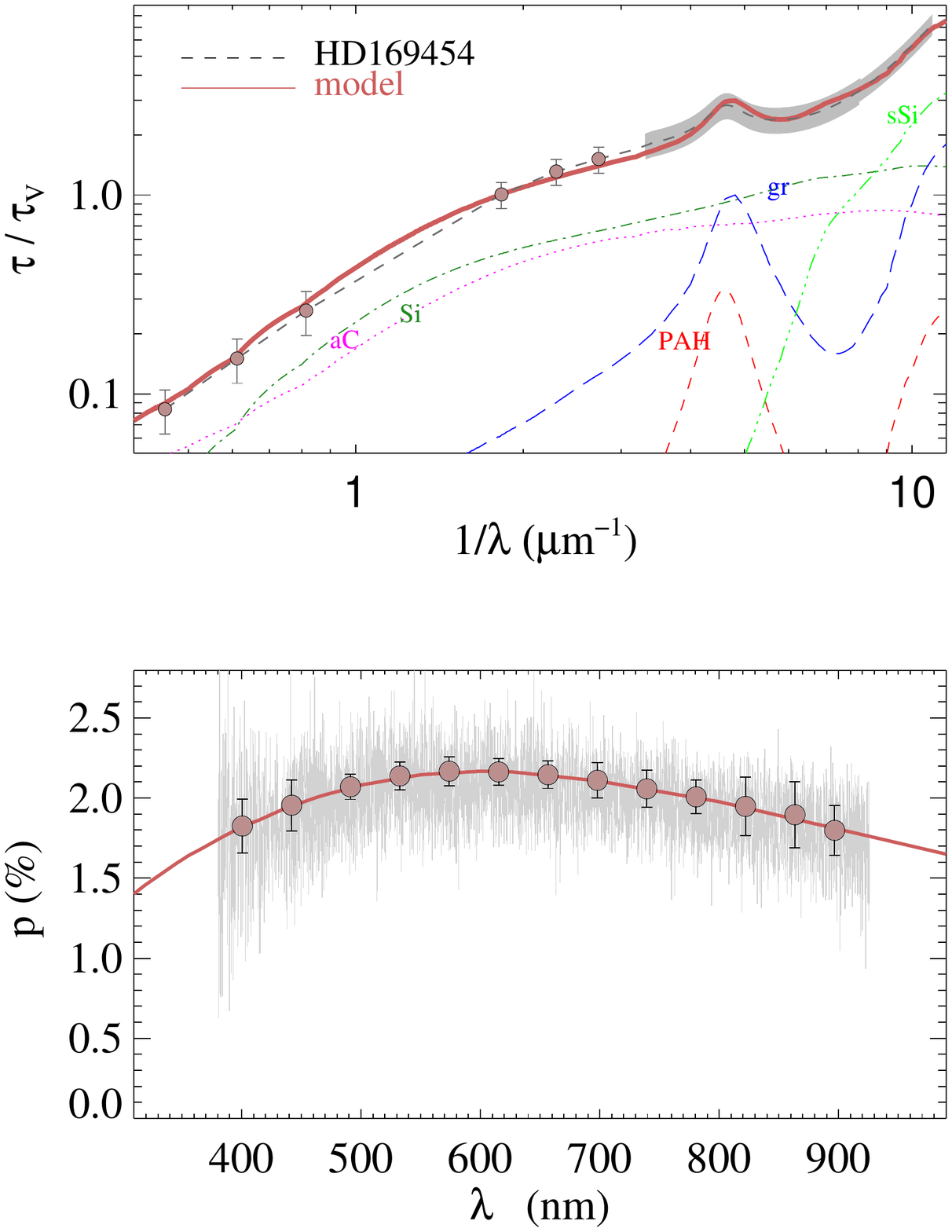}
\caption{Notation same as in Fig.~\ref{appstart.fig}  for HD~169454.}
\end{figure}

\begin{figure} [h!tb]
\includegraphics[width=9.0cm,clip=true,trim=2.1cm 2.7cm 1cm 3.0cm]{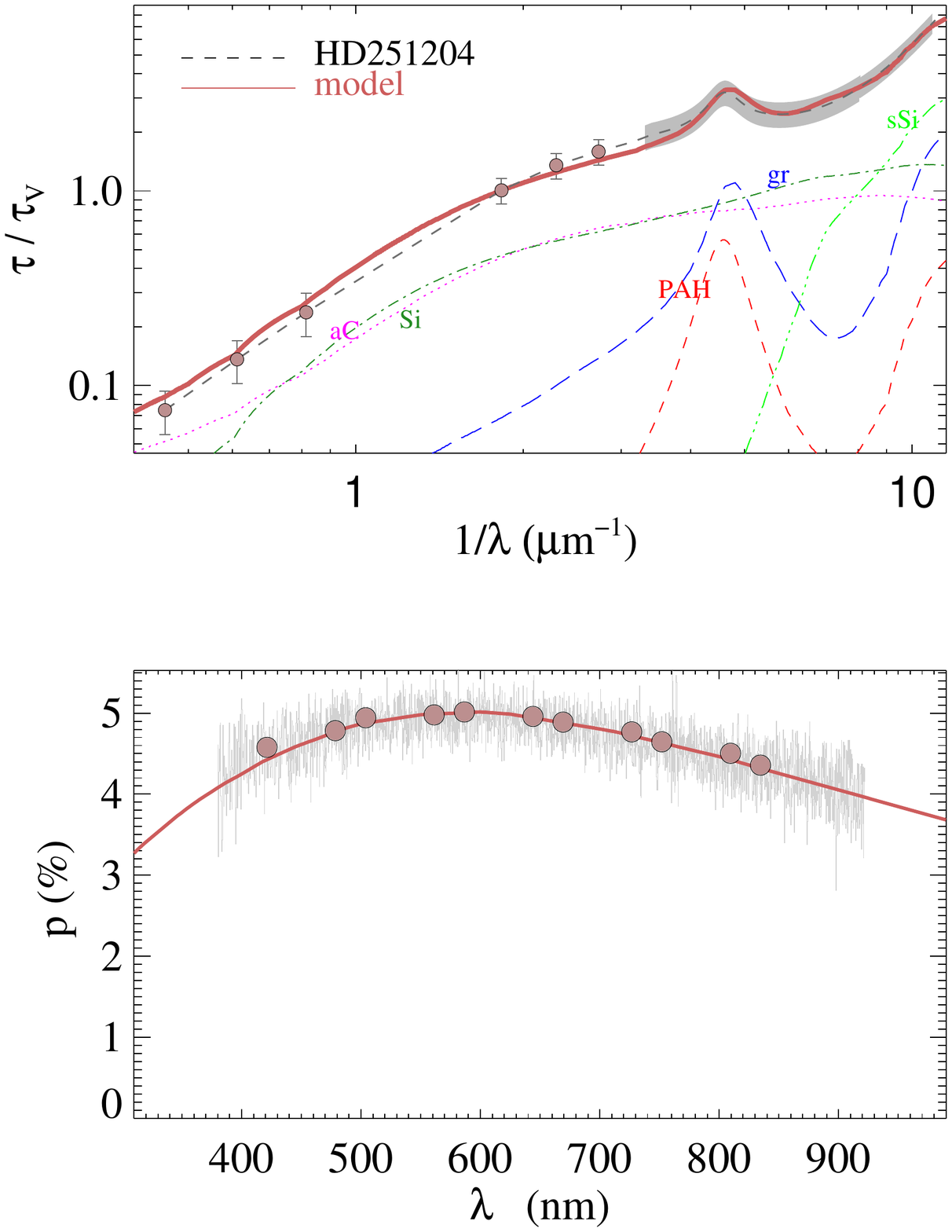}
\caption{Notation same as in Fig.~\ref{appstart.fig}  for HD~251204.}
\end{figure}

\begin{figure} [h!tb]
\includegraphics[width=9.0cm,clip=true,trim=2.1cm 2.7cm 1cm 3.0cm]{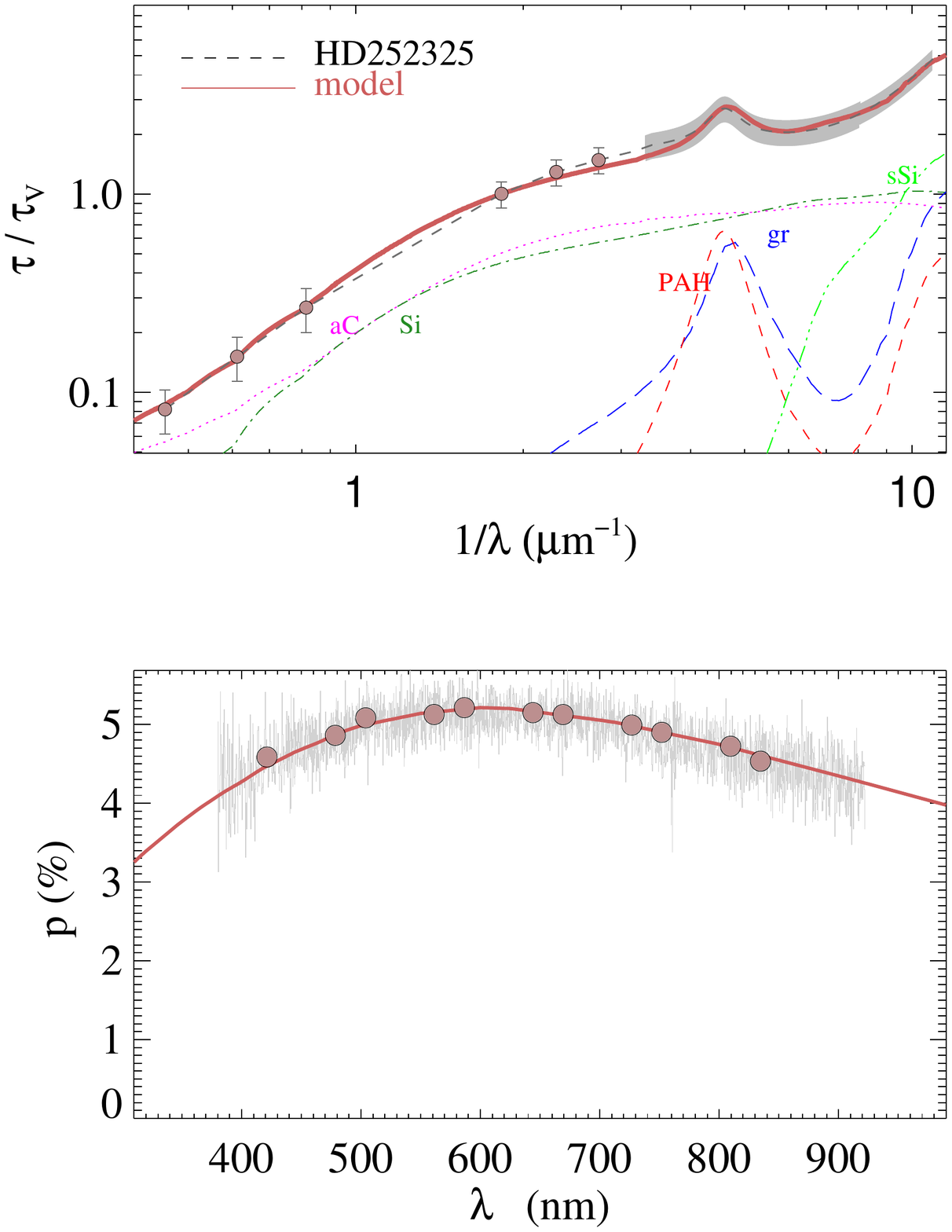}
\caption{Notation same as in Fig.~\ref{appstart.fig}  for HD~252325.}
\end{figure}

\clearpage
\begin{figure} [h!tb]
\includegraphics[width=9.0cm,clip=true,trim=2.1cm 2.7cm 1cm 3.0cm]{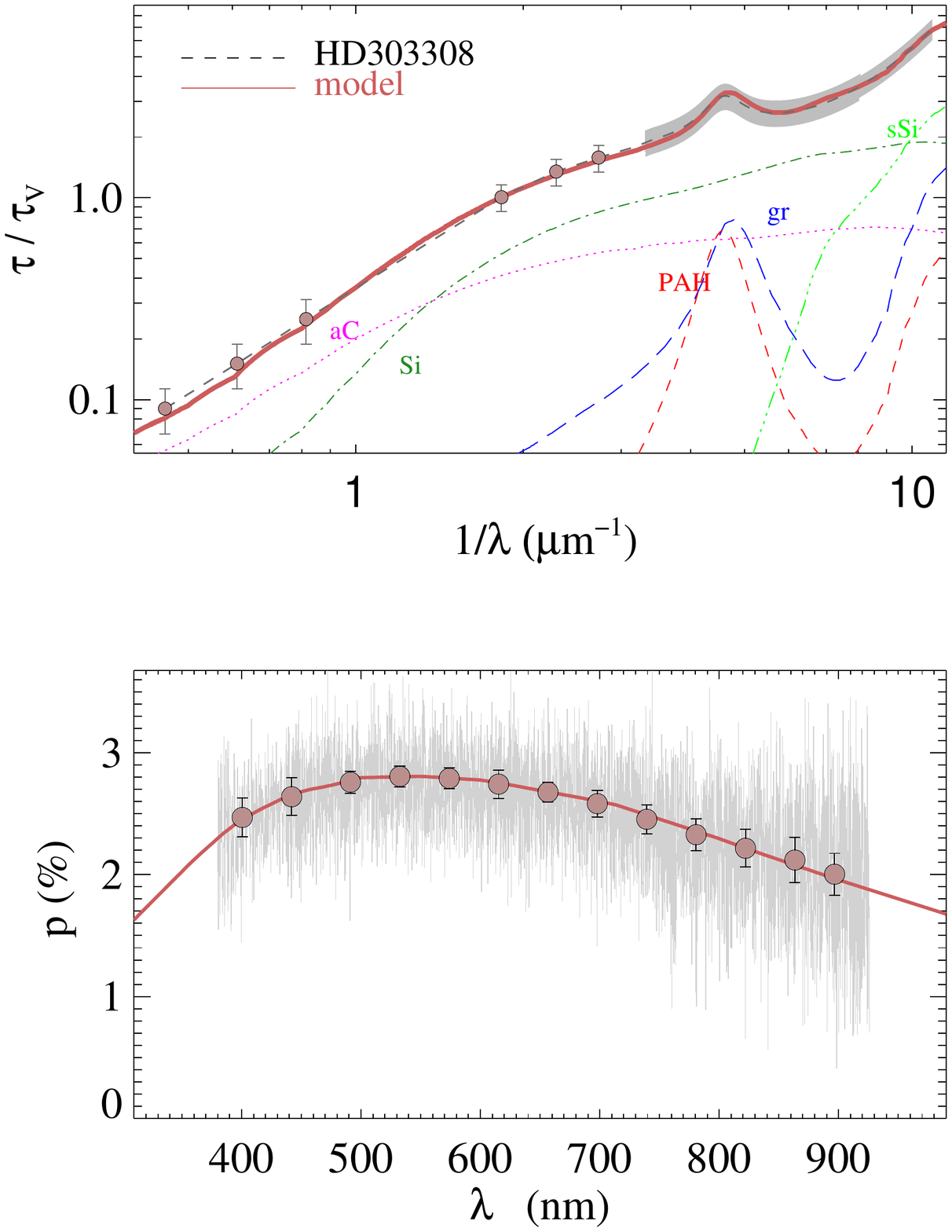}
\caption{Notation same as in Fig.~\ref{appstart.fig}  for HD~303308.}
\end{figure}

\begin{figure} [h!tb]
\includegraphics[width=9.0cm,clip=true,trim=2.1cm 2.7cm 1cm 3.0cm]{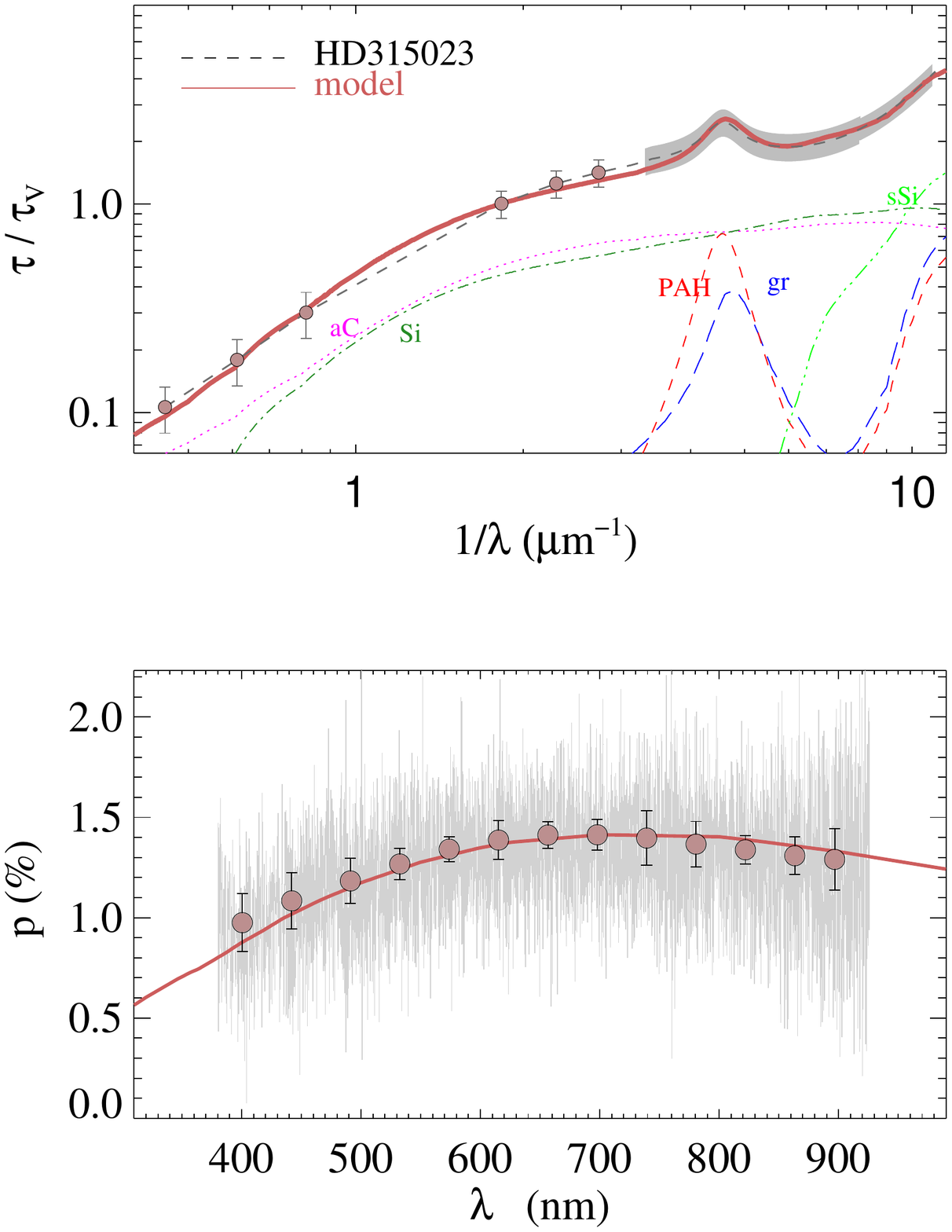}
\caption{Notation same as in Fig.~\ref{appstart.fig}  for HD~315023.}
\end{figure}

\begin{figure} [h!tb]
\includegraphics[width=9.0cm,clip=true,trim=2.1cm 2.7cm 1cm 3.0cm]{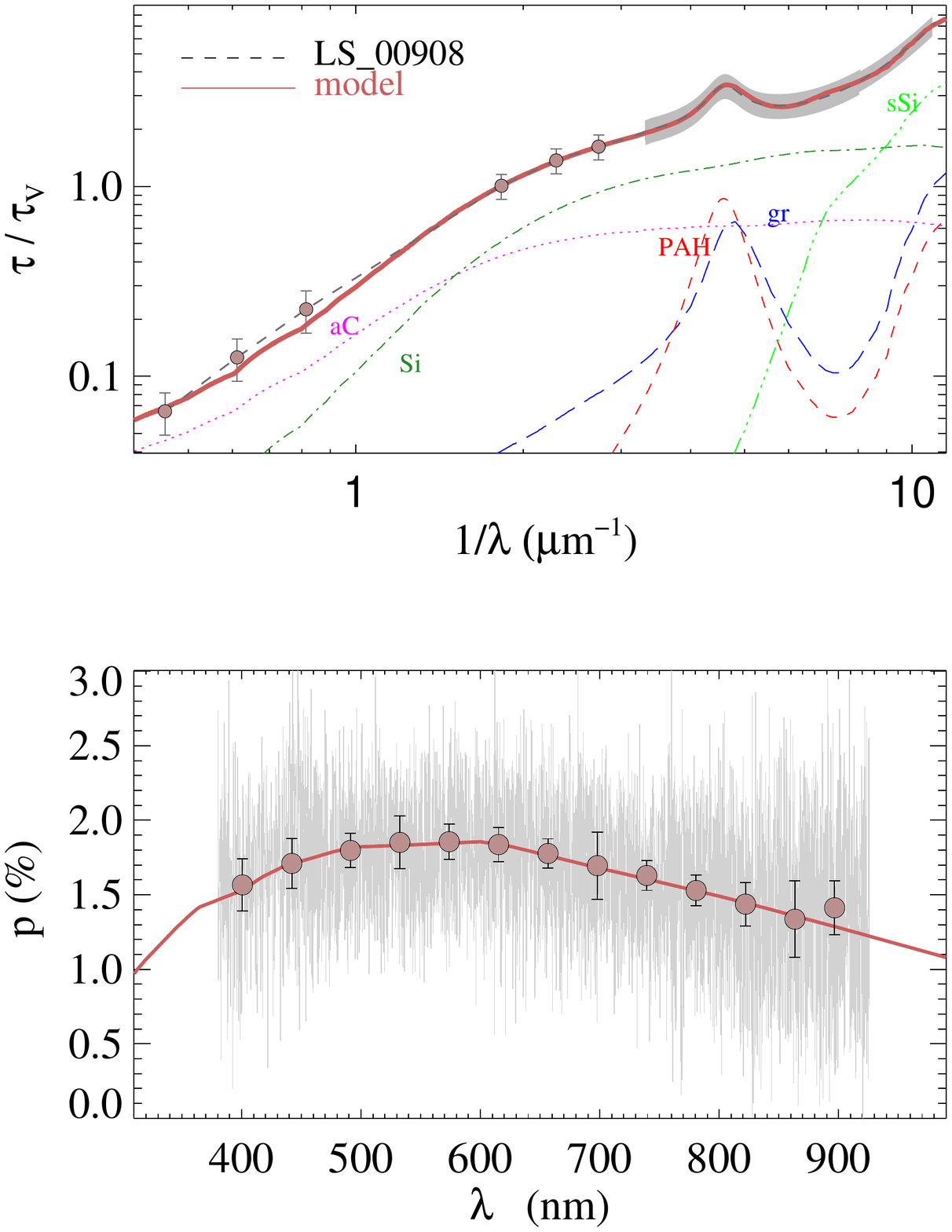}
\caption{Notation same as in Fig.~\ref{appstart.fig}  for LS--908.}
\end{figure}


\end{appendix} 
\end{document}